\documentstyle[12pt]{report}
\newcommand{\bm}[1]{\mbox{\boldmath$#1$}}
 
\renewcommand{\theequation}
{\arabic{chapter}.\arabic{section}.\arabic{equation}}

\begin{document}

\begin{center}

\vspace*{2cm}

\vspace{2cm}

{\Huge  Stability of Bose-Einstein Condensates Confined in Traps}

\vspace{2cm}

{\Large TAKEYA TSURUMI, HIROFUMI MORISE and MIKI WADATI}

\vspace{1.5cm}

{\large\it Department of Physics, Graduate School of Science, 
University of Tokyo, Hongo 7-3-1, Bunkyo-ku, Tokyo 113-0033, \\
Japan}

\vspace{0.5cm}

\vspace{2.5cm}

{\bf Abstract}

\end{center}

Bose-Einstein condensation has been realized in dilute atomic
vapors. This achievement has generated immerse interest in this field.
Presented is a review of recent theoretical research into the properties 
of trapped dilute-gas Bose-Einstein condensates. Among them, stability 
of Bose-Einstein condensates confined in traps is mainly
discussed. Static properties of the ground state are investigated by
use of the variational method. The anlysis is extended to the
stability of two-component condensates. Time-development of the
condensate is well-described by the Gross-Pitaevskii equation which is 
known in nonlinear physics as the nonlinear Schr\"odinger equation.
For the case that the inter-atomic potential is effectively
attractive, a singularity of the solution emerges in a
finite time. This phenomenon which we call collapse explains the upper 
bound for the number of atoms in such condensates under traps.

\tableofcontents

\chapter{Introduction}
\label{chap:Introduction}
\setcounter{equation}{0}

    The Bose-Einstein condensation (BEC)~\cite{Bose}--\cite{GriSnoStr} 
is one of the most remarkable phenomena 
in quantum many-body systems. The condensation is a logical consequence 
of the Bose-Einstein statistics: a gas of non-interacting bosonic 
atoms, below a certain temperature, suddenly develops a macroscopic 
population in the lowest energy quantum mechanical state.
However, BEC occurs at such low temperature 
that the dimensionless phase-space
density, $\rho_{{\rm ps}} \equiv n (\lambda_{{\rm db}})^{3}$, is
larger than 2.612, where $n$ is the number density of atoms, 
and $\lambda_{{\rm db}}$ is the thermal de Broglie wavelength,
$
\lambda_{{\rm db}} = h/(2\pi m k_{{\rm B}} T)^{1/2},
$
with $h$  Planck's constant, $m$  the mass of the atom, 
$k_{{\rm B}}$ Boltzmann's constant and $T$ the temperature,
respectively (see, for instance, \cite{Huang} as a text book).
Accordingly, for many years, BEC had been  regarded as a mathematical
artifact: In December 1924, Einstein wrote to
P. Ehrenfest, `From a certain temperature on, the molecules
``condense'' without attractive forces, that is, they accumulate at
zero velocity. The theory is pretty, but is there also some truth to
it?'~\cite{Pais}.  
Keesom observed in 1928 the phase transition between He-I (the nomal
fluid phase) 
and He-II (the superfluid phase) in liquid helium, $^{4}{\rm
  He}$~\cite{Keesom}.
F. London interpreted this phenomenon as BEC in 1938~\cite{London}. 
In liquid helium, however, the nature of the BEC transition is difficult to 
observe clearly because of strong inter-atomic interactions.
There exists a non-negligible difference between the observed
transition temperature $T_{\lambda} = 2.18 {\rm K}$ and the
estimated transition temperature $T_{{\rm c}} = 3.14 {\rm K}$
by an ideal gas approximation.
In 1993, the evidence of BEC in a gas of excitons in a semi-conductor
host has been observed~\cite{LinWolfe}. Although the interactions in these
systems are weak, little information about them is known, and thus 
it is difficult to understand the net effect of BEC in the exciton gas. 
In 1970's, efforts started to observe BEC in a dilute vapor of
spin-polarized hydrogen~\cite{GriSnoStr}, whose inter-atomic
interaction is sufficiently weak and well understood.
However,  BEC in the hydrogen gas was not achieved for a long time, 
because of the existence 
of inelastic inter-atomic collisions, which cause trap loss and heating.

   Recently, 
combinations of various new technologies developed in atomic physics 
such as laser cooling and trapping~\cite{ChHoBjCaAs}, 
evaporation cooling and magnetic
trap have made it possible to increase the phase-space density 
of a vapor of alkali atoms,
and to observe BEC of such atoms in controllable situations. 
In 1995, BEC was observed in a series of experiments on vapors 
of $^{87}{\rm  Rb}$~\cite{AEMWC}, $^{7}{\rm  Li}$~\cite{BSTH,BSH2} 
and $^{23}{\rm  Na}$~\cite{DMADDKK} in which the atoms were confined 
in magnetic traps and cooled down to extremely low temperature 
of the order of $10^{-6}\sim 10^{-7}{\rm K}$ 
(Table~\ref{tab:BEC})~\cite{PW}--\cite{InStWi}.
\begin{table}[htb]
\begin{center}
\framebox[55mm]{\rule[-21mm]{0mm}{43mm}}
\end{center}
\caption{Experimental values of parameters. $a$; $s$-wave scattering
length, $T_{{\rm c}}$; critical temperature, $N_{{\rm t}}$; 
the total number of 
atoms in trap, $\omega$; trap frequency. Note that the total number of the
atoms is the sum of normal and condensed ones.
The values of $T_{{\rm c}}$, $N_{{\rm t}}$ and $\omega$ are taken from 
the reference in the right most column.
}
\label{tab:BEC}
\end{table}
Rigorously speaking, for an ideal gas of $N$ bosonic atoms in a harmonic
potential, the condition $n (\lambda_{{\rm db}})^{3} > 2.612$ is
replaced by $N (\hbar \bar{\omega})^{3}/(k_{{\rm B}} T)^{3} > 1.202$ 
where $\bar{\omega} \equiv (\omega_{x}\omega_{y}\omega_{z})^{1/3}$ is
the geometric mean of the harmonic trap frequencies (Appendix~\ref{chap:App3}).
In 1998, BEC of spin-polarized hydrogen atoms 
was finally observed~\cite{Hydrogen}.
Now, more than twenty groups have succeeded in observing BEC 
~\cite{HWCH}--\cite{HaBuLiDuBuGo}.

      The experimental realization of BEC of alkali-atoms
has stimulated experimental and theoretical research 
in their physical properties.
One of the advantages using confined atomic vapors is the
various choices of atomic systems.  
The low energy properties are characterized by the $s$-wave 
scattering length $a$. 
While $^{87}{\rm Rb}$, $^{23}{\rm Na}$ and $^{1}{\rm H}$ atoms have
positive $s$-wave scattering lengths~\cite{GCMHBV}--\cite{FrEt}, 
it is known that $^{7}{\rm Li}$ and $^{85}{\rm Rb}$ atoms have
negative $s$-wave scattering lengths,  
corresponding to low-energy attractive interactions~\cite{GCMHBV,AMSH}.
In the homogeneous case, it was predicted that the condensate
is unstable in three-dimension when the inter-particle interactions are
attractive~\cite{NP}. 
The stability of the condensate under magnetic trap has been 
studied numerically and analytically~\cite{RHBE}--\cite{KiZu}.
The condensate was predicted to be (meta)stable under the magnetic trap, 
only when the number of atoms is below some critical number. 
The estimated critical number for 
$^{7}{\rm Li}$ atoms is about $10^3$, which agrees with the 
observed upper bound of the number of atoms in the condensate~\cite{BSH2}.

The overlapping condensates 
of the two different spin states of $^{87}{\rm Rb}$ in a magnetic 
trap were observed~\cite{MBGCW}--\cite{Hall1998a},
which has also stimulated theoretical research 
of the multi-component condensates~\cite{HoSh}--\cite{MTW}.
Because of the repulsive interaction between the condensates 
and the gravitational force, two condensates were observed to be separated.
In the study of dynamics of component separation 
of this system~\cite{Hall1998a}, it was observed that    
the time-evolution is rather complex; the initial configuration 
quickly damps out and tends to preserve the total density profile. 
With respect to damping, 
effects of the excitations are not small and it is not yet 
certain what mechanism is responsible.
The time-independence of the total density profile  
is due to  the similarity in intraspecies and interspecies scattering lentghs  
in $^{87}{\rm Rb}$.

Further, 
optical trapping of the condensates of $^{23}{\rm Na}$ in several hyperfine 
states was demonstrated~\cite{StAnChInMiStKe}. 
In a magnetic trap, spins of atoms are polarized.
Thus, Bose-Einstein condensates 
are described in terms of scalar wavefunctions (order parameters).
In an optical trap, on the other hand, 
degrees of freedom of atomic spin are recovered,
and consequently the condensate wavefunction behaves as spinor.
By applying weak magnetic fields with spatial gradient, 
the degeneracy of the internal degree of freedom is lifted,  
and the freedom of spin orientation gives 
the formation of spin domains~\cite{StInStMiChKe,MiStStInChKe}. 
The domains can also be easily miscible by turn off the magnetic fields,
which is unlike the  experiments of $^{87}{\rm Rb}$~\cite{MBGCW}.
For this system, several theoretical works have also been 
done~\cite{Ho}--\cite{Yip}, where Skirmion~\cite{Ho},
spin texture~\cite{OhMa}, the dynamics of spin mixing~\cite{LaPuBi},
spin domain formation~\cite{IsMaOh}
and internal vortex structure~\cite{Yip} are studied.

\bigskip
\bigskip

    Keeping  the developments in mind, 
we study  static and dynamical properties 
of the Bose-Einstein condensates of atomic gases confined in traps.
In Chap.~\ref{chap:GPG}, a system of 
bosons with inter-particle interactions of finite range, 
trapped in an external potential and at very low temperature, are considered.
Applying the pseudopotential method~\cite{Huang,HY} 
and the mean field approach to the system, the Ginzburg-Pitaevskii-Gross
equation~\cite{GinzPita}--\cite{Gross2} is derived.
The time-dependent case, which we call the Gross-Pitaevskii equation,
is essentially equivalent to the nonlinear Schr\"odinger equation.
In Chap.~\ref{chap:Static Properties},
by use of the variational approach~\cite{Fetter,BP}, 
static properties of the Bose-Einstein condensates are discussed. 
The ground state properties 
of a condensate with repulsive or attractive 
inter-atomic interaction confined in axially symmetric magnetic trap
are investigated (Sec.~\ref{sec:AsymBec}). 
The analysis is extended to that of the stability of a
two-component Bose-Einstein condensate under magnetic traps,
where the possibility of phase separations is also discussed
(Sec.~\ref{sec:2com}).
While the inter-particle interaction is assumed to be of finite range
in the first two sections in this chapter, the properties of the Bose-Einstein 
condensate of long-ranged interacting bosons confined in traps
are considered in (and only in) Sec.~\ref{sec:LR}.
In Chap.~\ref{chap:Dynamical Properties},
dynamical properties of Bose-Einstein condensates are considered. 
A key idea is the extension of the Zakharov's theory~\cite{TW1}.
The stability of the wavefunction of the
$D$-dimensional nonlinear Schr\"odinger equation 
with harmonic potential terms is analyzed (Sec.~\ref{sec:StabNLS})
and is applied to investigate the instability of condensates with effectively 
attractive inter-atomic interaction (Sec.~\ref{sec:BecPL}).
There, a formula for the critical number of atoms, 
above which the collapse of the condensate occurs, is derived. 
The last chapter is devoted to summary and concluding remarks.

\clearpage

\chapter{Ginzburg-Pitaevskii-Gross Equation}
\label{chap:GPG}
\setcounter{equation}{0}

In this chapter, we present the basic equation 
to describe Bose-Einstein condensates.
In Sec.~\ref{sec:Time-Independent Case},
following Ref.~\cite{Gross2} based on the Hartree approximation,
we derive the Ginzburg-Pitaevskii-Gross
equation~\cite{GinzPita}--\cite{Gross2},
which is essentially equivalent to 
the time-independent nonlinear Schr\"odinger equation.
In Sec.~\ref{sec:Time-Dependent Case},
we consider the time-dependent case. 
To derive the equation of motion for the condensate, 
we begin with the second-quantized formulation.   
By employing the mean field approach,
we obtain the so-called Gross-Pitaevskii
equation~\cite{P}--\cite{Gross2} or equivalently  
the (time-dependent) nonlinear Schr\"odinger equation.

\section{Time-independent case}
\label{sec:Time-Independent Case}
\setcounter{equation}{0}

  We consider $N$ identical bosonic particles 
with inter-particle interactions of finite range, 
trapped in an external potential, $V(\bm{r})$.
We assume that the gas is sufficiently dilute and at very low temperature.
In such a situation, two-body interaction is dominant and the $s$-wave
part plays a central role.
We may replace the scattering from an inter-particle potential 
of finite range by that from a hard-sphere of diameter $a$, 
which is identical  to  $s$-wave scattering length in this case. 
Then, the  Hamiltonian operator of the system of hard-spheres can be given
in certain approximations, the detail of which we explain 
in the Appendix~\ref{chap:App1}, 
by the pseudopotential Hamiltonian operator~\cite{Huang,HY},
\begin{equation} 
H = \sum_{i=1}^{N}
\left(
-\frac{\hbar^{2}}{2m}\Delta_{i} + V(\bm{r}_{i})
\right)
+\frac{1}{2}\sum_{i \neq j} 
U_{0} \, \delta (\bm{r}_{i}-\bm{r}_{j}) \,
\frac{\partial}{\partial r_{ij}}
\left(r_{ij} \, \cdot\right),
\label{eqn:Hamiltonian of N-bosons in GPG}
\end{equation}
where $m$ is the mass of a bosonic particle, and 
\begin{eqnarray}
\Delta_{i} & \equiv & 
\frac{\partial^{2}}{\partial x_{i}^{2}}
+\frac{\partial^{2}}{\partial y_{i}^{2}}
+\frac{\partial^{2}}{\partial z_{i}^{2}},
\label{eqn:def of Deltai in GPG}
\\
r_{ij} & \equiv & \left| \bm{r}_{i}-\bm{r}_{j}\right|,
\label{eqn:def of rij in GPG}
\\
U_{0} & \equiv & 4 \pi \hbar^{2}a/m.
\label{eqn:def of U0 in GPG}
\end{eqnarray}
The magnetic trap is well approximated by a harmonic potential,
\begin{equation}
V (\bm{r}) =  
\frac{m}{2}\left(\omega_{x}^{2}x^{2}+\omega_{y}^{2}y^{2}+\omega_{z}^{2}z^{2}
\right), 
\label{eqn:harmonic potential in GPG}
\end{equation}
with $(\omega_{x},\omega_{y},\omega_{z})$ being trap frequencies.

    In the ground state of the system, almost all bosons 
may occupy the lowest single-particle state because 
of sufficiently weak inter-particle interactions.
Thus, following the Hartree approximation, 
we write the ground state wavefunction, 
$\Phi_{0} (\bm{r}_{1},\bm{r}_{2},\cdot \cdot \cdot
,\bm{r}_{N})$, in terms of the product of 
$N$ single-particle state wavefunctions, $g (\bm{r})$,
\begin{equation}
\Phi_{0} (\bm{r}_{1},\bm{r}_{2},\cdot \cdot \cdot ,\bm{r}_{N})
=
\prod_{i=1}^{N}g (\bm{r}_{i}),
\label{eqn:wavefunc. of system in GPG}
\end{equation}
where $g (\bm{r})$ is normalized as
\begin{equation} 
\langle g|g \rangle
\equiv
\int {\rm d}\bm{r}
|g (\bm{r})|^{2}
=1,
\label{eqn:normalization of g in GPG}
\end{equation}
and thus the norm of the wavefunction $\Phi_{0}$, defined by 
$\langle\Phi_{0} | \Phi_{0}\rangle$,
\begin{equation}
\langle\Phi_{0} | \Phi_{0}\rangle
\equiv 
\int {\rm d}\bm{r}_{1} \cdot \cdot \cdot \int {\rm d}\bm{r}_{N}
|\Phi_{0} (\bm{r}_{1} \cdot \cdot \cdot \bm{r}_{N})|^{2}
=
\left(
\int {\rm d}\bm{r}
|g (\bm{r})|^{2}
\right)^{N},
\label{eqn:normalization cond. for Phi in GPG}
\end{equation}
is equal to unity.
From Eqs.~(\ref{eqn:Hamiltonian of N-bosons in GPG}) 
and (\ref{eqn:wavefunc. of system in GPG}), we have
\begin{eqnarray}
\lefteqn{
\langle \Phi_{0} |H| \Phi_{0} \rangle
}
\nonumber \\
& = & 
\sum_{i=1}^{N}
\int {\rm d}\bm{r}_{1} \cdot \cdot \cdot \int {\rm d}\bm{r}_{N}
\Phi_{0}^{*} 
\left(
-\frac{\hbar^{2}}{2m}\Delta_{i} + V(\bm{r}_{i})
\right)
\Phi_{0} 
\nonumber \\
& & 
+ 
\frac{U_{0}}{2}
\sum_{i \neq j} 
\int {\rm d}\bm{r}_{1} \cdot \cdot \cdot \int {\rm d}\bm{r}_{N}
\Phi_{0}^{*} 
\delta (\bm{r}_{i}-\bm{r}_{j}) \,
\frac{\partial}{\partial r_{ij}}
\left(r_{ij} 
\Phi_{0} 
\right)
\nonumber \\
& = & 
N
\int {\rm d}\bm{r}
g^{*}(\bm{r}) 
\left(
-\frac{\hbar^{2}}{2m}\Delta + V(\bm{r})
\right)
g (\bm{r}) 
\nonumber \\
& & + 
\frac{N(N-1)}{2} U_{0}
\int {\rm d}\bm{r}_{1} \int {\rm d}\bm{r}_{2}
g^{*} (\bm{r}_{2}) g^{*} (\bm{r}_{1})
\delta (\bm{r}_{1}-\bm{r}_{2}) \,
\frac{\partial}{\partial r_{12}}
\left[r_{12} 
g (\bm{r}_{1}) g (\bm{r}_{2})
\right],
\label{eqn:def of H in GPG}
\end{eqnarray}
where the superscript $*$ means the complex conjugate and
\begin{equation}
\Delta \equiv  
\frac{\partial^{2}}{\partial x^{2}}
+\frac{\partial^{2}}{\partial y^{2}}
+\frac{\partial^{2}}{\partial z^{2}}.
\label{eqn:def of Delta in GPG}
\end{equation}
In Eq.~(\ref{eqn:def of H in GPG}), if the product of 
two single-state wavefunctions 
$g (\bm{r}_{1}) g (\bm{r}_{2})$ is not singular  
at $r_{12} = 0$, the operator $\left( \partial /\partial r_{12}\right)
\left(r_{12} \, \cdot \right)$ can be set equal to unity, which gives
\begin{eqnarray}
\langle \Phi_{0} |H| \Phi_{0} \rangle
& = & 
N
\int {\rm d}\bm{r}
g^{*}(\bm{r}) 
\left(
-\frac{\hbar^{2}}{2m}\Delta + V(\bm{r})
\right)
g (\bm{r}) 
\nonumber \\
& & + 
\frac{N(N-1)}{2} U_{0}
\int {\rm d}\bm{r} \int {\rm d}\bm{r}'
g^{*} (\bm{r}') g^{*} (\bm{r})
\delta (\bm{r}-\bm{r}') \,
g (\bm{r}) g (\bm{r}')
\nonumber \\
& = & 
N \int {\rm d}\bm{r}
\left[
g^{*}(\bm{r}) 
\left(
-\frac{\hbar^{2}}{2m}\Delta + V(\bm{r})
\right)
g (\bm{r}) 
+ \frac{N-1}{2} U_{0}
\left| g (\bm{r}) \right|^{4}
\right].
\label{eqn:H No.2 in GPG}
\end{eqnarray}

      We minimize the functional $\langle \Phi_{0}|H|\Phi_{0} 
\rangle$ (\ref{eqn:H No.2 in GPG}) under the constraint:
\begin{equation}
\langle\Phi_{0} | \Phi_{0} \rangle = 1.
\label{eqn:constraint in GPG}
\end{equation}
To find the constrained extremum of $\langle \Phi_{0} |H|
\Phi_{0} \rangle$, 
we set the variation of a functional $\langle \Phi_{0} |H| \Phi_{0} \rangle 
-\mu \langle\Phi_{0} | \Phi_{0}\rangle$ equal to  zero,
\begin{equation}
\frac{\delta}{\delta g^{*}(\bm{r})}
\left(\langle \Phi_{0} |H| \Phi_{0} \rangle
-\mu \, \langle\Phi_{0} | \Phi_{0}\rangle\right) = 0,
\label{eqn:constrained extremum cond in GPG}
\end{equation}
where $\mu$ is a Lagrange's multiplier.
Substituting Eqs.~(\ref{eqn:normalization cond. for Phi in GPG}) and 
(\ref{eqn:H No.2 in GPG}) 
into Eq.~(\ref{eqn:constrained extremum cond in GPG}), 
we obtain
\begin{equation}
-\frac{\hbar^{2}}{2m}\Delta g (\bm{r}) 
+ V(\bm{r})g(\bm{r}) 
+ (N-1)U_{0} \left| g (\bm{r}) \right|^{2} g (\bm{r})
=
\mu g (\bm{r})
.
\label{eqn:GPG eq in GPG}
\end{equation}
Introducing a wavefunction $\Psi (\bm{r})$,
\begin{equation}
\Psi (\bm{r}) \equiv N^{1/2} g (\bm{r}),
\label{eqn:def of Psi in GPG}
\end{equation}
we get
\begin{equation}
-\frac{\hbar^{2}}{2m}\Delta \Psi (\bm{r}) 
+ V(\bm{r})\Psi (\bm{r}) 
+ \left(1-\frac{1}{N}\right)
U_{0} \left| \Psi (\bm{r}) \right|^{2} \Psi (\bm{r})
=
\mu \Psi (\bm{r})
,
\label{eqn:GPG No.2 eq in GPG}
\end{equation}
with
\begin{equation} 
\langle \Psi | \Psi \rangle
\equiv
\int {\rm d}\bm{r}
|\Psi (\bm{r})|^{2}
=N.
\label{eqn:normalization of psi in GPG}
\end{equation}
The wavefunction $\Psi$ has many names; an order parameter, 
a macroscopic wavefunction and  a Bose-Einstein condensate wavefunction. 
For sufficiently large $N$, we can neglect the $1/N$-order term
appearing 
in the left hand side of Eq.~(\ref{eqn:GPG No.2 eq in GPG}).
The result is
\begin{equation}
-\frac{\hbar^{2}}{2m}\Delta \Psi (\bm{r}) 
+ V(\bm{r})\psi (\bm{r}) 
+ U_{0} \left| \Psi (\bm{r}) \right|^{2} \Psi (\bm{r})
=
\mu \Psi (\bm{r})
.
\label{eqn:GPG No.3 eq in GPG}
\end{equation}
We refer to  Eq.~(\ref{eqn:GPG No.3 eq in GPG})
as the Ginzburg-Pitaevskii-Gross
equation~\cite{GinzPita}--\cite{Gross2} with an external potential term.
This equation may also be referred to as 
the time-independent nonlinear Schr\"odinger equation.

    From Eqs.~(\ref{eqn:H No.2 in GPG}) and (\ref{eqn:def of Psi in GPG}), 
the ground state energy of the system, $E$, is
\begin{eqnarray}
E 
& = &
N \int {\rm d}\bm{r}
\left[
\frac{\hbar^{2}}{2m}\left|\nabla g(\bm{r})\right|^{2}
+ V(\bm{r}) \left|g(\bm{r})\right|^{2}
+ \left(N-1\right)\frac{U_{0}}{2} 
\left| g(\bm{r}) \right|^{4}
\right]
\nonumber \\
& = &
\int {\rm d}\bm{r}
\left[
\frac{\hbar^{2}}{2m}\left|\nabla \Psi (\bm{r}) \right|^{2}
+ V(\bm{r}) \left|\Psi (\bm{r})\right|^{2}
+ \left(1-\frac{1}{N}\right)\frac{U_{0}}{2} 
\left| \Psi (\bm{r})\right|^{4}
\right],
\label{eqn:ground state energy in GPG}
\end{eqnarray}
which we call the Ginzburg-Pitaevskii-Gross energy
functional~\cite{GinzPita}--\cite{Gross2} 
with an external potential term.
On the other hand,  multiplying  the both sides of 
Eq.~(\ref{eqn:GPG No.2 eq in GPG}) by $\Psi^{*}(\bm{r})$
and integrating it, we have
\begin{eqnarray}
N \mu 
& = &
\int {\rm d}\bm{r}
\left[
\frac{\hbar^{2}}{2m}\left|\nabla \Psi\right|^{2}
+ V(\bm{r}) \left|\Psi\right|^{2}
+ \left(1-\frac{1}{N}\right)U_{0} 
\left| \Psi  \right|^{4}
\right]
\nonumber \\
& = &
E + 
\left(1-\frac{1}{N}\right)\frac{U_{0}}{2} 
\int {\rm d}\bm{r}
\left| \Psi  \right|^{4}.
\label{eqn:mu in GPG}
\end{eqnarray}

     Another derivation of Eq.~(\ref{eqn:mu in GPG}) may be instructive. 
We give a small change to the single-particle
wavefunction $g(\bm{r})$, and denote the resultant by $\tilde{g}(\bm{r})$,
\begin{equation}
\tilde{g} (\bm{r})
\equiv
(1+\epsilon)^{1/2}g (\bm{r}),
\label{eqn:def of tilde g}
\end{equation}
where $\epsilon$ is a sufficiently small constant and $g(\bm{r})$
satisfies Eqs.~(\ref{eqn:normalization of g in GPG}) 
and (\ref{eqn:GPG eq in GPG}). Similar to  (\ref{eqn:def of tilde g}), 
we also define
\begin{equation}
\tilde{\Psi} (\bm{r})
\equiv
(1+\epsilon)^{1/2}\Psi (\bm{r}),
\label{eqn:def of tilde Psi}
\end{equation}
where $\Psi$ satisfies Eq.~(\ref{eqn:GPG No.2 eq in GPG}).
From the normalization condition of $\Psi (\bm{r})$ 
(\ref{eqn:normalization of psi in GPG}),
we have 
\begin{equation} 
\langle \tilde{\Psi} | \tilde{\Psi} \rangle
\equiv
\int {\rm d}\bm{r}
|\tilde{\Psi} (\bm{r})|^{2}
=N + \epsilon \, N.
\label{eqn:normalization of tilde Psi in GPG}
\end{equation}
In (\ref{eqn:normalization of tilde Psi in GPG}), 
we can regard $\epsilon \, N$ as a virtual change of the number of
particles, $\delta N$,
\begin{equation}
\delta N \equiv \epsilon \, N.
\label{eqn:def of delta N}
\end{equation}
From (\ref{eqn:wavefunc. of system in GPG})
with (\ref{eqn:def of tilde g}), 
we construct a  transformed 
wavefunction of the system, $\tilde{\Phi}_{0}$,
\begin{equation}
\tilde{\Phi}_{0} 
\equiv
\prod_{i=1}^{N}\tilde{g} (\bm{r}_{i}),
\label{eqn:tilde Phi0 of system in GPG}
\end{equation}
Then, the norm of the wavefunction $\tilde{\Phi}_{0}$ is 
\begin{eqnarray}
\langle\tilde{\Phi}_{0} | \tilde{\Phi}_{0}\rangle
& = &  
\left(
\int {\rm d}\bm{r}
|\tilde{g} (\bm{r})|^{2}
\right)^{N}
\nonumber \\
& = & 
(1 + \epsilon)^{N}
\nonumber \\
& \approx &
1 + \epsilon \, N = 1 + \delta N.
\label{eqn:normalization cond. for tildePhi in GPG}
\end{eqnarray}
From (\ref{eqn:constraint in GPG})
and (\ref{eqn:normalization cond. for tildePhi in GPG}),
we obtain a variation of the norm of $\Phi_{0}$, 
\begin{equation}
\delta \langle \Phi_{0} | \Phi_{0}\rangle
\equiv
\langle\tilde{\Phi}_{0} | \tilde{\Phi}_{0}\rangle
-
\langle \Phi_{0} | \Phi_{0}\rangle
= \delta N.
\label{eqn:variation of Norm in GPG}
\end{equation}
The corresponding change of the energy is calculated as follows.
By replacing $\Psi$ with $\tilde{\Psi}$
in (\ref{eqn:ground state energy in GPG}), we get 
\begin{equation}
\int {\rm d}\bm{r}
\left[
\frac{\hbar^{2}}{2m}\left|\nabla \tilde{\Psi} (\bm{r}) \right|^{2}
+ V(\bm{r}) \left|\tilde{\Psi} (\bm{r})\right|^{2}
+ \left(1-\frac{1}{N}\right)\frac{U_{0}}{2} 
\left| \tilde{\Psi} (\bm{r})\right|^{4}
\right]
\approx 
E + \delta E,
\label{eqn:E + delta E in GPG}
\end{equation}
where
\begin{equation}
\delta E
\equiv
\frac{\delta N}{N}
\int {\rm d}\bm{r}
\left[
\frac{\hbar^{2}}{2m}\left|\nabla \Psi\right|^{2}
+ V(\bm{r}) \left|\Psi\right|^{2}
+ \left(1-\frac{1}{N}\right)U_{0} 
\left| \Psi  \right|^{4}
\right].
\label{eqn:def of delta E}
\end{equation}

Because $\Psi$ satisfies Eq.~(\ref{eqn:GPG No.2 eq in GPG}),
the variations $\delta N$ (\ref{eqn:variation of Norm in GPG})
and $\delta E$ (\ref{eqn:def of delta E}) satisfy
the extremum condition 
(\ref{eqn:constrained extremum cond in GPG}), and thus we have
\begin{equation}
\delta E - \mu \, \delta N = 0.
\label{eqn:delta E - mu delta N = 0 in GPG}
\end{equation}
Substituting (\ref{eqn:def of delta E}) 
into (\ref{eqn:delta E - mu delta N = 0 in GPG}), we get
\begin{equation}
\mu  =  \frac{\delta E}{\delta N}
= 
\frac{1}{N}
\int {\rm d}\bm{r}
\left[
\frac{\hbar^{2}}{2m}\left|\nabla \Psi\right|^{2}
+ V(\bm{r}) \left|\Psi\right|^{2}
+ \left(1-\frac{1}{N}\right)U_{0} 
\left| \Psi  \right|^{4}
\right],
\label{eqn:mu No. 2 in GPG}
\end{equation}
which is the same as (\ref{eqn:mu in GPG}).
From (\ref{eqn:mu No. 2 in GPG}), it is clear that $\mu$, 
introduced as a Lagrange's multiplier, 
has a meaning of the chemical potential of the system.


\section{Time-dependent case}
\label{sec:Time-Dependent Case}
\setcounter{equation}{0}

     In this sub-section, 
we consider the time-dependent case, $\Psi = \Psi (\bm{r},t)$. 
To derive the equation of motion of $\Psi$, 
we use the second-quantized formulation.   
Let $\hat{\Psi}(\bm{r},t)$ and $\hat{\Psi}^{\dag}(\bm{r},t)$
denote the bosonic annihilation and creation operators, respectively.
The (equal-time) commutation relations among the operators are
\begin{eqnarray}
\left[\hat{\Psi}(\bm{r},t),\hat{\Psi}(\bm{r}',t)\right]
& \equiv & 
\hat{\Psi}(\bm{r},t) \hat{\Psi}(\bm{r}',t)
-\hat{\Psi}(\bm{r}',t) \hat{\Psi}(\bm{r},t) = 0,
\label{eqn:comm rel between Psi and Psi in GPG}
\\
\left[\hat{\Psi}(\bm{r},t),\hat{\Psi}^{\dag}(\bm{r}',t)\right] 
& = & \delta (\bm{r} - \bm{r}'),
\label{eqn:comm rel between Psi and PsiDag in GPG}
\end{eqnarray}
and the second-quantized Hamiltonian of the system, $\hat{H}$, 
can be written as 
\begin{equation}
\hat{H}
\equiv 
\int
{\rm d} \bm{r}
\left[
\frac{\hbar^{2}}{2m}\nabla \hat{\Psi}^{\dag} \cdot \nabla \hat{\Psi} 
+ V (\bm{r}) \hat{\Psi}^{\dag} \hat{\Psi} 
+ \frac{U_{0}}{2}
\hat{\Psi}^{\dag}\hat{\Psi}^{\dag}\hat{\Psi}\hat{\Psi}
\right].
\label{eqn:second-quantized Hamiltonian in GPG}
\end{equation}

      The time-evolution of the operator $\hat{\Psi}(\bm{r},t)$ 
 obeys the Heisenberg equation, 
\begin{equation}
{\rm i} \hbar \frac{\partial }{\partial t} \hat{\Psi}
 =  \left[\hat{\Psi},\hat{H}\right].
\label{eqn:Heisenberg eq in GPG}
\end{equation}
Substituting the Hamiltonian~(\ref{eqn:second-quantized Hamiltonian in GPG})
into Eq.~(\ref{eqn:Heisenberg eq in GPG}) 
and using the commutation 
relations~(\ref{eqn:comm rel between Psi and Psi in GPG}) 
and (\ref{eqn:comm rel between Psi and PsiDag in GPG}), we get
\begin{equation}
{\rm i} \hbar \frac{\partial }{\partial t} \hat{\Psi}
= 
- \frac{\hbar^{2}}{2m} \Delta \hat{\Psi} 
+ V (\bm{r}) \hat{\Psi} 
+ U_{0} \hat{\Psi}^{\dag}\hat{\Psi}\hat{\Psi}.
\label{eqn:Heisenberg eq  No2 in GPG}
\end{equation}

      We denote an expectation value by $\left\langle \,\cdot\, \right\rangle$.
The Heisenberg equation for
the bosonic operator (\ref{eqn:Heisenberg eq No2 in GPG}) gives 
\begin{equation}
{\rm i} \hbar \frac{\partial }{\partial t}
\left\langle\hat{\Psi}\right\rangle 
= 
- \frac{\hbar^{2}}{2m} \Delta \left\langle\hat{\Psi}\right\rangle 
+ V (\bm{r}) \left\langle\hat{\Psi}\right\rangle 
+ U_{0} \left\langle\hat{\Psi}^{\dag}\hat{\Psi}\hat{\Psi}\right\rangle 
.
\label{eqn:expectation value of Heisenberg eq in GPG}
\end{equation}
According to the mean field theory~\cite{PW,DaGiPiSt}, we may replace 
expectation values of the bosonic annihilation and creation operators 
by the condensate wavefunction $\Psi (\bm{r},t)$ and its complex conjugate 
$\Psi^{*} (\bm{r},t)$ respectively, 
\begin{equation}
\left\langle \hat{\Psi} (\bm{r},t) \right\rangle = \Psi (\bm{r},t),
\,\,\,\,\,\,\,
\left\langle \hat{\Psi}^{\dag} (\bm{r},t) \right\rangle = \Psi^{*} (\bm{r},t).
\label{eqn:def of order parameter in GPG}
\end{equation}
For the third  term in the right hand side of 
Eq.~(\ref{eqn:expectation value of Heisenberg eq in GPG}),
we take the following approximation,  
\begin{equation}
\left\langle\hat{\Psi}^{\dag}\hat{\Psi}\hat{\Psi}\right\rangle
\approx 
\left\langle\hat{\Psi}^{\dag}\right\rangle
\left\langle\hat{\Psi}\right\rangle
\left\langle\hat{\Psi}\right\rangle
= 
\left|\Psi\right(\bm{r},t)|^{2}\Psi (\bm{r},t).
\label{eqn:approx for PsiDag Psi Psi in GPG}
\end{equation}
Substituting (\ref{eqn:def of order parameter in GPG}) and
(\ref{eqn:approx for PsiDag Psi Psi in GPG}) 
into Eq.~(\ref{eqn:expectation value of Heisenberg eq in GPG}), 
we  have
\begin{equation}
{\rm i} \hbar \frac{\partial}{\partial t} \Psi (\bm{r},t) 
=
-\frac{\hbar^{2}}{2m}\Delta \Psi (\bm{r},t) 
+ V(\bm{r}) \Psi (\bm{r},t)
+U_{0} \left| \Psi (\bm{r},t)\right|^{2} \Psi (\bm{r},t),
\label{eqn:GP eq in GPG}
\end{equation}
which is called the Gross-Pitaevskii
equation~\cite{P}--\cite{Gross2} with an external potential
or the (time-dependent) nonlinear Schr\"odinger equation.
The one-dimensional nonlinear  Schr\"odinger equation
is known to be integrable and has been studied extensively related to
various areas of physics~\cite{ZakharovShabat,AS}. 
The equation~(\ref{eqn:GP eq in GPG}) can be written in a variational form,
\begin{equation}
{\rm i} \hbar \frac{\partial \Psi}{\partial t} 
= \frac{\delta }{\delta \Psi^{*}}E[\Psi],
\label{eqn:eq of mn for Psi(r,t) in GPG}
\end{equation}
where the functional $E[\,\cdot\,]$ is defined 
by Eq.~(\ref{eqn:ground state energy in GPG})
with the $1/N$-order term deleted.
We note that, by setting
\begin{equation}
\Psi (\bm{r},t) = \exp (-{\rm i}\,\mu \,t) \, \Psi (\bm{r}),
\label{eqn:stationary setting for Psi in GPG}
\end{equation}
in Eq.~(\ref{eqn:GP eq in GPG}), 
we obtain the Ginzburg-Pitaevskii-Gross
equation~(\ref{eqn:GPG No.3 eq in GPG}) again.

   The existence of non-zero $\Psi (\bm{r},t) 
= \left\langle \hat{\Psi} (\bm{r},t) \right\rangle$ 
has an important meaning in physics, 
the breakdown of the gauge symmetry. One of the consequences is that 
the condensate has a definite phase $\theta (\bm{r},t)$ as defined by
$\Psi (\bm{r},t) = f(\bm{r},t) \exp [{\rm i} \, \theta (\bm{r},t)]$ 
where $f$ and $\theta$ are real functions.

\clearpage

\chapter{Static Properties of Bose-Einstein Condensates}
\label{chap:Static Properties}
\setcounter{equation}{0}
\setcounter{figure}{0}

In this chapter, we consider the static properties of Bose-Einstein
condensates.
In Sec.~\ref{sec:AsymBec}, by using the variational approach,
we examine the ground state properties 
of a Bose-Einstein condensate with repulsive or attractive 
inter-atomic interaction confined in axially symmetric magnetic trap~\cite{WT2}.
Employing a gaussian trial wavefunction to describe the condensate, 
we derive the minimum conditions of
the Ginzburg-Pitaevskii-Gross energy
functional~(\ref{eqn:ground state energy in GPG}).
In the repulsive inter-atomic interaction case, it is shown that, 
if the trap has a high asymmetry and thus the equi-potential surface
of the trap is  ``cigar" or ``pancake" shaped,
we can assume different approximations to each of 
axial and transverse directions.
We also compare the energy with the
one obtained from the Thomas-Fermi approximation~\cite{BP}.
In the attractive interaction case, different from the repulsive case, 
the energy function has only a local minimum. The local minimum of 
the energy no longer exists above some critical number of atoms.
This can be regarded as a static theory of the collapse.
We calculate the critical number for three shapes of 
the equi-potential surface
of the trap: almost sphere, cigar and pancake.

    In Sec.~\ref{sec:2com},
we extend the analysis in Sec.~\ref{sec:AsymBec} 
to discuss the stability of a
two-component Bose-Einstein condensate under magnetic traps~\cite{MTW}.
We first investigate the case where the numbers of two 
species of atoms are the same.  
The effect of interspecies interaction on the critical number
 of atoms is explicitly shown. 
We also present phase diagrams for various combinations of 
the scattering lengths.  
We further develop the variational approach 
and consider the phase separation of the condensate.

Inter-particle interactions of finite range are well approximated by the delta function.
We may think of the long-range interaction
like a Coulomb potential, because of the possibility 
of Bose-Einstein condensation for 
all bosonic atoms and molecules with and without charges in future.
Then, in Sec.~\ref{sec:LR} (and only in this section), 
we consider the properties of the Bose-Einstein 
condensate of long-ranged interacting bosons confined in traps.
As in Sec.~\ref{sec:AsymBec}, we derive the minimum conditions of the
energy, and obtain the ground state energy.
It is interesting to see that the condensate of long-range
interacting bosons under traps is stable
both for repulsive and attractive cases.

\section{Ground state of a Bose-Einstein condensate under axially
symmetric magnetic trap}
\label{sec:AsymBec}

The ground state properties 
of a Bose-Einstein condensate with repulsive or attractive 
inter-atomic interaction confined in axially symmetric magnetic trap
are studied through the variational approach~\cite{WT2}.
It is shown that, 
if the trap has a high asymmetry and thus the equi-potential surface
of the trap is  of ``cigar" or ``pancake" shape,
the properties of the condensate are drastically changed
from the case of isotropic (spherically symmetric) trap.


\subsection{Formulation}
\label{sec:Formulation in AsymBec}

We consider the ground state properties 
of a Bose-Einstein condensate confined in axially symmetric magnetic trap,
\begin{equation}
V(\bm{r})
= 
\frac{m}{2}  
(\omega_{\bot}^{2}r_{\bot}^{2}
+\omega_{z}^{2}z^{2}). 
\label{eqn:magnetic potential in AsymBec}
\end{equation}
Here, the axis of the symmetry is chosen to be the $z$-axis,
and $r_{\bot}$ denotes the radius of the projection 
of the position vector $\bm{r}$ on the $xy$-plane,
$m$ the atomic mass, and $\omega_{z}$ and $\omega_{\bot}$ 
the trap (angular) frequencies along the $z$-axis and in the $xy$-plane,
respectively.

    Following Baym and Pethick~\cite{BP} and Fetter~\cite{Fetter}
for isotropic case, 
we employ as the ground state wavefunction a gaussian trial function,
\begin{equation} 
\Psi (r_{\bot},z) 
 =  \left(\frac{N}{\pi^{3/2}d_{\bot}^{2} d_{z}}\right)^{1/2}
\exp [-(r_{\bot}^{2}/d_{\bot}^{2} + z^{2}/d_{z}^{2})/2],
\label{eqn:trial gaussian wavefuncion}
\end{equation} 
where $d_{\bot}$ and $d_{z}$ are variational parameters.
The macroscopic wavefunction of the condensate, $\Psi (\bm{r})$,
is assumed to be axially symmetric, 
as the magnetic trap is. 
This assumption is reasonable as far as the
system is near the ground state.
We have normalized the trial function (\ref{eqn:trial gaussian wavefuncion}) 
to the number of the particles in the condensate, $N$,
\begin{equation}
2 \pi 
\int_{0}^{\infty}  \, r_{\bot}{\rm d}r_{\bot}
\int_{-\infty}^{\infty}{\rm d}z
| \Psi (r_{\bot},z) |^{2}
=N.
\label{eqn:normalization of trial function in AsymBec}
\end{equation}
For the weakly interacting system at zero temperature,
the number of the non-condensate particles is considered to be negligible.
The parameters $d_{\bot}$ and $d_{z}$ measure the extent of the
wavefunction in the radial and axial directions.
If there is no interaction between atoms, 
the exact ground state wavefunction is obtained by setting 
$d_{\bot}$ and $d_{z}$ equal to characteristic oscillator lenghs, 
$l_{\bot}$ and $l_{z}$ respectively, which are defined by
\begin{equation}
l_{\bot} \equiv \left(\frac{\hbar}{m \omega_{\bot}}\right)^{1/2},
\,\,\,\,\,\,
l_{z} \equiv \left(\frac{\hbar}{m \omega_{z}}\right)^{1/2}.
\label{eqn:no inter-atomic interaction case in AsymBec}
\end{equation}
In the Hartree approximation (see Sec.~\ref{sec:Time-Independent Case}),
the ground state energy of the system is given 
by a Ginzburg-Pitaevskii-Gross energy functional~\cite{GinzPita}--\cite{Gross2}
with the harmonic potential terms,
\begin{equation}
E[\Psi ]
= 
\int
{\rm d} \bm{r}
\left[
\frac{\hbar^{2}}{2m}|\nabla \Psi |^{2} 
+ \frac{m}{2}  
(\omega_{\bot}^{2}r_{\bot}^{2}
+\omega_{z}^{2}z^{2})|\Psi|^{2} 
+ \frac{2\pi \hbar^{2}a}{m}  |\Psi |^{4}
\right],
\label{eqn:GPG functional in AsymBec}
\end{equation}
where $a$ is the $s$-wave scattering length.
We note that the minimization of the functional $E - \mu N$,
where $\mu$ is the chemical potential, gives the 
Ginzburg-Pitaevskii-Gross
equation~\cite{GinzPita}--\cite{Gross2} with the harmonic potential terms,
\begin{equation}
-\frac{\hbar^{2}}{2m}\Delta \Psi 
+ \frac{m}{2}  
(\omega_{\bot}^{2}r_{\bot}^{2}
+\omega_{z}^{2}z^{2})\Psi 
+ 
\frac{4\pi \hbar^{2}a}{m}  |\Psi |^{2} \Psi
=
\mu \Psi.
\label{eqn:GP eq in AsymBec}
\end{equation}

     Substituing (\ref{eqn:trial gaussian wavefuncion}) 
into (\ref{eqn:GPG functional in AsymBec}), we obtain
\begin{equation}
E(d_{\bot},d_{z})
= 
N \left(
\frac{\hbar^{2}}{2m d_{\bot}^{2}} 
+ \frac{m}{2} \omega_{\bot}^{2}d_{\bot}^{2}
\right)
+
\frac{N}{2} \left(
\frac{\hbar^{2}}{2md_{z}^{2}} 
+ \frac{m}{2} \omega_{z}^{2}d_{z}^{2}
\right)
+\frac{\hbar^{2}a N^{2}}{(2 \pi)^{1/2}m}\frac{1}{d_{\bot}^{2} d_{z}}. 
\label{eqn:GPG enegy after substituting gaussian in AsymBec}
\end{equation}
The kinetic energy and the potential energy are respectively
proportional to $d^{-2}$ and $d^{2}$.
We introduce dimensionless variational parameters
\begin{equation}
s_{\bot} \equiv l_{\bot}/d_{\bot},   
\,\,\,\,\, s_{z} \equiv l_{z}/d_{z},
\label{eqn:def of lambdas in AsymBec}
\end{equation}
and dimensionless experimental constants
\begin{equation}
\delta \equiv \omega_{z}/\omega_{\bot},   
\,\,\,\,\, G_{\bot} \equiv (2/\pi)^{1/2} N a/l_{\bot}.
\label{eqn:def of delta and G1 in AsymBec}
\end{equation}  
The constant $\delta$ measures the anisotropy of the trap.
The constant $G_{\bot}$ represents the strength of the interaction 
and is positive (negative) for the repulsive (attractive) 
interaction case.
We may use other constants such as $G_{z} \equiv (2/\pi)^{1/2} N a/l_{z}$,
but those can be expressed in terms of $\delta$ and $G_{\bot}$.
Substituting (\ref{eqn:def of lambdas in AsymBec}) 
and (\ref{eqn:def of delta and G1 in AsymBec}) into
(\ref{eqn:GPG enegy after substituting gaussian in AsymBec}) yields
\begin{equation}
E(s_{\bot},s_{z})
=
\frac{N\hbar \omega_{\bot}}{2} 
\left[
(s_{\bot}^{2} + s_{\bot}^{-2})
+
\frac{\delta}{2}
(s_{z}^{2} + s_{z}^{-2})
+
\delta^{1/2}G_{\bot}s_{\bot}^{2}s_{z}
\right].
\label{eqn:nondimensionalized GPG enegy after substituting gaussian in AsymBec}
\end{equation}
We note that the large (small) $s$ corresponds to the
contraction (spread) of the condensate.
We investigate the existence of a minimum of 
$E(s_{\bot},s_{z})$ as a function of the two 
variational parameters $s_{\bot}$ and $s_{z}$.
The minimum conditions,
\begin{eqnarray} 
\frac{\partial E}{\partial s_{\bot}} & = & 0,
\label{eqn:minimal cond for bot in AsymBec}
\\
\frac{\partial E}{\partial s_{z}} & = & 0,
\label{eqn:minimal cond for z in AsymBec}
\end{eqnarray}
determine the location of the minimum.
Substituting (\ref{eqn:nondimensionalized GPG enegy after substituting
  gaussian in AsymBec})  
into (\ref{eqn:minimal cond for bot in AsymBec}), we get
\begin{equation}
(s_{\bot} - s_{\bot}^{-3}) + \delta^{1/2}G_{\bot}s_{\bot}s_{z} = 0.
\label{eqn:minimal cond for bot No 2 in AsymBec}
\end{equation}
Similarly, substitution
of (\ref{eqn:nondimensionalized GPG enegy after substituting
  gaussian in AsymBec}) into
(\ref{eqn:minimal cond for z in AsymBec}) gives
\begin{equation}
(s_{z} - s_{z}^{-3}) + \delta^{-1/2}G_{\bot} s_{\bot}^{2} = 0.
\label{eqn:minimal cond for z No 2 in AsymBec}
\end{equation}
We also calculate the second derivatives
$\frac{\partial^{2} E}{\partial s_{\bot}^{2}}$, 
$\frac{\partial^{2} E}{\partial s_{z}^{2}}$
and $\frac{\partial^{2} E}{\partial s_{\bot} \partial s_{z}}$ 
of $E(s_{\bot},s_{z})$,
\begin{eqnarray}
\frac{\partial^{2} E}{\partial s_{\bot}^{2}}
& = &
N \hbar \omega_{\bot} 
(1 + 3 s_{\bot}^{-4} + \delta^{1/2}G_{\bot} s_{z}),
\label{eqn:par2E/parbot2 in AsymBec}
\\
\frac{\partial^{2} E}{\partial s_{z}^{2}}
& = &
\frac{\delta}{2} N \hbar \omega_{\bot} (1 + 3 s_{z}^{-4}),
\label{eqn:par2E/parz2 in AsymBec}
\\
\frac{\partial^{2} E}{\partial s_{\bot} \partial s_{z}}
& = &
\delta^{1/2} N \hbar \omega_{\bot} G_{\bot} s_{\bot},
\label{eqn:par2E/parbot parz in AsymBec}
\end{eqnarray}
from which we have the determinant of the Hessian matrix, $\Delta$,
\begin{eqnarray} 
\Delta
& \equiv &
\frac{\partial^{2} E}{\partial s_{\bot}^{2}}
\frac{\partial^{2} E}{\partial s_{z}^{2}}
-
\left( \frac{\partial^{2} E}{\partial s_{\bot} \partial s_{z}} \right)^{2}
\nonumber
\\
&=&
\delta (N \hbar \omega_{\bot})^{2} 
\left[
(1 + 3 s_{\bot}^{-4} + \delta^{1/2}G_{\bot} s_{z})
(1 + 3 s_{z}^{-4})/2
-G_{\bot}^{2} s_{\bot}^{2}
\right].
\label{eqn:Det of Hessian}
\end{eqnarray}

    In the following, we deal with the minimum conditions 
(\ref{eqn:minimal cond for bot No 2 in AsymBec}) 
and (\ref{eqn:minimal cond for z No 2 in AsymBec}) analytically 
through appropriate approximations depending on the values of 
$\delta$ and $G_{\bot}$.


\subsection{Repulsive case}
\label{sec:Repulsive Case in AsymBec}

We consider the case where the inter-atomic interaction is repulsive.
In this case, the $s$-wave scattering length $a$
and accordingly the constant $G_{\bot}$ 
defined by (\ref{eqn:def of delta and G1 in AsymBec}) are positive.
From (\ref{eqn:minimal cond for bot No 2 in AsymBec}), we have
\begin{equation}
(s_{\bot}^{4} - 1)
=
-\delta^{1/2}G_{\bot}s_{\bot}^{4}s_{z}
< 0.
\label{eqn:minimal cond for bot No 3 in AsymBec}
\end{equation}
Then, the parameter $s_{\bot}$ is smaller than 1 at the minimum point.
In a similar manner, 
we can show from (\ref{eqn:minimal cond for z No 2 in AsymBec})
that $s_{z}$ is smaller than 1 at the minimum point.
Those indicate that a repulsive 
inter-atomic interaction always acts to expand the condensate size~\cite{Fetter}.
From (\ref{eqn:par2E/parbot2 in AsymBec}), we see that 
$\frac{\partial^{2} E}{\partial s_{\bot}^{2}}$ is positive, and
by substituting (\ref{eqn:minimal cond for bot No 2 in AsymBec}) 
and (\ref{eqn:minimal cond for z No 2 in AsymBec}) 
into (\ref{eqn:Det of Hessian}), 
we have
\begin{equation}
\Delta
= 
\delta (N \hbar \omega_{\bot})^{2} 
(3 s_{\bot}^{-4} + s_{z}^{-4} 
+ 5 s_{\bot}^{-4} s_{z}^{-4} - 1),
\label{eqn:evaluation of Hessian at minimum point in AsymBec}
\end{equation}
which gives
\begin{equation}
\Delta > 
\delta (N \hbar \omega_{\bot})^{2} (s_{z}^{-4} - 1)
>
0.
\label{eqn:Positivity of Hessian at minimum point in AsymBec}
\end{equation}
Here, we have used the fact that $s_{\bot}$ and 
$s_{z}$ are smaller than 1.
Therefore, we confirm that a point $(s_{\bot},s_{z})$ 
which satisfies (\ref{eqn:minimal cond for bot No 2 in AsymBec}) 
and (\ref{eqn:minimal cond for z No 2 in AsymBec}) is indeed
a minimum point.


\noindent{\it 1) Weak interaction approximation}

\begin{figure}[htb]
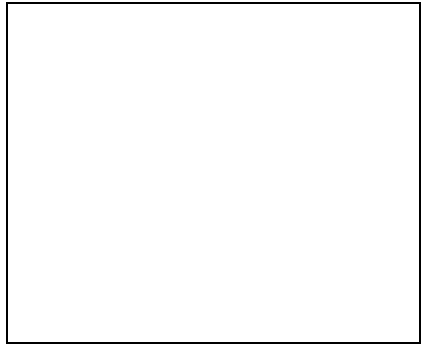

\begin{center}
\framebox[55mm]{\rule[-21mm]{0mm}{43mm}}
\end{center}
\caption{Approximations for the repulsive inter-atomic interaction:
I. weak interaction case (\ref{eqn:cond for both gaussian approx No 2
in Asym Bec}), II. strong interaction case
(\ref{eqn:cond for both TFA approx No 2 in Asym Bec}), 
III. intermediate case-1 
(\ref{eqn:cond for bot gaussian and z TFA approx No 2 in Asym Bec}), 
IV. intermediate case-2 
(\ref{eqn:cond for bot TFA and z gaussian approx No 2 in Asym Bec}).
The abscissa and ordinate represent the anisotropy of the trap, 
$\delta$, and the strength of the interaction, $G_{\bot}$, respectively.}
\label{fig:PD}
\end{figure}

We first consider the case that the third terms 
in both (\ref{eqn:minimal cond for bot No 2 in AsymBec}) 
and (\ref{eqn:minimal cond for z No 2 in AsymBec}) 
are negligible, which means the following conditions, 
\begin{equation} 
s_{\bot}  \sim  s_{\bot}^{-3} 
\gg \delta^{1/2} G_{\bot} s_{\bot} s_{z}, 
\,\,\,\,\,
s_{z} \sim  s_{z}^{-3} 
\gg \delta^{-1/2} G_{\bot} s_{\bot}^{2}.
\label{eqn:cond for both gaussian approx in Asym Bec}
\end{equation}
This approximation corresponds to the situation where 
the inter-atomic interaction is sufficiently weak 
in both $xy$- and $z$-directions.
In this case, the approximate solutions of 
(\ref{eqn:minimal cond for bot No 2 in AsymBec}) 
and (\ref{eqn:minimal cond for z No 2 in AsymBec}) are
\begin{equation}
s_{\bot} = 1 - \frac{1}{4} \delta^{1/2} G_{\bot},
\,\,\,\,\,\,
s_{z}  =  1 - \frac{1}{4} \delta^{-1/2} G_{\bot}.
\label{eqn:sol of LambdaBot in both gaussian approx in Asym Bec}
\end{equation}
Recall that $s_{\bot} = s_{z} = 1$ recovers the exact
result for non-interaction case.
By substituting (\ref{eqn:sol of LambdaBot in both gaussian approx in
  Asym Bec}) 
into  (\ref{eqn:cond for both gaussian approx in Asym Bec}),
we obtain a criterion of the approximation for $\delta$ and $G_{\bot}$,
\begin{equation}
G_{\bot} \ll \min \{\delta^{-1/2},\delta^{1/2}\}.
\label{eqn:cond for both gaussian approx No 2 in Asym Bec}
\end{equation}
We refer to (\ref{eqn:cond for both gaussian approx No 2 in Asym Bec})
as the weak interaction case (region I in Fig.~\ref{fig:PD}).
From (\ref{eqn:def of lambdas in AsymBec}), 
we have the original variational parameters as 
\begin{equation} 
d_{\bot}  =  \left(1 + \frac{1}{4} \delta^{1/2} G_{\bot}\right) 
l_{\bot} ,
\,\,\,\,\,
d_{z}  =   \left(1 + \frac{1}{4} \delta^{-1/2} G_{\bot}\right) l_{z},
\label{eqn:sol of dBot in both gaussian approx in Asym Bec}
\end{equation}
which give the aspect ratio of the condensate, $A \equiv d_{z}/d_{\bot}$,
\begin{equation}
A = \delta^{-1/2}
\left[
1 + \frac{1}{4}(\delta^{-1/2} - \delta^{1/2}) G_{\bot}
\right].
\label{eqn:A in in both gaussian approx in Asym Bec}
\end{equation}
We note that, if $\delta < 1$ ($\delta > 1$), 
the aspect ratio (\ref{eqn:A in in both gaussian approx in Asym Bec})
is larger (smaller) than that for the non-interaction 
case, $G_{\bot} = 0$.
The energy $E$ of the condensate~(\ref{eqn:nondimensionalized GPG enegy
  after substituting gaussian in AsymBec}) is estimated as
\begin{equation}
E = \frac{N \hbar \omega_{\bot}}{2}(2 + \delta + \delta^{1/2} G_{\bot}).
\label{eqn:energyin both gaussian approx in Asym Bec}
\end{equation}


\noindent{\it 2) Strong interaction approximation}

      We consider the case that the first terms 
in both (\ref{eqn:minimal cond for bot No 2 in AsymBec}) 
and (\ref{eqn:minimal cond for z No 2 in AsymBec}) are small. 
The conditions are
\begin{equation} 
s_{\bot}  \ll  \delta^{1/2} G_{\bot} s_{\bot} s_{z} \sim s_{\bot}^{-3},
\,\,\,\,\,
s_{z}  \ll  \delta^{-1/2} G_{\bot} s_{\bot}^{2} \sim s_{z}^{-3}.
\label{eqn:cond for both TFA approx in Asym Bec}
\end{equation}
This is similar to the Thomas-Fermi approximation~\cite{BP}  
where the effect of the inter-atomic interaction is assumed to be
dominant over the kinetic one.
In this case, the approximate solutions of $s_{\bot}$ and $s_{z}$ are given by
\begin{eqnarray}
s_{\bot} & = & (\delta G_{\bot})^{-1/5}
\left[
1 + \frac{1}{10}(\delta^{2} - 3)(\delta G_{\bot})^{-4/5}
\right],
\label{eqn:sol of LambdaBot in both TFA approx in Asym Bec}
\\
s_{z} & = & \delta^{1/2}(\delta G_{\bot})^{-1/5}
\left[
1 - \frac{2}{5}\left(\delta^{2} - \frac{1}{2}\right)(\delta G_{\bot})^{-4/5}
\right].
\label{eqn:sol of LambdaZ in both TFA approx in Asym Bec}
\end{eqnarray}
By substituting (\ref{eqn:sol of LambdaBot in both TFA approx in
  Asym Bec}) and (\ref{eqn:sol of LambdaZ in both TFA approx in
  Asym Bec}) into  (\ref{eqn:cond for both TFA approx in Asym Bec}),
we obtain a criterion of the approximation for $\delta$ and $G_{\bot}$,
\begin{equation}
G_{\bot} \gg \max \{\delta^{-1},\delta^{3/2}\}.
\label{eqn:cond for both TFA approx No 2 in Asym Bec}
\end{equation}
We call (\ref{eqn:cond for both TFA approx No 2 in Asym Bec})
the strong interaction case (region II in Fig.~\ref{fig:PD}).
The corresponding values of $d_{\bot}$, $d_{z}$, $A$ and $E$ are calculated as
\begin{eqnarray}
d_{\bot}
&=&
(\delta G_{\bot})^{1/5}
\left[
1 - \frac{1}{10}(\delta^{2} - 3)(\delta G_{\bot})^{-4/5}
\right]
 l_{\bot},
\label{eqn:sol of dBot in both TFA approx in Asym Bec}
\\
d_{z} &= & \delta^{-1/2}(\delta G_{\bot})^{1/5}
\left[
1 + \frac{2}{5}\left(\delta^{2} - \frac{1}{2}\right)(\delta G_{\bot})^{-4/5}
\right]
l_{z},
\label{eqn:sol of dZ in both TFA approx in Asym Bec}
\\
A &= & 
\delta^{-1}
\left[
1 + \frac{1}{2}(\delta^{2} - 1) (\delta G_{\bot})^{-4/5}
\right],
\label{eqn:A in both TFA approx in Asym Bec}
\\
E 
& = & 
\frac{5 N \hbar \omega_{\bot}}{4}
(\delta G_{\bot})^{2/5}
\left[
1 + \frac{2}{5}\left(1 + \frac{\delta^{2}}{2}\right) (\delta G_{\bot})^{-4/5}
\right]
\nonumber
\\
& = & 
\frac{5}{8 \pi^{3/5}} N \hbar \omega_{\bot}
\delta^{2/5}
\left(\frac{8 \pi N a}{l_{\bot}}\right)^{2/5}
\left[
1 + \frac{2}{5}\left(1 + \frac{\delta^{2}}{2}\right) (\delta G_{\bot})^{-4/5}
\right].
\label{eqn:energy in both TFA approx in Asym Bec}
\end{eqnarray}
The energy~(\ref{eqn:energy in both TFA approx in Asym Bec})
is equal to Eq.~(7) in Ref.~\cite{BP} to the leading order.
It is  interesting to compare the leading order term of
the energy~(\ref{eqn:energy in both TFA approx in Asym Bec})
with the energy obtained from the three-dimensional Thomas-Fermi approximation
(see Appendix~\ref{chap:App2}),
\begin{equation}
E_{\rm{3D}} 
= 
\frac{5}{7} \left(\frac{15 g}{8\pi}\right)^{2/5} 
\left(\frac{m \omega_{x}^{2}}{2}\right)^{1/5} 
\left(\frac{m \omega_{y}^{2}}{2}\right)^{1/5} 
\left(\frac{m \omega_{z}^{2}}{2}\right)^{1/5} 
N^{7/5}.
\label{eqn:energy in 3D TFA in Asym Bec}
\end{equation}
We see that both of which are proportional to $N^{7/5}$.


\noindent{\it 3) Mixed approximations}

   In both weak and strong interaction cases,
we have treated similarly the two minimum conditions,
(\ref{eqn:minimal cond for bot No 2 in AsymBec}) 
and (\ref{eqn:minimal cond for z No 2 in AsymBec}).
However, there exist  situations where 
we can give different approximations to 
each condition.

       First, we consider the case that the third term 
in (\ref{eqn:minimal cond for bot No 2 in AsymBec}) 
and the first term in (\ref{eqn:minimal cond for z No 2 in AsymBec})
are small, satisfying
\begin{equation} 
s_{\bot}  \sim  s_{\bot}^{-3} \gg \delta^{1/2} G_{\bot} s_{\bot} s_{z}, 
\,\,\,\,\,
s_{z}  \ll  \delta^{-1/2} G_{\bot} s_{\bot}^{2} \sim s_{z}^{-3}.
\label{eqn:cond for bot gaussian and z TFA approx in Asym Bec}
\end{equation}
This implies that 
the effect of the inter-atomic interaction is negligible
in $xy$-direction, but important in $z$-direction like
the Thomas-Fermi approximation.
In this case, $s_{\bot}$ and $s_{z}$ are approximated as
\begin{eqnarray}
s_{\bot} & = & 1 - \frac{1}{4} (\delta G_{\bot})^{2/3},
\label{eqn:sol of LambdaBot in bot gaussian and z TFA approx in Asym Bec}
\\
s_{z} & = & (\delta^{-1/2} G_{\bot})^{-1/3}
\left[
1 - \frac{1}{3}(\delta^{-1/2}G_{\bot})^{-4/3} 
+ \frac{1}{6}(\delta G_{\bot})^{2/3}
\right].
\label{eqn:sol of LambdaZ in bot gaussian and z TFA approx in Asym Bec}
\end{eqnarray}
By use of 
(\ref{eqn:sol of LambdaBot in bot gaussian and z TFA approx in Asym Bec}) 
and 
(\ref{eqn:sol of LambdaZ in bot gaussian and z TFA approx in Asym Bec}) 
in (\ref{eqn:cond for bot gaussian and z TFA approx in Asym Bec}), 
a criterion of the approximation is given by
\begin{equation}
\delta^{1/2} \ll G_{\bot} \ll \delta^{-1},
\label{eqn:cond for bot gaussian and z TFA approx No 2 in Asym Bec}
\end{equation}
which requires 
\begin{equation}
\delta \ll 1.
\label{eqn;delta << 1 in AsymBec}
\end{equation}
Therefore, this approximation is allowed only when the equi-potential surface
is elongated in the $z$-direction and thus  ``cigar" shaped. 
We call (\ref{eqn:cond for bot gaussian and z TFA approx No 2 in Asym Bec})
the intermediate case-1 (region III in Fig.~\ref{fig:PD}).
Correspondingly, the original parameters
$d_{\bot}$ and $d_{z}$, the aspect ratio $A$ 
and the energy $E$ are calculated as
\begin{eqnarray} 
d_{\bot} & = & 
\left[1 + \frac{1}{4} (\delta G_{\bot})^{2/3}\right] l_{\bot},
\label{eqn:sol of dBot in bot gaussian and z TFA approx in Asym Bec}
\\
d_{z} & = &
(\delta^{-1/2} G_{\bot})^{1/3}
\left[
1 + \frac{1}{3}(\delta^{-1/2}G_{\bot})^{-4/3} 
- \frac{1}{6}(\delta G_{\bot})^{2/3}
\right] 
l_{z},
\label{eqn:sol of dZ in bot gaussian and z TFA approx in Asym Bec}
\\
A & = &
\delta^{-1/2} 
(\delta^{-1/2} G_{\bot})^{1/3}
\left[
1 + \frac{1}{3}(\delta^{-1/2}G_{\bot})^{-4/3} 
- \frac{5}{12}(\delta G_{\bot})^{2/3}
\right] 
 \gg 1,
\label{eqn:A in bot gaussian and z TFA approx in Asym Bec}
\\
E & = &
\frac{N \hbar \omega_{\bot}}{2}
\left[
2 + \frac{3}{2}(\delta G_{\bot})^{2/3}
\right].
\label{eqn:energy in bot gaussian and z TFA approx in Asym Bec}
\end{eqnarray}
The energy can be rewritten as
\begin{equation}
E = 
N \hbar \omega_{\bot} 
+
\frac{3 \cdot 2^{1/3}}{8 \pi}
(m \omega_{\bot} g/\hbar)^{2/3} 
\left(\frac{m \omega_{z}^{2}}{2}\right)^{1/3}
N^{5/3},
\label{eqn:energy in bot gaussian and z TFA approx No 2 in Asym Bec}
\end{equation}
where
\begin{equation}
g \equiv 4 \pi \hbar^{2}a/m.
\label{eqn:def of g in Asym Bec}
\end{equation}
The first term is the ground state energy of two-dimensional 
harmonic oscillators. We note that the second term 
is proportional to $N^{5/3}$, in the same way as the energy obtained 
from the one-dimensional Thomas-Fermi approximation 
(see Appendix~\ref{chap:App2}),
\begin{equation}
E_{\rm{1D}} 
=
\frac{3}{5} \left(\frac{3 g}{4}\right)^{2/3} 
\left(\frac{m \omega^{2}}{2}\right)^{1/3} N^{5/3}.
\label{eqn:energy in 1D TFA in Asym Bec}
\end{equation}

      Second, we consider the opposite case;
the first term in (\ref{eqn:minimal cond for bot No 2 in AsymBec}) 
and the third term in (\ref{eqn:minimal cond for z No 2 in AsymBec})
can be treated perturbatively, which means the following conditions,
\begin{equation} 
s_{\bot}  \ll  \delta^{1/2} G_{\bot} s_{\bot} s_{z} \sim s_{\bot}^{-3}, 
\,\,\,\,\,
s_{z} \sim s_{z}^{-3} \gg \delta^{-1/2} G_{\bot} s_{\bot}^{2}.
\label{eqn:cond for bot TFA and z gaussian approx in Asym Bec}
\end{equation}
In this case, the effect of the inter-atomic interaction is
dominant in $xy$-direction but negligible in $z$-direction.
The parameters $s_{\bot}$ and $s_{z}$ are given by
\begin{eqnarray}
s_{\bot} &= & (\delta^{1/2} G_{\bot})^{-1/4}
\left[
1  
+ \frac{1}{16}(\delta^{-3/2} G_{\bot})^{1/2}
- \frac{1}{4}(\delta^{1/2}G_{\bot})^{-1}
\right],
\label{eqn:sol of LambdaBot in bot TFA and z gaussian approx in Asym Bec}
\\
s_{z} & = & 1 - \frac{1}{4} (\delta^{-3/2} G_{\bot})^{1/2}.
\label{eqn:sol of LambdaZ in bot TFA and z gaussian approx in Asym Bec}
\end{eqnarray}
Substitution of
(\ref{eqn:sol of LambdaBot in bot TFA and z gaussian approx in Asym Bec}) 
and 
(\ref{eqn:sol of LambdaZ in bot TFA and z gaussian approx in Asym Bec}) 
into (\ref{eqn:cond for bot TFA and z gaussian approx in Asym Bec}) gives 
a criterion of the approximation for $\delta$ and $G_{\bot}$ as 
\begin{equation}
\delta^{-1/2} \ll G_{\bot} \ll \delta^{3/2},
\label{eqn:cond for bot TFA and z gaussian approx No 2 in Asym Bec}
\end{equation}
which requires
\begin{equation}
\delta \gg 1.
\label{eqn;delta >> 1 in AsymBec}
\end{equation}
Therefore, this approximation is allowed only when
the equi-potential surface of the trap is an extremely flat spheroid, 
like a ``pancake".
The condition (\ref{eqn:cond for bot TFA and z gaussian approx No 2 in
  Asym Bec}) is referred to as the intermediate case-2
(region IV in Fig.~\ref{fig:PD}). 
The expressions of $d_{\bot}$, $d_{z}$, $A$ and $E$ are given by
\begin{eqnarray} 
d_{\bot} &= & 
(\delta^{1/2} G_{\bot})^{1/4}
\left[
1  
- \frac{1}{16}(\delta^{-3/2} G_{\bot})^{1/2}
+ \frac{1}{4}(\delta^{1/2}G_{\bot})^{-1}
\right] l_{\bot},
\label{eqn:sol of dBot in bot TFA and z gaussian approx in Asym Bec}
\\
d_{z} & = &
\left[
1 + \frac{1}{4} (\delta^{-3/2} G_{\bot})^{1/2}
\right]
l_{z},
\label{eqn:sol of dZ in bot TFA and z gaussian approx in Asym Bec}
\\
A & = &
\delta^{-1/2} 
(\delta^{1/2} G_{\bot})^{-1/4}
\left[
1 
+ \frac{5}{16}(\delta^{-3/2} G_{\bot})^{1/2}
- \frac{1}{4}(\delta^{1/2}G_{\bot})^{-1}
\right] \ll 1,
\label{eqn:A in bot TFA and z gaussian approx in Asym Bec}
\\
E & = &
\frac{N \hbar \omega_{z}}{2}
\left[
1 + 2 (\delta^{-3/2} G_{\bot})^{1/2}
\right]
\nonumber
\\
& = &
\frac{N \hbar \omega_{z}}{2}
+
\frac{1}{2^{1/4} \pi^{3/4}}
\left(\frac{m \omega_{\bot}^{2}}{2}\right)^{1/2}
\left(\frac{m \omega_{z}}{\hbar}\right)^{1/4}
g^{1/2}
N^{3/2},
\label{eqn:energy in bot TFA and z gaussian approx in Asym Bec}
\end{eqnarray}
with $g$ defined by (\ref{eqn:def of g in Asym Bec}).
We note that the second term in the last expression of
(\ref{eqn:energy in bot TFA and z gaussian approx in Asym Bec})
is proportional to $N^{3/2}$,
just like the two-dimensional Thomas-Fermi approximation energy 
(see Appendix~\ref{chap:App2}),
\begin{equation}
E_{\rm{2D}} 
=
\frac{2}{3}
\left(\frac{2}{\pi}\right)^{1/2}
\left(\frac{m \omega_{x}^{2}}{2}\right)^{1/4}
\left(\frac{m \omega_{y}^{2}}{2}\right)^{1/4}
g^{1/2}
N^{3/2},
\label{eqn:energy in 2D TFA in Asym Bec}
\end{equation}
while the first term is the ground state energy of one-dimensional harmonic 
oscillators.


\subsection{Attractive case}
\label{sec:Attractive Case in AsymBec}

We investigate
the case that the inter-atomic interaction is effectively attractive;
the $s$-wave scattering length $a$
and accordingly the constant $G_{\bot}$ 
defined by (\ref{eqn:def of delta and G1 in AsymBec}) are assumed to
be negative.
There, the energy function~(\ref{eqn:nondimensionalized GPG enegy 
after substituting gaussian in AsymBec}) has only a local minimum,
different from the previous section~\cite{Fetter}.
Since $G_{\bot}$ is negative,
we have from (\ref{eqn:minimal cond for bot No 2 in AsymBec})
\begin{equation}
(s_{\bot}^{4} - 1)
=
-\delta^{1/2}G_{\bot} s_{\bot}^{4} s_{z}
> 0,
\label{eqn:minimal cond for bot No 4 in AsymBec}
\end{equation}
which shows that $s_{\bot}$ is larger than 1 at the minimum point.
Similarly, $s_{z}$ is larger than 1 at the minimum point.
We thus observe that an attractive 
inter-atomic interaction works to shrink the condensate size~\cite{Fetter}.

      By use of  the minimum 
condition~(\ref{eqn:minimal cond for bot No 2 in AsymBec}) in
(\ref{eqn:par2E/parbot2 in AsymBec}), 
we find that $\frac{\partial^{2} E}{\partial s_{\bot}^{2}}$
is always positive at a local minimum point, 
\begin{equation}
\frac{\partial^{2} E}{\partial s_{\bot}^{2}}
= 4 N \hbar \omega_{\bot} s_{\bot}^{-4} > 0.
\label{eqn:positivity of par2E/parbot2 at minimum point in AsymBec}
\end{equation}
If $G_{\bot}$ is sufficiently small, the minimization of the energy
(\ref{eqn:nondimensionalized GPG enegy 
after substituting gaussian in AsymBec}) is attained at
$(s_{\bot},s_{z}) \approx (1,1)$,
and we see that, by substituting $s_{\bot} \approx 1$ and $s_{z} \approx 1$  
into (\ref{eqn:evaluation of Hessian at minimum point in AsymBec}), 
the determinant of the Hessian matrix $\Delta$
is positive at this point.
However,  it can be shown that $\Delta$ becomes zero
for some value of $G_{\bot}$, at which the local minimum
of the energy disappears, leading to the instability of the
condensate. 
This instability will be discussed in Chap.~\ref{chap:Dynamical Properties}
as a dynamical problem.
The critical value of $G_{\bot}$ is determined by 
taking the value of $\Delta$ at the minimum point to be zero,
\begin{equation}
\Delta = 
\delta (N \hbar \omega_{\bot})^{2} 
(3 s_{\bot}^{-4} + s_{z}^{-4} + 5 s_{\bot}^{-4} s_{z}^{-4} - 1) = 0,
\label{eqn:Hessian at minimum point equal to 0 in AsymBec}
\end{equation}
with the minimum conditions
(\ref{eqn:minimal cond for bot No 2 in AsymBec}) 
and (\ref{eqn:minimal cond for z No 2 in AsymBec}). 
From (\ref{eqn:minimal cond for bot No 2 in AsymBec}),
(\ref{eqn:minimal cond for z No 2 in AsymBec})
and (\ref{eqn:Hessian at minimum point equal to 0 in AsymBec}),
we get a set of three equations which determine the stability condition,
\begin{equation}
(s_{\bot}^{4} + 5)(s_{\bot}^{4} - 3)
(s_{\bot}^{4} - 1)^{2} = (8\delta)^{2} s_{\bot}^{4},
\label{eqn:eq for LambdaBot in AsymBec}
\end{equation}
\begin{equation}
s_{z}^{4} = \frac{s_{\bot}^{4} + 5}{s_{\bot}^{4} - 3},
\label{eqn:eq for LambdaZ in AsymBec}
\end{equation}
\begin{equation}
G_{\bot} = -\delta^{-1/2} s_{z}^{-1} (1 - s_{\bot}^{-4}).
\label{eqn:eq for Gbot in AsymBec}
\end{equation}

    We can solve (\ref{eqn:eq for LambdaBot in AsymBec}) analytically
for $s_{\bot}^{4}$. However, the expression is too complicated to 
get useful information.
In the following, we investigate Eqs.~(\ref{eqn:eq for LambdaBot in AsymBec})--
(\ref{eqn:eq for Gbot in AsymBec}), by using some approximations to them.


\noindent {\it 1) Almost spherical trap}

  Here we consider the case that the trap is almost spherically 
symmetric, namely
\begin{equation}
\delta \equiv \omega_{z}/\omega_{\bot} = 1 + \epsilon , 
\,\,\,\,\,\,
|\epsilon | \ll 1.
\label{eqn:delta in spherical approx in AsymBec}
\end{equation}
In this case, Eq.~(\ref{eqn:eq for LambdaBot in AsymBec}) 
and then Eq.~(\ref{eqn:eq for LambdaZ in AsymBec})
are approximately solved as
\begin{equation}
s_{\bot} = 5^{1/4} \left(1 + \frac{\epsilon}{9} \right),
\,\,\,\,\,\,
s_{z} = 5^{1/4} \left(1 - \frac{2 \epsilon}{9} \right).
\label{eqn:sol of LamdaBot in spherical approx in AsymBec}
\end{equation}
From (\ref{eqn:sol of LamdaBot in spherical approx in AsymBec}),
we get the expressions of the original variational parameters, 
\begin{equation}
d_{\bot} = 
5^{-1/4} \left(1 - \frac{\epsilon}{9} \right) l_{\bot},
\,\,\,\,\,\,
d_{z}  = 
5^{-1/4} \left(1 + \frac{2 \epsilon}{9} \right) l_{z},
\label{eqn:sol of dBot in spherical approx in AsymBec}
\end{equation}
which give the aspect ratio $A = d_{z}/d_{\bot}$,
\begin{equation}
A = 
1 - \frac{\epsilon}{6}.
\label{eqn:A in in spherical approx in AsymBec}
\end{equation}
Using (\ref{eqn:sol of LamdaBot in spherical approx in AsymBec}) 
in (\ref{eqn:eq for Gbot in AsymBec}), 
we obtain the critical value of $G_{\bot}$, denoted 
by $G_{\bot, \rm{c}}$,
\begin{equation}
G_{\bot, \rm{c}} =
-\frac{4}{5^{5/4}}
\left(1 - \frac{\epsilon}{6}\right).
\label{eqn:GbotC in spherical approx in AsymBec}
\end{equation}
The result agrees with the one obtained by Fetter (Eq.~(7)
in Ref.~\cite{Fetter}), to the leading order.
Further, by substituting
(\ref{eqn:def of delta and G1 in AsymBec})
into (\ref{eqn:GbotC in spherical approx in AsymBec}), 
we have the critical number of particles, $N_{\rm{c}}$,
at which the condensate becomes unstable, 
\begin{equation}
N_{\rm{c}}  
= 
\frac{2^{3/2}\pi^{1/2}}{5^{5/4}}
\frac{l_{\bot}}{|a|}\left(1 - \frac{\epsilon}{6}\right)
=
0.671
\frac{l_{\bot}}{|a|}\left(1 - \frac{\epsilon}{6}\right)
=
\frac{2^{3/2}\pi^{1/2}}{5^{5/4}}
\frac{l_{\bot}^{2/3}l_{z}^{1/3}}{|a|}.
\label{eqn:Nc in spherical approx in AsymBec}
\end{equation}
We also calculate the (locally) minimized energy at $G_{\bot} =
G_{\bot,{\rm c}}$, defined as $E_{\rm{c}}$,
\begin{equation}
E_{\rm{c}} =
\frac{5^{1/2}}{2} \left(1 + \frac{\epsilon}{3}\right) N \hbar \omega_{\bot} 
=
\frac{5^{1/2}}{2} N \hbar \omega_{\bot}^{2/3} \omega_{z}^{1/3}.
\label{eqn:Ec in spherical approx in AsymBec}
\end{equation}

      In an experiment for $^{7}{\rm Li}$~\cite{BSTH},
the values of frequencies $\omega_{z}$ and $\omega_{\bot}$ were
$\omega_{z} /2 \pi \approx 117 \, {\rm Hz}$ and $\omega_{\bot} /2 \pi
\approx 163 \, {\rm Hz}$.
The $s$-wave scattering length of $^{7}{\rm Li}$ was 
observed to be $a = -27.3 a_{0}$ where $a_{0}$ is the Bohr radius~\cite{AMSH}.
Using the data in the formula~(\ref{eqn:Nc in spherical approx in AsymBec}),
we have $N_{{\rm c}} = 1450$. This value agrees with the experimental
result, $N_{{\rm c}} = 650 \sim 1300$~\cite{BSH2}.


\noindent {\it 2) Cigar shaped trap}

     We consider the case that the equi-potential surface
of the magnetic trap is ``cigar'' shaped,
\begin{equation}
\delta = \epsilon \ll 1.
\label{eqn: delta << 1 in AsymBec}
\end{equation}
In this case, 
Eqs.~(\ref{eqn:eq for LambdaBot in AsymBec}) 
and (\ref{eqn:eq for LambdaZ in AsymBec})
are approximately solved to give
\begin{equation}
s_{\bot}  = 3^{1/4} \left(1 + \frac{\epsilon^{2}}{2} \right),
\,\,\,\,\,\,
s_{z} = \left(\frac{4}{3}\right)^{1/4} \epsilon^{-1/2}
\left(1 + \frac{3 \epsilon^{2}}{16} \right).
\label{eqn:sol of LamdaBot in cigar approx in AsymBec}
\end{equation}
From (\ref{eqn:sol of LamdaBot in cigar approx in AsymBec}),
we get $d_{\bot}$, $d_{z}$ and $A$ as follows,
\begin{equation}
d_{\bot} =
3^{-1/4} \left(1 - \frac{\epsilon^{2}}{2} \right) l_{\bot},
\,\,\,\,\,\,
d_{z}  = 
\left(\frac{3}{4}\right)^{1/4}
\left(1 - \frac{3 \epsilon^{2}}{16} \right) l_{\bot},
\label{eqn:sol of dBot in cigar approx in AsymBec}
\end{equation}
\begin{equation}
A =  
\left(\frac{3}{2}\right)^{1/2}
\left(1 + \frac{5 \epsilon^{2}}{16} \right).
\label{eqn:A in cigar approx in AsymBec}
\end{equation}
We observe that $d_{z}$ is about the order of $l_{\bot}$ rather than $l_{z}$, 
and accordingly the aspect ratio of the condensate $A$ is about the unity.
It is intriguing that, near $G_{\bot} = G_{\bot,\rm{c}}$,
the condensate shape does not reflect the anisotropy of the trap, 
which is assumed to be highly elongated in $z$-direction. 

        From (\ref{eqn:eq for Gbot in AsymBec})
and (\ref{eqn:sol of LamdaBot in cigar approx in AsymBec}), 
we obtain $G_{\bot, \rm{c}}$,
\begin{equation}
G_{\bot, \rm{c}}=
-\frac{2^{1/2}}{3^{3/4}}
\left(1 + \frac{13 \epsilon^{2}}{16}\right),
\label{eqn:GbotC in cigar approx in AsymBec}
\end{equation}
and then the critical number of particles $N_{\rm{c}}$,
\begin{equation}
N_{\rm{c}}  = 
\frac{\pi^{1/2}}{3^{3/4}}
\left(1 + \frac{13 \epsilon^{2}}{16}\right)
\frac{l_{\bot}}{|a|}
 =
0.778
\left(1 + \frac{13 \epsilon^{2}}{16}\right)
\frac{l_{\bot}}{|a|}.
\label{eqn:Nc in cigar approx in AsymBec}
\end{equation}
This critical number is larger than the one obtained in 
the almost spherical case
(\ref{eqn:Nc in spherical approx in AsymBec}),
if we use the same $l_{\bot}$ and $a$ in both cases.
The critical (locally) minimized energy $E_{\rm{c}}$ is obtained as 
\begin{equation}
E_{\rm{c}} = 
\frac{3^{1/2}}{2} 
\left(1 - \frac{7 \epsilon^{2}}{24}\right)
 N \hbar \omega_{\bot}.
\label{eqn:Ec in cigar approx in AsymBec}
\end{equation}


\noindent {\it 3) Pancake shaped trap}

     We consider the opposite case to the previous one;
the equi-potential surface of the trap is an extremely flat spheroid, 
like a ``pancake", namely
\begin{equation}
\delta = \epsilon^{-1} \gg 1 \,\,\,\,\,\, (\epsilon \ll 1).
\label{eqn: delta >> 1 in AsymBec}
\end{equation}
In this case, the approximate solutions of
Eqs.~(\ref{eqn:eq for LambdaBot in AsymBec}) 
and (\ref{eqn:eq for LambdaZ in AsymBec}) are
\begin{equation}
s_{\bot} = 4^{1/4} \epsilon^{-1/6},
\,\,\,\,\,\,
s_{z} = 1 + \frac{\epsilon^{2/3}}{2}.
\label{eqn:sol of LamdaBot in pancake approx in AsymBec}
\end{equation}
From (\ref{eqn:sol of LamdaBot in pancake approx in AsymBec}),
we get
\begin{equation}
d_{\bot} =
4^{-1/4} \epsilon^{1/6} l_{\bot}
= 
4^{-1/4} \epsilon^{-1/3} l_{z},
\,\,\,\,\,\,
d_{z} = 
\left(1 - \frac{\epsilon^{2/3}}{2}\right)
l_{z},
\label{eqn:sol of dBot in pancake approx in AsymBec}
\end{equation}
\begin{equation}
A = 
4^{1/4} \left(1 - \frac{\epsilon^{2/3}}{2}\right) \epsilon^{1/3}.
\label{eqn:A in pancake approx in AsymBec}
\end{equation}
Note that $d_{\bot}$ is the order of $\epsilon^{1/6} l_{\bot}$
or equivalently $\epsilon^{-1/3} l_{z}$, meaning that 
$d_{\bot}$ is much larger than $l_{z}$ but much smaller than $l_{\bot}$.
On the other hand, $d_{z}$ remains the order of $l_{z}$.
Therefore, the aspect ratio $A = d_{z}/d_{\bot}$, 
which is the order of $\epsilon^{1/3}$,
is much larger than the ratio between the characteristic oscillator
lengths, $l_{z}/l_{\bot} = \epsilon^{1/2}$.
We observe again that the condensate shape near the instability is not 
highly anisotropic.

       Substitution of 
(\ref{eqn:sol of LamdaBot in pancake approx in AsymBec}) 
into (\ref{eqn:eq for Gbot in AsymBec}) gives 
\begin{equation}
G_{\bot, \rm{c}} =
-\epsilon^{1/2}
\left(1 - \frac{3}{4} \epsilon^{2/3} \right),
\label{eqn:GbotC in pancake approx in AsymBec}
\end{equation}
from which we get
\begin{equation}
N_{\rm{c}}  = 
\frac{\pi^{1/2}}{2^{1/2}}
\left(1 - \frac{3}{4} \epsilon^{2/3} \right)
\frac{l_{z}}{|a|}
\approx 
1.25
\left(1 - \frac{3}{4} \epsilon^{2/3} \right)
\frac{l_{z}}{|a|}.
\label{eqn:Nc in pancake approx in AsymBec}
\end{equation}
We see that the critical number 
(\ref{eqn:Nc in pancake approx in AsymBec})
is larger than the ones obtained in the previous cases
(\ref{eqn:Nc in spherical approx in AsymBec})
and (\ref{eqn:Nc in cigar approx in AsymBec}),
if we set $l_{z}$ 
in (\ref{eqn:Nc in pancake approx in AsymBec})
equal to $l_{\bot}$ in (\ref{eqn:Nc in spherical approx in AsymBec}) and 
(\ref{eqn:Nc in cigar approx in AsymBec}), for a fixed $a$.
The critical minimized energy $E_{\rm c}$ is calculated as
\begin{equation}
E_{\rm c} =
\frac{1}{2} (1 + \epsilon^{4/3}) N \hbar \omega_{z}.
\label{eqn:Ec in pancake approx in AsymBec}
\end{equation}

\section{Stability of a two-component Bose-Einstein condensate}
\label{sec:2com}
\setcounter{equation}{0}
\setcounter{figure}{0}

        The stability of a two-component Bose-Einstein condensate is
an interesting subject experimentally and theoretically.
Using the  
variational method, we investigate the stable and unstable
conditions as functions of  
the particle numbers and the $s$-wave scattering lengths~\cite{MTW}.  
We find that interaction between  
the different species of atoms (simply, interspecies
interaction) has significant effects on  
the stability of each condensate.
We also consider the phase separation of the condensate.


\subsection{Formulation}
\label{sec:Formulation in 2com}

        We consider a two-component Bose-Einstein condensate in
        isotropic harmonic trap potentials. 
        We denote by $m_i$, $\omega_i$, $N_i$, and $a_{ii}$
        the mass, the trap frequency, the number of atoms, and the
        $s$-wave scattering length of  
        the {\it i}-th component respectively.
        The energy functional of the macroscopic wavefunctions
        $\Psi_i(\bm{r})$, which are normalized to $N_i$,  
        can be written as 
        \begin{eqnarray}
                E[\Psi _1,\Psi _2]&=&\int {\rm d}\bm{r} \left[
                \sum_{i=1,2} \left(\frac{\hbar^2}{2m_i}
                |\nabla\Psi_i(\bm{r})|^2+ 
                V_i(r)|\Psi_i(\bm{r})|^2 \right) \right.
\nonumber \\
                &+&\left. \sum_{i=1,2} \frac{2\pi \hbar ^2
                    a_{ii}}{m_i}  |\Psi_i(\bm{r})|^4  
                +\frac{2\pi \hbar ^2 a_{12}}{m_{12}}|\Psi_1(\bm{r})|^2|
                \Psi_2(\bm{r})|^2 \right],
                \label{eq:hamiltonian}
        \end{eqnarray}
        where $V_i(r)$ $(i=1,2)$ stand for the isotropic potentials, 
        \begin{equation}
                V_i(r)=\frac{1}{2}m_i \omega_i^2 r^2.
                \label{eq:trap}
        \end{equation}
        Minimizing the energy functional (\ref{eq:hamiltonian}) with
        (\ref{eq:trap}) gives the Ginzburg-Pitaevskii-Gross  
        equation for binary mixtures of the condensates under magnetic traps.
        The last term in (\ref{eq:hamiltonian}) represents the
        interactions between  
        the species-1 and the species-2 atoms,
        whose strength is proportional to the 
        $s$-wave scattering length $a_{12}$
        between them and is anti-proportional to the reduced mass
        $m_{12}=m_1 m_2/(m_1+m_2)$.  
        The positive (negative) value of $a_{ij}$ means the
        effectively repulsive (attractive) interaction. 
        To perform variational calculations, we use trial functions of
        the gaussian form  
        \begin{equation}
                \Psi_i(r)=N_i^{\frac{1}{2}}
\pi^{-\frac{3}{4}}d_i^{-\frac{3}{2}}\exp\left(-\frac{r^2}{2d_i^2}\right).   
                \label{eq:psi_i}
        \end{equation}
        The parameter $d_i$ measures the extent of spreading of $\Psi_i(r)$.
        By substituting (\ref{eq:psi_i}) into (\ref{eq:hamiltonian}),
        one gets the energy as a function of  
        variational parameters $d_1$ and $d_2$
        \begin{eqnarray}
                E(d_1,d_2)&=&\sum_i\frac{3}{4}N_i
\left(\frac{\hbar^2}{m_i}d_i^{-2}+m_i \omega_i^2 d_i^2\right) 
+\sum_i \frac{2\pi \hbar ^2 a_{ii}}{m_i}
                (2\pi)^{-\frac{3}{2}} N_i^2 d_i^{-3} \nonumber \\ 
                &+&\frac{2\pi \hbar ^2 a_{12}}{m_{12}}
                \pi^{-\frac{3}{2}} N_1 N_2   
                \left(d_1^2+d_2^2\right)^{-\frac{3}{2}}.
                \label{E}
        \end{eqnarray}
        To investigate the effects of all the inter-atomic
        interactions as rigorous as possible,  
        we concentrate on a case where $m_1=m_2=m$ and
        $\omega_1=\omega_2=\omega$.  
        We introduce the dimensionless energy $\varepsilon$,
        scattering length $\alpha_{ij}$, and 
        width parameter $\lambda_i$ as
        \begin{equation}
                \varepsilon=\frac{E}{\hbar \omega},\enskip
                \alpha_{ij}=\frac{a_{ij}}{l}, \enskip
                \lambda_i=\frac{d_i}{l}, 
                \label{DefOfAlpha}
        \end{equation}
        where $l=\sqrt{\hbar/(m\omega)}$ 
(the characteristic oscillator length). 
        Using $\varepsilon$, $\alpha_{ij}$, and $\lambda_i$ instead of
        $E$, $a_{ij}$, and $d_i$, 
        we rewrite (\ref{E}) as   
        \begin{equation}
                \label{energy}
                \varepsilon=\frac{3}{4}\sum_i N_i (\lambda_i^{-2}+\lambda_i^2) 
                +\frac{1}{\sqrt{2\pi}}\sum_i N_i^2 \alpha_{ii}
                \lambda_i^{-3}
                +\frac{4}{\sqrt{\pi}} N_1 N_2 \alpha_{12}
                (\lambda_1^2+\lambda_2^2)^{-\frac{3}{2}}. 
        \end{equation}
        With respect to the global properties of the function
        $\varepsilon=\varepsilon(\lambda_1,\lambda_2)$, 
        we remark the following:\\
        (i) A global maximum does not exist because 
        $\varepsilon \to \infty$ as $\lambda_1,\lambda_2 \to \infty$.\\
        (ii) The existence of a minimum depends on the signs of the
        scattering lengths $\alpha_{ij}$.\\ 
        The energy $\varepsilon=\varepsilon(\lambda_1,\lambda_2)$
        takes a local minimum  
        when $\lambda_1$ and $\lambda_2$ satisfy
        \begin{eqnarray}
                \label{eq:d1}
                \frac{\partial \varepsilon }{ \partial \lambda_1}=0&:& \enskip
                1-\lambda_1^{-4}-\sqrt{\frac{2}{\pi}}N_1 \alpha_{11}
                \lambda_1^{-5} 
-\frac{8}{\sqrt{\pi}} N_2 \alpha_{12}
                (\lambda_1^2+\lambda_2^2)^{-\frac{5}{2}}=0, \\ 
                \label{eq:d2}
                \frac{\partial \varepsilon }{ \partial \lambda_2}=0&:& \enskip
                1-\lambda_2^{-4}-\sqrt{\frac{2}{\pi}}N_2 \alpha_{22}
                \lambda_2^{-5} 
-\frac{8}{\sqrt{\pi}} N_1 \alpha_{12}
                (\lambda_1^2+\lambda_2^2)^{-\frac{5}{2}}=0, 
        \end{eqnarray}
        and the eigenvalues of the Hessian matrix are positive:
        \begin{equation}
                \Delta \equiv \frac{\partial ^2 \varepsilon}{\partial
                  \lambda_1^2}  
                \frac{\partial ^2 \varepsilon}{\partial \lambda_2^2}
                -\left(\frac{\partial ^2 \varepsilon}{\partial
                    \lambda_1 \partial \lambda_2}\right)^2 > 0, 
                \enskip
                \frac{\partial ^2 \varepsilon}{\partial \lambda_1^2}>0.
                \label{eq:hessian}
        \end{equation}
        Therefore we can examine the existence of a stable state with
        gaussian wavefunctions using  
        (\ref{eq:d1})--(\ref{eq:hessian}).
        We analyze these conditions 
        for various physical situations described by $\alpha_{ij}$ and
        $N_i$, as follows.


\subsection{$N_1=N_2$ case}
\label{sec:N_1=N_2 Case in 2com}

        We begin with the case where $N_1=N_2=N$ 
and $\alpha_{11}=\alpha_{22}=\alpha$.  
We have set $\alpha_{11}=\alpha_{22}$ 
to express the results in analytic forms. 
 The conditions (\ref{eq:d1}) and (\ref{eq:d2}) give
\begin{equation}
\label{eq:31}
\lambda_1=\lambda_2,\quad
\lambda_1^5-\lambda_1=\sqrt{\frac{2}{\pi}}N(\alpha+\alpha_{12}).   
\end{equation}
\begin{figure}
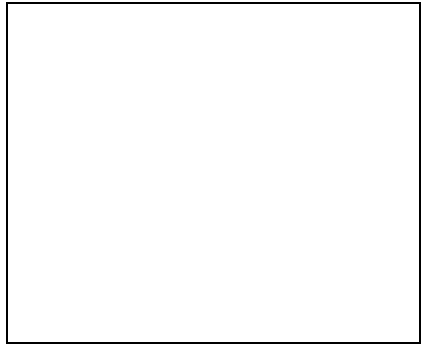

\begin{center}
\framebox[55mm]{\rule[-21mm]{0mm}{43mm}}
\end{center}
\caption{
Stability of the two-component condensate:
$N_1=N_2$ case for (a) $\alpha > 0$, (b) $\alpha < 0$.
The boundaries of stable and unstable regions (the solid lines) are
determined by Eqs.~(\protect\ref{eq:hessian}) and (\protect\ref{eq:31}).
The dashed lines are discussed in \protect\ref{sec:Phase Separation in 2com}.}
\label{fig:31}
\end{figure}
        Figure \ref{fig:31} illustrates the stable region for $\alpha$,
        $\alpha_{12}$ and $N$, where $(\lambda_1, \lambda_2)$ 
        satisfy (\ref{eq:hessian}) and (\ref{eq:31}).
        The unstable region implies the instability (collapse) of the condensates.
        From the analysis of the boundary curve, we obtain a critical number 
        $N_c (= N_{1c} = N_{2c})$, which gives an upper limit to the
        number of atoms,  
        and the parameter $\lambda_c (= \lambda_{1c} = \lambda_{2c})$,
        \begin{eqnarray}
                N_c = -\frac{2\sqrt{2\pi}}{5^\frac{5}{4}}
\frac{1}{\alpha+\alpha_{12}}, \enskip \lambda_c=5^{-\frac{1}{4}}  
                \enskip 
                (\alpha_{12} < 0, \,\,\, \alpha+\alpha_{12} < 0),
                \label{Nc1}
        \end{eqnarray}
        \begin{eqnarray}
                N_c = \frac{2\sqrt{2\pi}}{5^\frac{5}{4}}
|\alpha|^{-1}\left(1+4\frac{\alpha_{12}}{|\alpha|}\right)^{\frac{1}{4}}, 
                \enskip 
                \lambda_c=5^{-\frac{1}{4}}
\left(1+4\frac{\alpha_{12}}{|\alpha|}\right)^{\frac{1}{4}}  
                \enskip 
                \!(\alpha_{12} > 0, \,\,\, \alpha < 0).
                \label{Nc2}
        \end{eqnarray}
        The conditions (\ref{Nc1}) and (\ref{Nc2}) are worth to be discussed. 
        First, if there is no interaction between two species of
        atoms, that is,  
        $\alpha_{12}\equiv 0$, we recover
        $N_c=2\sqrt{2\pi}/{5^\frac{5}{4}}|\alpha|$ for the
        one-component case~\cite{Fetter}. 
Recall that the isotropic limit of (\ref{eqn:Nc in spherical approx in AsymBec}) 
gives this $N_{{\rm c}}$.        
Second, when the interspecies interaction is effectively
        attractive (repulsive), that is,  
        $\alpha_{12}<0$ $(\alpha_{12}>0)$, the critical number of
        atoms $N_c$ is reduced (enhanced). 
Thirdly, it is remarkable that the instability occurs 
even for $\alpha > 0$ if $\alpha_{12} < -\alpha$.       


\subsection{Phase diagrams}
\label{eqn:Phase Diagrams in 2com}
  
             We construct phase diagrams for $N_1$ and $N_2$:
        find the boundary line $N_2=f(N_1)$ 
        which divides $N_1 N_2$-plane into stable and unstable
        regions. In other words,  
        we pursue the maximum of $N_2$ 
        as a function of $N_1$.

\noindent {\it 1) $\alpha_{11}>0$, $\alpha_{22}>0$}

        This case corresponds to a situation that both intraspecies
        interactions are repulsive. 
        If interspecies interaction is not strongly attractive, the
        condensate is stable for any $N_1$ and  
        $N_2$ because of their intraspecies repulsions. 
        Therefore we consider here the case $\alpha_{12}<0$ 
        and $|\alpha_{12}|$ is sufficiently large. 
        We also consider the case that $\{N_1,N_2\}$,
        $\{\lambda_1,\lambda_2\}$, and  
        $\{\alpha_{11},\alpha_{22},|\alpha_{12}|\}$ are sets of the
        same order quantities. 
        We represent as $N$, $\lambda$ and $\alpha$  the quantities which
        are the same order as  
        $\{N_1,N_2\}$, $\{\lambda_1,\lambda_2\}$ and
        $\{\alpha_{11},\alpha_{22},|\alpha_{12}|\}$  
        respectively.
        We start from the following assumption for $(N_1,N_2)$ on the boundary 
        $N_2=f(N_1)$ and other variables, and we check the validity of 
        the assumption at the end. \\ 
        The assumption: on the boundary, $f(N_1) / N_1$ remains
        finite as $N_1$ tends to infinity,  
        and $\lambda_1$ and $\lambda_2$ do not get too large when
        $N_1,N_2 \gg 1$, so as to satisfy 
        \begin{equation}
                1 \approx O(\lambda^{-4}) \ll O(N \alpha \lambda^{-5}).
                \label{eq:assumption}
        \end{equation}
This assumption is reasonable since we expect that  the attractive interspecies
interaction suppresses the growth of $\lambda$ as $N$ gets large. 
      
        Under the assumption (\ref{eq:assumption}), neglecting the
        first and second terms  
        in (\ref{eq:d1}) and (\ref{eq:d2}), we obtain the ratios $N_2/N_1$
        and $\lambda_2/\lambda_1$  
        in the case $\alpha_{12}<-\sqrt{\alpha_{11} \alpha_{22}}$,
        \begin{eqnarray}
                \frac{N_2}{N_1}=\gamma \equiv
                \frac{|\alpha_{12}|}{\alpha_{22}}  
                \left(1\pm \sqrt{1-\left(\frac{\alpha_{11}
                        \alpha_{22}}{|\alpha_{12}|^2}\right) 
                ^{\frac{2}{5}}}\right)^{\frac{5}{2}} ,
                \label{gamma} 
        \end{eqnarray}
        \begin{equation}
                \frac{\lambda_2}{\lambda_1}=\beta \equiv   
                \left(\frac{|\alpha_{12}|^2}{\alpha_{11}
                    \alpha_{22}}\right)^{\frac{1}{5}}  
                \left(1\pm \sqrt{1-\left(\frac{\alpha_{11}
                        \alpha_{22}}{|\alpha_{12}|^2}\right) 
                ^{\frac{2}{5}}}\right).
        \end{equation}
        Equation (\ref{gamma}) determines asymptotic lines of the boundary
        curve for sufficiently large  
        $N_1$ and $N_2$.
        Here we remark that $\partial \varepsilon / \partial
        \lambda_i=0$ do not give a useful 
        information about the boundary line (or the critical number),
        but with the 
        approximation used here, 
        $\partial \varepsilon / \partial \lambda_i=0$ imply
        $\Delta=0$. That is the reason why the ratio      
        $f(N_1)/N_1$ has been pursued in (\ref{gamma}). 
        To calculate $\lambda_1 $ (not $\lambda_2/\lambda_1$), we must
        take account of terms  
        in the first order which are neglected here.
        By substituting $\lambda_2/\lambda_1=\beta (1+\delta)$ and 
        $N_2/N_1=\gamma (1+\epsilon)$, where $|\delta|,|\epsilon|\ll 1$, 
        into (\ref{eq:d1}), (\ref{eq:d2}) and $\Delta=0$, we obtain
        \begin{equation}
                \lambda_1=5^{-\frac{1}{4}}\left(\frac{\alpha_{11}
                    \beta^3+\alpha_{22} \gamma} 
                {\alpha_{11} \beta^7+\alpha_{22} \gamma}\right)^{\frac{1}{4}}.
        \end{equation}
        This shows that $\lambda_1$ is a quantity of $O(1)$ and
        therefore the assumption  
        (\ref{eq:assumption}) is confirmed. 
       
           Next we consider the case of small $N_1$.
        For simplicity, we consider the case where
        $\alpha_{11}=\alpha_{22}=\alpha$. 
        The boundary curve $N_2=f(N_1)$ passes through
        $(N_1,N_2)=(N_c,N_c)$, where  
        $N_c=2\sqrt{2\pi}/5^{\frac{5}{4}}(|\alpha_{12}|-\alpha)$ (\ref{Nc1}).
        We examine the behavior of the boundary curve
in a neighborhood of this point.
        We substitute $N_i=N_c+\delta N_i$ and
        $\lambda_i=\lambda_c+\delta \lambda_i$ into  
        $\partial \varepsilon / \partial \lambda_i=0$ and $\Delta=0$. 
        The zeroth order terms give $\lambda_c=5^{-\frac{1}{4}}$ and 
        the first order terms give
        \begin{eqnarray}
                \frac{|\alpha_{12}|}{|\alpha_{12}|-\alpha} (\delta
                \lambda_1-\delta \lambda_2) 
                =\frac{1}{\sqrt{2\pi}}(\alpha \delta N_1+\alpha_{12}
                \delta N_2) \nonumber \\ 
                =-\frac{1}{\sqrt{2\pi}}(\alpha_{12} \delta N_1+\alpha
                \delta N_2). 
        \end{eqnarray}
        Therefore, we get
        \begin{equation}
                \frac{{\rm d}N_2}{{\rm d}N_1}=-1 \quad{\rm
                  at}\quad(N_1,N_2)=(N_c,N_c). 
                \label{eq:46}
        \end{equation}
        \begin{figure}
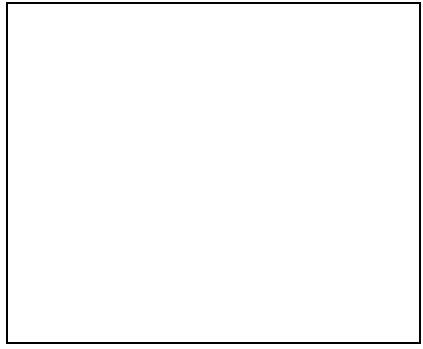

\begin{center}
\framebox[55mm]{\rule[-21mm]{0mm}{43mm}}                
\end{center}                
\caption{
                                Phase diagram for
                                $\alpha_{11}=\alpha_{22}\equiv\alpha>0$ and 
                                $\alpha_{12}<-\alpha$.
                        }
                \label{fig:41}
        \end{figure}
        Figure \ref{fig:41} illustrates the phase diagram based on the
        above discussion. 
        We see that the large attractive interspecies interaction overcomes
        the repulsive intraspecies interaction and leads to the collapse.
        We expect the similar tendencies for $\alpha_{11} \neq \alpha_{22}$.

\noindent {\it 2) $\alpha_{11}<0$, $\alpha_{22}<0$}

        This case corresponds to a situation that both intraspecies
        interactions are attractive. 
        The curve $N_2=f(N_1)$ intersects with the $N_2$-axis at
        $(N_1,N_2)=(0,N_{2c}^{(0)})$,  
        where $N_{2c}^{(0)}=2\sqrt{2\pi}/5^\frac{5}{4}|\alpha_{22}|$ is
        obtained from the one-component theory. 
        We first consider the behavior of $N_2=f(N_1)$ near this point.
        Setting 
        $\lambda_i^{(0)}=\lim_{N_1 \to 0} \lambda_i$ and substituting
        $\lambda_i=\lambda_i^{(0)}+\delta \lambda_i$ into 
        (\ref{eq:d1}), (\ref{eq:d2}) and $\Delta=0$, one gets from the zeroth 
        order terms
        \begin{eqnarray}
1-{\lambda_1^{(0)}}^{-4}-2^{\frac{9}{2}} \frac{\alpha_{12}}{|\alpha_{22}|}
\left(5^\frac{1}{2}{\lambda_1^{(0)}}^2+1\right)^{-\frac{5}{2}}=0,    
\label{eq321}\\ 
\lambda_2^{(0)}=5^{-\frac{1}{4}},  \label{eq322}
        \end{eqnarray}
        and from the first order terms
        \begin{equation}
                \frac{{\rm d}N_2}{{\rm d}N_1}=2^{\frac{5}{2}}
                \frac{\alpha_{12}}{|\alpha_{22}|}  
                \left(5^\frac{1}{2}{\lambda_1^{(0)}}^2+1\right)^{-\frac{5}{2}}
                =\frac{1-{\lambda_1^{(0)}}^{-4}}{4}
\,\,\,\,\,\,\,\,\,\,                {\rm at}\quad (N_1,N_2)=(0,N_{2c}^{(0)}),
                \label{eq:49}
        \end{equation}
        where $\lambda_1^{(0)} $ is determined by (\ref{eq321}).
      
             Next, in the case  
        $\alpha_{11}=\alpha_{22}=\alpha < 0$, 
we consider the neighborhood of the point
        $(N_1,N_2)=(N_c,N_c)$, where 
         $N_c$ is given by (\ref{Nc1}) and (\ref{Nc2}).
        If $\alpha_{12}<0 $, $N_c$ is the same as the case where
        $\alpha > 0$ and  
        $\alpha_{12}<-\alpha$.
        Therefore,
        \begin{equation}
                \frac{{\rm d}N_2}{{\rm d}N_1}=-1 \quad{\rm at}\quad
                (N_1,N_2)=(N_c,N_c). 
                \label{eq:410}
        \end{equation} 
        In the case $\alpha_{12}>0$, substituting $N_i=N_c+\delta N_i$ and 
        $\lambda_i=\lambda_c+\delta \lambda_i$ into
        (\ref{eq:d1}), (\ref{eq:d2}) and $\Delta=0$ gives 
        \begin{eqnarray}
                \frac{\alpha_{12}}{\alpha}
\left(-\delta\lambda_1-\delta\lambda_2\right)= 
                \frac{1}{\sqrt{2\pi}}\left(\alpha \delta
                  N_1+\alpha_{12} \delta N_2\right) \nonumber \\  
                =\frac{1}{\sqrt{2\pi}} \left(\alpha_{12} \delta
                  N_1+\alpha \delta N_2\right).  
        \end{eqnarray}
        Thus, we get
        \begin{equation}
                \frac{{\rm d}N_2}{{\rm d}N_1}=1  \quad {\rm at}
                \quad(N_1,N_2)=(N_c,N_c). 
                \label{eq:412}
        \end{equation}
        \begin{figure}
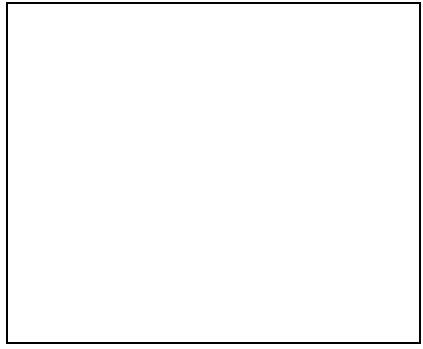

\begin{center}
\framebox[55mm]{\rule[-21mm]{0mm}{43mm}}                
\end{center}                
                \caption{
                        Phase diagram for
                        $\alpha_{11}=\alpha_{22}\equiv\alpha<0$. 
                        (a) $\alpha_{12}>0$, (b)
                        $\alpha<\alpha_{12}<0$, (c)
                        $\alpha_{12}<\alpha<0$. 
                }
                \label{fig:42}
        \end{figure}
        Figure \ref{fig:42} illustrates the phase diagrams for various
        $\alpha_{12}$. 
        We can see that the repulsive (attractive) interspecies
        interaction enlarges (narrows) the stable  
        region.

\noindent {\it 3) $\alpha_{11}>0$, $\alpha_{22}<0$}

        In this case, the species-1 atoms have repulsive interactions 
        and the species-2 atoms have attractive interactions.
        In a neighborhood of
        $(N_1,N_2)=(0,2\sqrt{2\pi}/5^\frac{5}{4}|\alpha_{22}|)$,  
        the behavior of the boundary $N_2=f(N_1)$ is the same as that
        in the previous case. 
        We investigate here $\lim_{N_1 \to \infty} f(N_1)$.
        First we assume $N_2$ is sufficiently smaller than $N_1$.
        Setting $N_2=0$ in (\ref{eq:d1}) and (\ref{eq:d2}) gives
        \begin{eqnarray}
                \lambda_1^5-\lambda_1-\sqrt{\frac{2}{\pi}}N_1 \alpha_{11}=0 ,
                \label{eq331} \\
                1-\lambda_2^{-4}
                -\frac{8}{\sqrt{\pi}} N_1 \alpha_{12}
                (\lambda_1^2+\lambda_2^2)^{-\frac{5}{2}}=0 . 
                \label{eq332}
        \end{eqnarray}
 As seen from the condition (\ref{eq331}), the width $\lambda_1$
        gets larger as $N_1$ increases because of  
        the repulsive intraspecies interaction.
        If $\alpha_{12} <0$ (attractive interspecies interaction), we
        have $\lambda_2<1$ 
        from (\ref{eq332}) and  $\lambda_2 \ll \lambda_1$ is satisfied.
        This condition is also satisfied for positive and small $\alpha_{12}$.
        When $1 \ll \lambda_1$, $\lambda_2 \ll \lambda_1 $, and $N_2 \ll N_1 $,
        one gets from (\ref{eq:d1}) and (\ref{eq:d2})
        \begin{eqnarray}
                \label{eq:433}
                1-\sqrt{\frac{2}{\pi}}N_1 \alpha_{11} \lambda_1^{-5}=0 ,\\
                \label{eq:434}
                1-\lambda_2^{-4}-\sqrt{\frac{2}{\pi}}N_2 \alpha_{22}
                \lambda_2^{-5} 
                -\frac{8}{\sqrt{\pi}} N_1 \alpha_{12} \lambda_1^{-5}=0.
        \end{eqnarray}
        Since
        \begin{equation}
                \frac{\partial^2 \varepsilon}{\partial \lambda_1^2}=
                 \frac{3}{2}N_1\left(1+4\sqrt{\frac{2}{\pi}}N_1
                   \alpha_{11} \lambda_1^{-5}\right)=\frac{15}{2}N_1>0, 
                 \end{equation}
        and 
        \begin{equation}
                \frac{\partial^2 \varepsilon}{\partial \lambda_1
                  \partial \lambda_2}= 
                 \frac{60}{\sqrt{\pi}}N_1 N_2
                 \alpha_{12}\lambda_1^{-6}\lambda_2 \approx 0, 
        \end{equation}
        we can use $ \partial^2 \varepsilon/\partial \lambda_2^2=0 $
        instead of $\Delta =0$. 
        The condition $\partial^2 \varepsilon / \partial \lambda_2^2=0$ gives
        \begin{equation}
                1+3\lambda_2^{-4}+4\sqrt{\frac{2}{\pi}}N_2 \alpha_{22}
                \lambda_2^{-5} 
                -\frac{8}{\sqrt{\pi}} N_1 \alpha_{12} \lambda_1^{-5}=0.
                \label{eq:436}
        \end{equation}
        From (\ref{eq:433}), (\ref{eq:434}) and (\ref{eq:436}), we
        obtain the critical  
        number of $N_2$,
        \begin{equation}
                N_{2c}=\frac{2\sqrt{2\pi}}{5^\frac{5}{4}}
\frac{1}{|\alpha_{22}|}  
                \left(1-\frac{2^{\frac{5}{2}}\alpha_{12}}{\alpha_{11}} \right)
                ^{-\frac{1}{4}}.
        \end{equation}
        This implies that $N_{2c}$ is independent of $N_1$ and thus 
        $\lim_{N_1 \to \infty} N_{2c}/N_1 =0 $. 
       
\begin{figure}
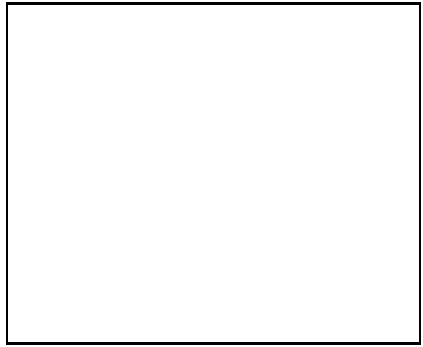

\begin{center}
\framebox[55mm]{\rule[-21mm]{0mm}{43mm}}                
\end{center}                
                \caption{
                        $\lim_{N_1 \to \infty} f(N_1)/N_1$ as a function of 
                        $2^{\frac{5}{2}}\alpha_{12}/\alpha_{11}$.
                        We set $\alpha_{11} : \alpha_{22} = 1 : -4$.
                }
                \label{fig:44}
        \end{figure}

           If $\alpha_{12}$ is positive and large enough, we expect that
        the species-2 condensate
        broadens in its size and $N_{2c}/N_1$ does not converge to
        zero as $N_1$ tends to infinity. 
        We can pursue $N_{2c}$ for a fixed large $N_1$ without using
        $\Delta=0$ 
directly, as follows. 
        We assume $\lambda_2^4 \gg 1$ for sufficiently large
        $\alpha_{12}$ and obtain from  
        Eqs.~(\ref{eq:d1}) and (\ref{eq:d2})
        \begin{eqnarray}
                1=\sqrt{\frac{2}{\pi}}N_1 \alpha_{11}
                \lambda_1^{-5}+\frac{8}{\sqrt{\pi}} N_2 \alpha_{12}  
                \left(\lambda_1^2+\lambda_2^2 \right)^{-\frac{5}{2}} ,
                \label{eq338}\\
                1=\sqrt{\frac{2}{\pi}}N_2 \alpha_{22}
                \lambda_2^{-5}+\frac{8}{\sqrt{\pi}} N_1 \alpha_{12}  
                \left(\lambda_1^2+\lambda_2^2 \right)^{-\frac{5}{2}} .
                \label{eq339}
        \end{eqnarray}
        Equations (\ref{eq338}) and (\ref{eq339}) give a relation for
        $N_2/N_1$  
        as a function of the ratio $\beta=\lambda_2/\lambda_1$,
        \begin{equation}
                \frac{N_2}{N_1}=\gamma(\beta)\equiv
                \frac{-\alpha_{11}+\alpha_{12} 
                \left(\frac{1+\beta^2}{2}\right)^{-\frac{5}{2}}}
                {|\alpha_{22}|\beta^{-5}+\alpha_{12}
\left(\frac{1+\beta^2}{2}\right)^{-\frac{5}{2}}} . 
                \label{eq:4310}
        \end{equation}
        \begin{figure}
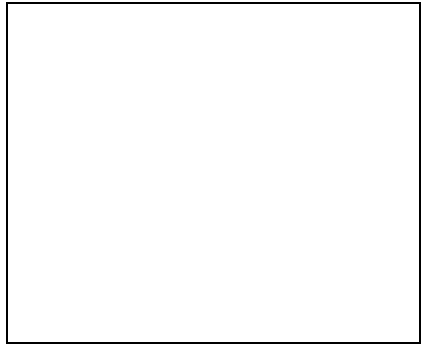

\begin{center}
\framebox[55mm]{\rule[-21mm]{0mm}{43mm}}                
\end{center}                
                \caption{
                        Function $\gamma(\beta)$ defined 
in (\protect\ref{eq:4310}).
                        We set $\alpha_{11} : \alpha_{22} :
                        \alpha_{12} = 1 : -4 :2^{-\frac{3}{2}}$. 
                }
                \label{fig:43}
        \end{figure}
        Figure \ref{fig:43} illustrates a function $\gamma (\beta)$ in
        (\ref{eq:4310}). 
        In the region $ \beta \sim 0 $, $\gamma (\beta) \sim
        \{(-\alpha_{11}+2^{\frac{5}{2}}\alpha_{12})/|\alpha_{22}|\}\beta^5 $.
        Therefore when $\alpha_{12}>2^{-\frac{5}{2}} \alpha_{11}$, 
        the maximum of $\gamma(\beta)$ determines the critical number of $N_2$.
        Explicit form of  $\max_\beta \gamma(\beta)$ can be given 
in two special cases where 
        $\zeta \equiv 2^{\frac{5}{2}} \alpha_{12}/\alpha_{11}$ is
        $1+$ and $\infty$. \\ 
        (i) Setting $\zeta=1+\epsilon $ and expanding in $\epsilon$ give
        \begin{equation}
                \frac{N_{2c}}{N_1}=\max_\beta \gamma(\beta) \simeq 
                \left(\frac{2}{7}\right)^{\frac{7}{2}}
                \frac{\alpha_{11}}{|\alpha_{22}|}\epsilon^{\frac{7}{2}}. 
        \end{equation}
        (ii) Supposing $\zeta$ is large and expanding in $1/\zeta$ give
        \begin{equation}
                \frac{N_{2c}}{N_1}=\max_\beta \gamma(\beta) \simeq 
                1-\frac{1}{\zeta}
\left\{1+\left(\frac{|\alpha_{22}|}{\alpha_{11}}\right)^{\frac{2}{7}}      
                \right\}^{\frac{7}{2}} .
        \end{equation}
        Summarizing the above discussion, $\lim_{N_1 \to
          \infty}f(N_1)/N_1$ as a function of  
        $2^{\frac{5}{2}}\alpha_{12}/\alpha_{11}$ is illustrated in
        Fig. \ref{fig:44}. 
        \par
        \begin{figure}
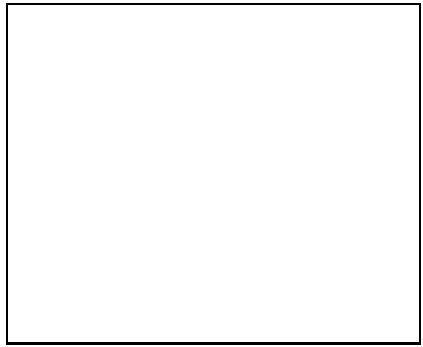

\begin{center}
\framebox[55mm]{\rule[-21mm]{0mm}{43mm}}                
\end{center}                
                \caption{
                        Phase diagram for $\alpha_{11}>0$ and $\alpha_{22}<0$.
                        (a) $\alpha_{12}>2^{-\frac{5}{2}}\alpha_{11}$, 
                        (b)
                        $0<\alpha_{12}<2^{-\frac{5}{2}}\alpha_{11}$,
                        (c) $\alpha_{12}<0$. 
                }
                \label{fig:45}
        \end{figure}
        Figure \ref{fig:45} shows phase diagrams for various $\alpha_{12}$.
         If $\alpha_{12}>2^{-\frac{5}{2}} \alpha_{11}$,  
        $N_{2c}$ gets large proportional to  $N_1$ by enlarging the width
        $\lambda_2$. 
On the other hand, in the case
         $\alpha_{12}<2^{-\frac{5}{2}} \alpha_{11}$, the extent of 
        the species-2 condensate
        remains small because of the attractive intraspecies
        interaction and its critical number converges to  
        a finite number 
        as $N_1$ increases.


\subsection{Phase separation}
\label{sec:Phase Separation in 2com}

In the experiments of two-component BEC~\cite{MBGCW}--\cite{Hall1998a},
phase separation due to the gravitation and the repulsive interspecies
interaction was observed.
In this sub-section, we discuss the possibility of phase separation caused
by the latter. 
For simplicity, we consider the case where
\begin{equation}
                N_1=N_2\equiv N, \enskip a_{11}=a_{22}\equiv a, \enskip 
                \omega_x=\omega_y\equiv \omega_\bot, \enskip
                \omega_z=\delta\omega_\bot. 
\end{equation}

     First, we assume that the anisotropic parameter $\delta$ satisfies 
$\delta \geq 1$ (pancake-like trap).\\
        To discuss a possibility of phase separation,
        we generalize the trial functions into
        \begin{equation}
                \Psi _i(\bm{r})=N^\frac{1}{2}\pi^{-\frac{3}{4}} (d_x
                d_y d_z)^{-\frac{1}{2}} 
                \exp
\left(-\frac{x^2}{2d_x^2}-\frac{y^2}{2d_y^2}-\frac{z^2}{2d_z^2}\right)
                \left(P_0 \pm \sqrt{2} P_1 \frac{x}{d_x}\right).
                \label{trialfunction}
        \end{equation}
Here $+$ ($-$) corresponds to $i=1$ ($i=2$), and real parameters
        $P_0$ and $P_1$  
        satisfy the relation $P_0^2+P_1^2=1$, which is  required by the
        normalization condition,  
        \begin{equation}
                \int \mathrm{d}\bm{r} |\Psi _i(\bm{r})|^2 = N.
        \end{equation}
The non-zero value of $P_{1}$ detects the phase separation.        
Substitution of the trial functions (\ref{trialfunction}) into
        (\ref{eq:hamiltonian}) 
        gives the energy as a function of the parameters
        $d_x$, $d_y$, $d_z$ and $P_1$.
In terms of  dimensionless variables, 
        \begin{equation}
                \varepsilon=\frac{E}{\hbar \omega _\bot}, \enskip
                \alpha=\frac{a}{l_\bot}, \enskip
                \alpha _{12}=\frac{a _{12}}{l_\bot}, \enskip
                \lambda _x=\frac{d_x}{l_\bot}, \enskip
                \lambda _y=\frac{d_y}{l_\bot}, \enskip
                \lambda _z=\frac{d_z}{l_z}, 
        \end{equation}
        where $l_\bot=\sqrt{\hbar/(m \omega _\bot)}$, 
        $l_z=\sqrt{\hbar/(m \omega _z)}$,
        the energy becomes
        \begin{eqnarray}
\varepsilon&=&\frac{N}{2}\left
                  [ (1+2P_1^2)(\lambda_x^2+\lambda_x^{-2})+ 
                \lambda_y^2+\lambda_y^{-2}+\delta
                ( \lambda_z^2+\lambda_z^{-2} ) \right. 
                \nonumber \\
                & &
+2\sqrt{\frac{2}{\pi}}\lambda_x^{-1}
                  \lambda_y^{-1} \lambda_z^{-1} 
                \delta^{1/2}
                \left\{\alpha N \left(-\frac{5}{4}P_1^4+P_1^2+1\right)
\right.
\nonumber \\
& &\left. \left.
+ \alpha_{12}N\left(\frac{11}{4}P_1^4-3P_1^2+1\right)\right\}\right].
        \end{eqnarray}
        The necessary condition for phase separation is
        \begin{equation}
                \frac{\partial\varepsilon}{\partial P_1^2}=0,\enskip
                \frac{\partial\varepsilon}{\partial \lambda_k}=0 
                \enskip \textrm{at} \enskip 0<P_1^2\leq 1,\lambda_k>0
                \enskip (k=x,y,z). 
        \end{equation}
        Thus, the conditions:
        \begin{equation}
                \left.\frac{\partial\varepsilon}{\partial
                    P_1^2}\right|_{P_1^2=0}=0, 
\,\,\,\,\,                
\left.\frac{\partial\varepsilon}{\partial
                    \lambda_k}\right|_{P_1^2=0}=0 
                \enskip (k=x,y,z),
                \label{BoundaryOfPhaseSeparation}
        \end{equation}
        give the boundary curve of phase separation.
        Solving (\ref{BoundaryOfPhaseSeparation}) gives
        \begin{equation}
                \lambda_x=\lambda_y=
\left(\frac{2\alpha_{12}}{\alpha_{12}-\alpha}\right)^\frac{1}{4},
        \enskip
        \lambda_z=\delta^{\frac{1}{2}}
\sqrt{\frac{2}{\pi}}N(\alpha_{12}-\alpha),  
\end{equation}
\begin{equation}
                N=\frac{\sqrt{\pi}}{2}
                \frac{\sqrt{\alpha+\alpha _{12}+\sqrt{(\alpha+\alpha_{12})^2+
                8\delta^2 \alpha_{12}(\alpha_{12}-\alpha)}}}
                {\delta
                  (\alpha_{12}-\alpha)^\frac{5}{4}(2\alpha_{12})^\frac{1}{4}}. 
                \label{Npancake}
        \end{equation}
        If $\alpha _{12}$ is larger than $\alpha$ and 
        $N$ is larger than the value (\ref{Npancake}), phase separation occurs.

     Second, we consider the case, $\delta<1$ (cigar-like trap). Using trial functions
        \begin{equation}
                \Psi_i(\bm{r})=N^\frac{1}{2}\pi^{-\frac{3}{4}} (d_x
                d_y d_z)^{-\frac{1}{2}} 
                \exp
\left(-\frac{x^2}{2d_x^2}-\frac{y^2}{2d_y^2}-\frac{z^2}{2d_z^2}\right)
                \left(P_0 \pm \sqrt{2} P_1 \frac{z}{d_z}\right),
\label{eqn:trialfunction in cigar}        
\end{equation}
        we can calculate the critical number of phase separation,
        \begin{equation}
                N=\frac{\sqrt{2\pi}}{4}
                \frac{ \delta^\frac{3}{2}(\alpha +\alpha _{12})+
                        \sqrt{\delta ^3(\alpha+\alpha _{12})^2+
                        8\delta \alpha_{12} (\alpha _{12}-\alpha) }}
  {(\alpha_{12}-\alpha)^\frac{5}{4}(2\alpha_{12})^\frac{3}{4}}.    
\label{Ncigar}
        \end{equation}
        In the limit $\delta \to 1$, (\ref{Npancake}) and
        (\ref{Ncigar}) coincide. 
        The boundary curves of the phase separation for this case are shown 
        in Fig.~\ref{fig:31} (the dashed lines). We see that the larger
        interspecies interaction 
        causes phase separation for the smaller number of particles.

\section{Long-range interacting bosons confined in traps}
\label{sec:LR}
\setcounter{equation}{0}

   Throughout this paper except this section, we consider a system of
neutral bosonic atoms under magnetic traps. The interaction between
neutral atoms is well approximated by the delta function, 
$V (\bm{r}) = g \,\delta (\bm{r})$.
We may think of the other limit, that is, the long-range interaction
like a Coulomb potential, $V (\bm{r}) = g /|\bm{r}|$.
In principle, we believe that the Bose-Einstein condensation occurs for 
all bosonic atoms and molecules with and without charges.

  In this section, we consider the properties of the Bose-Einstein 
condensate of long-ranged interacting bosons confined in traps.
The problem is rather academic, but turns out to be quite interesting:
The condensates in the short-ranged and long-ranged cases have
different properties.


\subsection{Stability of the ground state}
\label{sec:St of GS in LR}
\setcounter{equation}{0}

We again start from the Ginzburg-Pitaevskii-Gross energy functional, 
\begin{equation}
E [\Psi]
 = 
\int {\rm d}\bm{r}
\left[
\frac{\hbar^{2}}{2m}\left|\nabla \Psi \right|^{2}
+ \frac{1}{2}m\omega^{2}r^{2}\left|\Psi \right|^{2}
\right]
+ 
\frac{1}{2}
\int {\rm d}\bm{r}
\int {\rm d}\bm{r}'
\, U(\bm{r}-\bm{r}')
\left|\Psi (\bm{r})\right|^{2}\left|\Psi (\bm{r}')\right|^{2}
,
\label{eqn:GP functional in LR}
\end{equation}
where the trap is assumed to be an isotropic harmonic potential, and 
\begin{equation}
U (\bm{r}) = g/\left|\bm{r}\right|.
\label{eqn:Coulomb potential in LR}
\end{equation}
The coupling constant $g$  can be positive or negative in this
sub-section. 

  We assume that the ground state wavefunction $\Psi (\bm{r})$ depends
only on $r = |\bm{r}|$. With this assumption and an integral formula
for a fixed $\bm{r}$,
\begin{equation}
\int_{0}^{\pi} \sin \theta  \, {\rm d}\theta
\left(r^{2} + r'^{2} -2 r r' \cos \theta \right)^{-1/2}
=
\left\{
\begin{array}{ll}
2/r & ({\rm for} \,\, r' < r) \\
2/r' & ({\rm for}\,\,  r' > r),
\end{array}
\right.
\label{eqn:integral formula in LR}
\end{equation}
the last term in (\ref{eqn:GP functional in LR}) is written as
\begin{equation}
\frac{g}{2}
\int {\rm d}\bm{r}
\int {\rm d}\bm{r}'
\, \frac{1}{\left|\bm{r}-\bm{r}'\right|}
\left|\Psi (\bm{r})\right|^{2}\left|\Psi (\bm{r}')\right|^{2}
=
16 \pi^{2} g
\int_{0}^{\infty} {\rm d}r \, r^{2}
\left|\Psi (\bm{r})\right|^{2}
\int_{r}^{\infty}  {\rm d}r' \, r'
\left|\Psi (\bm{r}')\right|^{2}.
\label{eqn:calculation of interaction term in LR}
\end{equation}

     We calculate the ground state energy by the variational method.
As a trial function, we choose
\begin{equation}
\Psi(\bm{r}) = C \exp \left(-\frac{r^{2}}{2d^{2}}\right),
\label{eqn:gaussian trial function in LR}
\end{equation}
where $C$ and $d$ are real constants to be determined.
The particle number $N$ and the ground state energy are calculated as
\begin{eqnarray}
N &=& C^{2} \pi^{3/2} d^{3},
\label{eqn:N in LR} 
\\
E &=& \frac{3\pi^{3/2}\hbar^{2}C^{2}}{4m}d 
+ \frac{3\pi^{3/2}m\omega^{2}C^{2}}{4}d^{5}
+ \frac{\sqrt{2}}{2}\pi^{5/2}g \,C^{4}d^{5}.
\label{eqn:E in LR}
\end{eqnarray}
By use of (\ref{eqn:N in LR}), we eliminate the normalization
constant $C$ in (\ref{eqn:E in LR}). The result is 
\begin{equation}
E (d) = \frac{3\hbar^{2}N}{4md^{2}} 
+ \frac{3m\omega^{2}Nd^{2}}{4}
+ \frac{N^{2}g}{(2\pi)^{1/2}d}.
\label{eqn:E of d in LR}
\end{equation}
It is convenient to introduce a dimensionless parameter $\lambda$,
\begin{equation}
\lambda \equiv d/d_{0},
\label{eqn:def of lambda in LR}
\end{equation}
where $d_{0} \equiv \left[\hbar/(m\omega)\right]^{1/2}$. 
Then, (\ref{eqn:E of d in LR}) is rewritten as 
\begin{equation}
E (\lambda) = 
\frac{1}{2} N\hbar \omega
\left[
\frac{3}{2}
(\lambda^{-2}+\lambda^{2}) 
+ \sigma \lambda^{-1}
\right],
\label{eqn:E of lambda in LR}
\end{equation}
where 
\begin{equation}
\sigma \equiv \sqrt{\frac{2}{\pi}}
\,
\frac{N g}{\hbar \omega}
\left(\frac{m\omega}{\hbar}\right)^{1/2}.
\label{eqn:def of sigma}
\end{equation}
We can show that $E(\lambda)$ has an absolute minimum irrespective of
the sign of $\sigma$, that is, the condensate of long-range
interacting bosons under harmonic traps is stable both for repulsive
and attractive interactions.

      It is interesting to compare the above result with that for the
delta function case. The same calculation with $U (\bm{r}) = g\, \delta 
(\bm{r})$ gives 
\begin{eqnarray}
E_{{\rm d}} (\lambda) 
& = &
\frac{1}{2}N \hbar \omega
\left[
\frac{3}{2}
(\lambda^{-2}+\lambda^{2}) 
+ \sigma_{{\rm d}} \lambda^{-3}
\right],
\label{eqn:Ed of lambda in LR}
\\
\sigma_{{\rm d}} & \equiv &  
\sqrt{\frac{2}{\pi}}
\,
\frac{m g N}{4\pi\hbar^{2}}
\left(\frac{m\omega}{\hbar}\right)^{1/2}.
\label{eqn:def of sigmad}
\end{eqnarray}
A clear difference between (\ref{eqn:E of lambda in LR})
and (\ref{eqn:Ed of lambda in LR}) is the $\lambda$-dependence of the
interaction term. The powers of $\lambda$, $\lambda^{-1}$ 
in (\ref{eqn:E of lambda in LR}) and 
$\lambda^{-3}$ in (\ref{eqn:Ed of lambda in LR}), are easily
understood from the scaling property of the interaction potential 
$U (\bm{r})$ under the transformation $\bm{r} \rightarrow \lambda \bm{r}$.
In the case of the long-range interaction, the kinetic term
proportional to $\lambda^{-2}$ is dominant over the interaction term
proportional to $\lambda^{-1}$ as $\lambda \rightarrow 0$. 
This explains why the collapse of the condensate does not occur for
the long-range interacting bosons confined in traps.


\subsection{Charged bosons confined in ion traps}
\label{Charged BEC in LR}

   We consider the Bose-Einstein condensate of charged bosons confined 
in ion traps. Based on the results 
obtained in the sub-section \ref{sec:St of GS in LR},
we further examine the ground state energy. 
Setting $g = e^{2}$, $e$ being the electric charge, we have
\begin{equation}
E (\lambda) = 
\frac{1}{2} N\hbar \omega
\left[
\frac{3}{2}
(\lambda^{-2}+\lambda^{2}) 
+ \sigma_{{\rm e}} \lambda^{-1}
\right],
\label{eqn:E of lambda for charged bosons in LR}
\end{equation}
where 
\begin{equation}
\sigma_{{\rm e}} \equiv \sqrt{\frac{2}{\pi}}
\,
\frac{N e^{2}}{\hbar \omega}
\left(\frac{m\omega}{\hbar}\right)^{1/2}.
\label{eqn:def of sigmae}
\end{equation}
We note that $\sigma_{{\rm e}}$ is positive.

  We minimize $E (\lambda)$ with respect to $\lambda$.
The condition, $\partial E / \partial \lambda = 0$, gives
\begin{equation}
3 \lambda^{4} - 3 - \sigma_{{\rm e}}\lambda = 0.
\label{eqn:extremum cond for charged BEC in LR}
\end{equation}
For a weak interaction case where $\sigma_{{\rm e}}$ is small, 
the approximate solution of (\ref{eqn:extremum cond for charged BEC in LR}) is
$\lambda \approx 1 + (\sigma/12)$. Using this 
in (\ref{eqn:E of lambda for charged bosons in LR}), we obtain
\begin{equation}
E =
\frac{3}{2} N \hbar \omega
\left(
1+ \frac{\sigma_{{\rm e}}}{3}
\right).
\label{eqn:approx of E for charged bosons for small sigmae in LR}
\end{equation}
In the non-interaction limit $\sigma_{{\rm e}} \rightarrow 0$, the
exact result $E = 3 N \hbar \omega /2$ for harmonic oscillators is recovered.

   For a strong interaction case, $\sigma_{{\rm e}} \gg 1$, 
(\ref{eqn:extremum cond for charged BEC in LR}) is approximately solved to have
\begin{equation}
\lambda = 3^{-1/3} \sigma_{{\rm e}}^{1/3} + \sigma_{{\rm e}}^{-1}.
\label{eqn:sol of extremum cond for charged BEC for large sigmae in LR}
\end{equation}
Then, the ground state energy is obtained as 
\begin{equation}
E =
\frac{1}{4} 3^{4/3} N \hbar \omega \sigma_{{\rm e}}^{2/3}
+ 
\frac{1}{4} 3^{5/3} N \hbar \omega \sigma_{{\rm e}}^{-2/3}.
\label{eqn:approx of E for charged bosons for large sigmae in LR}
\end{equation}
The dominant term 
in (\ref{eqn:approx of E for charged bosons for large sigmae in LR}) 
shows the $N$-dependence of the energy explicitly as 
\begin{equation}
E =
\frac{3^{4/3}2^{2/3}}{4(2\pi)^{1/3}} m^{1/3} \omega^{2/3}e^{4/3}N^{5/3}.
\label{eqn:domninant term of E for charged bosons for large sigmae in LR}
\end{equation}
This result corresponds to the Thomas-Fermi approximation in the sense 
that the kinetic energy is negligible.
There should be no confusion with the formula (\ref{eqn:E in 1D in AsymBec})
where the interaction is the delta function type.

\section{Summary}
\label{sec:Static Concl}
\setcounter{equation}{0}

  We have investigated the static properties of 
Bose-Einstein condensates.
In Sec.~\ref{sec:AsymBec}, 
we have studied the ground state properties 
of a Bose-Einstein condensate with repulsive or attractive 
inter-atomic interaction confined in axially symmetric magnetic trap.
The gaussian trial wavefunction has two variational parameters
which are determined by minimum conditions of the 
Ginzburg-Pitaevskii-Gross energy functional with harmonic potential terms.
    For the repulsive case (sub-section~\ref{sec:Repulsive Case in AsymBec}), 
we find four situations according to which terms appearing 
in (\ref{eqn:minimal cond for bot No 2 in AsymBec}) 
and (\ref{eqn:minimal cond for z No 2 in AsymBec}) can be neglected
in a first approximation.
The approximation conditions in each situation are summarized
in Fig.~\ref{fig:PD} with the anisotropy parameter of the trap 
$\delta = \omega_{z}/\omega_{\bot}$ 
and the relative strength of the inter-atomic interaction 
$G_{\bot} = (2/\pi)^{1/2} N a (m \omega_{\bot}/\hbar)^{1/2}$.
For all cases, we can estimate the variational parameters $s_{\bot}$
and $s_{z}$ or equivalently $d_{\bot}$ and $d_{z}$,
the aspect ratio of the condensate $A = d_{z}/d_{\bot}$
and the energy $E$ at the minimum point.
In the cases {\it 1)} and {\it 2)}, 
we have used common approximations to both of the two minimum conditions
(\ref{eqn:minimal cond for bot No 2 in AsymBec}) 
and (\ref{eqn:minimal cond for z No 2 in AsymBec}).
For the case that the inter-atomic interaction 
is sufficiently weak (resp. strong) in both $xy$- and $z$-directions,
the approximation condition is given by
(\ref{eqn:cond for both gaussian approx No 2 in Asym Bec})
(resp. (\ref{eqn:cond for both TFA approx No 2 in Asym Bec}))
corresponding to the region I, the weak interaction case
(resp. the region II, the strong interaction case).
The energy~(\ref{eqn:energy in both TFA approx in Asym Bec}) and
the one obtained from the three-dimensional Thomas-Fermi approximation
have the same $N$-dependence, proportinal to $N^{7/5}$.
It is interesting that, when the anisotropy parameter $\delta$ is
much larger or smaller than unity, there exist cases
where we can give different description to each direction.
In the case {\it 3)}, we have first considered the case that 
the effect of the inter-atomic interaction is negligible
in $xy$-direction but important in $z$-direction.
This occurs when the trap is cigar-shaped.
The approximation condition is obtained as
(\ref{eqn:cond for bot gaussian and z TFA approx No 2 in Asym Bec}),
corresponding to the region III, the intermediate case-1.
The energy (\ref{eqn:energy in bot gaussian and z TFA approx No 2 in Asym Bec})
consists of two parts.
The first term describes the two-dimensional harmonic oscillator.
The $N$-dependence of the second term is $N^{5/3}$ which is the same for
the one-dimensional Thomas-Fermi approximation.
Next, we have considered the case that
the effect of the inter-atomic interaction is
dominant in $xy$-direction but negligible in $z$-direction.
This is realized when the trap is pancake-shaped. 
We obtain the approximation condition 
as (\ref{eqn:cond for bot TFA and z gaussian approx No 2 in Asym Bec}), 
corresponding to the region IV, the intermediate case-2.  
We find here that the second term in the 
energy~(\ref{eqn:energy in bot TFA and z gaussian approx in Asym Bec})
is proportional to $N^{3/2}$, 
as the energy obtained from the two-dimensional Thomas-Fermi approximation.
To summarize, in the above results, the $N$-dependence of the energy
is consistent with the geometrical shape of the condensate and the
dimensionality of the Thomas-Fermi approximation.
    For the attractive case (sub-section~\ref{sec:Attractive Case in AsymBec}),
the energy function~(\ref{eqn:nondimensionalized GPG enegy 
after substituting gaussian in AsymBec}) has only a local (not an absolute) 
minimum. As we expect, the local minimum of 
the energy disappears for some critical value of $G_{\bot}$, 
which implies the instability (collapse) of the condensate. 
By setting the value of the determinant of the Hessian matrix
$\Delta$ at the minimum point to be zero
(\ref{eqn:Hessian at minimum point equal to 0 in AsymBec})
with the minimum conditions
(\ref{eqn:minimal cond for bot No 2 in AsymBec}) 
and (\ref{eqn:minimal cond for z No 2 in AsymBec}),
we calculate the critical value $G_{\bot,{\rm c}}$ 
and obtain the critical number of particles $N_{{\rm c}}$.
First, for  the case that the trap is nearly spherical, $N_{{\rm c}}$ is
estimated as (\ref{eqn:Nc in spherical approx in AsymBec}), which
is essentially the same as the one obtained by Fetter~\cite{Fetter}.
Next, for  the cigar-shaped trap, $N_{{\rm c}}$, estimated as 
(\ref{eqn:Nc in cigar approx in AsymBec}), is larger than the critical number 
for the nearly spherical trap. The aspect ratio of the condensate $A$ 
at $G_{\bot} = G_{\bot,{\rm c}}$
(\ref{eqn:A in cigar approx in AsymBec}) is the order of unity. 
Similarly, for  the pancake-shaped trap, 
we obtain the critical number as (\ref{eqn:Nc in pancake approx in AsymBec}), 
which is larger than the ones 
for the nearly spherical and the cigar-shaped traps.


                In Sec.~\ref{sec:2com},  we have
        investigated the stability of a two-component 
        Bose-Einstein condensate.
        For various choices of ``dimensionless'' $s$-wave scattering
        lengths $\alpha_{ij}$, we have 
        constructed the phase diagram in $N_1N_2$-plane where $N_i$ is
        the number of species-{\it i}  
        $(i=1,2)$ atoms. Recall that $\alpha_{ij}$ is related to the
        $s$-wave scattering length $a_{ij}$ by 
        (\ref{DefOfAlpha}). We summarize the salient points of the
        results, paying attention to the collapse.\\ 
        1) $a_{11}>0, a_{22}>0$. A sufficiently strong attractive
        interaction between the different species
        can overcome repulsive interactions within each condensate,
        which leads to their collapse.\\ 
        2) $a_{11}<0, a_{22}<0$. A repulsive (attractive) interaction
        between the different species makes  
        the critical number of 
        atoms larger (smaller) than the one for one-component case.\\
        3) $a_{11}>0, a_{22}<0$. For a sufficiently strong repulsive
        interspecies interaction, the critical  
        number of species-2 atoms, $N_{2c}$, increases as the number
        of species-1 atoms, $N_1$. For an  
        attractive interspecies interaction,
        $N_{2c}$ decreases and approaches a finite value as $N_1$ increases.

   A basic assumption of our analysis has been the gaussian trial
functions (\ref{eq:psi_i}). 
While the gaussian ansatz is reasonable for the coexisting
condensates in magnetic traps, 
there is no guarantee that (\ref{eq:psi_i}) exhausts all the possible stable states. 
Thus, in the sub-section~\ref{sec:Phase Separation in 2com}, 
we have modified the trial functions,   
and discussed the possibility of the phase separations.
In the pancake trap case, we generalize the trial
functions into (\ref{trialfunction}), and obtain
the critical number of atoms (\ref{Npancake}), 
above which the phase separation occurs.
In the cigar trap case, the trial functions are modified 
as (\ref{eqn:trialfunction in cigar}), and the critical number 
of atoms for the phase separation is (\ref{Ncigar}). 
The boundary curves of the phase separation in the isotropic limit are 
plotted by the dashed lines in Fig.~\ref{fig:31}.


  In Sec.~\ref{sec:LR},
we have considered the properties of the Bose-Einstein
condensate of long-ranged interacting bosons confined in traps.
As in the other sections in this chapter, we have chosen a gaussian 
function (\ref{eqn:gaussian trial function in LR}) as a trial
function, and obtained the energy function (\ref{eqn:E of lambda in LR}).
The energy~(\ref{eqn:E of lambda in LR}) 
has an absolute minimum irrespective of
the sign of the strength of the interaction, $\sigma$, 
and therefore the condensate of long-range
interacting bosons under traps is stable.
By use of  the minimum conditions of the
energy~(\ref{eqn:extremum cond for charged BEC in LR}), 
we have investigated the ground state energy.
For the weak interation case, 
the energy~(\ref{eqn:approx of E for charged bosons for small sigmae in LR}) 
is approximately equal to that for harmonic oscillators.
For the strong interaction case, 
the leading term of the energy is proportional to $N^{5/3}$.
The results obtained in this sub-section
may give useful information,
when the Bose-Einstein condensation for bosonic ions under ion traps
will be  observed in future.
 
\clearpage

\chapter{Dynamical Properties of Bose-Einstein Condensates}
\label{chap:Dynamical Properties}
\setcounter{equation}{0}

   In this chapter, we investigate the dynamical properties of 
Bose-Einstein condensates.
In Sec.~\ref{sec:StabNLS},  
we first investigate the stability of the wavefunction of the
$D$-dimensional nonlinear Schr\"odinger equation 
with harmonic potential terms~\cite{TW6}. 
For the repulsive case, we prove that the wavefunction is absolutely stable.
For the attractive case, by extending the Zakharov's theory~\cite{Z,ZS}, 
we show that the singularity of the wavefunction
surely emerges in a finite time when the total
energy of the system is negative, and even when the energy is positive, 
the wavefunction collapses in a finite time for a certain class of the
initial conditions.
In Sec.~\ref{sec:BecPL}, 
based on the results obtained in Sec.~\ref{sec:StabNLS},
we investigate the dynamics of the condensates analytically, 
and prove that the singularity of solution emerges in a finite time 
when the total energy of the system is both negative 
and positive~\cite{TW1}--\cite{WT}.  
In the analysis, 
the initial wavefunction is assumed to be gaussian with two parameters.
We present a formula for the critical number of atoms. 
Further, by improving our analysis quantitatively, we apply it to the
assembly of $^{7}{\rm Li}$ atoms.
Then, within a reasonable parameter region, the estimated critical number for
$^{7}{\rm Li}$  atoms is comparable with the upper bound for 
the number of atoms in the recent
experiment of BEC. 


\section{Stability of the $D$-dimensional nonlinear Schr\"odinger
equation under confined potential}
\label{sec:StabNLS}
\setcounter{equation}{0}

 We investigate the stability of the wavefunction of the
$D$-dimensional nonlinear Schr\"odinger equation with harmonic potential terms.
The analysis in this section is a mathematical introduction to
the problem of the collapse of the condensate in Sec.~\ref{sec:BecPL}.

  Let $x \equiv (x_{1},x_{2}, \cdot \cdot \cdot ,x_{D})$ be a point on
a $D$-dimensional Euclidian space ${\bf R}^{D}$. 
As usual, time is denoted by $t$. Then the equation 
for the wavefunction $\phi (x,t)$ is expressed as 
\begin{equation}
2 {\rm i} \frac{\partial \phi }{\partial t } 
= - \Delta \phi + V(x) \phi + g | \phi |^{2} \phi .
\label{eqn:NLS under external potential in StabNLS}
\end{equation}
Here $\Delta \equiv \sum_{n = 1}^{D} \partial^{2}/\partial x_{n}^{2}$
is the $D$-dimensional Laplace operator, $g = \pm 1$, and
\begin{equation}
V(x) = \sum_{n = 1}^{D} \nu_{n}^{2} x_{n}^{2}, \,\,\,\,\,\,
\nu_{n} > 0.
\label{eqn:def of harmonic potential in rep in StabNLS}
\end{equation}
The coupling constant $g = +1$ ($g = -1$) represents the repulsive (attractive)
self-interaction. The numerical factors 
in (\ref{eqn:NLS under external potential in StabNLS}) with
(\ref{eqn:def of harmonic potential in rep in StabNLS}) are chosen for 
convenience.

    Equation~(\ref{eqn:NLS under external potential in
StabNLS}) without external potential ($V(x) = 0$)
has been studied by many researchers~\cite{Z}--\cite{SulemSulem}.
Zakharov~\cite{Z} and Zakharov and Synakh~\cite{ZS} studied 
the multi-dimensional nonlinear Schr\"odinger equation ($D = 2, 3$)
for the attractive case ($g = -1$) 
as models of Langmuir waves in plasma and light waves in
nonlinear media. They showed that, when the total energy of the system is
negative, the singularity surely emerges in the wavefunction in a finite
time. The phenomenon is termed the collapse.  
Weinstein~\cite{Weinstein} derived the condition that the
collapse occurs even when the energy of the system is positive.
Then, it is extremely interesting 
to examine how the condition obtained by Weinstein
is changed when the harmonic potential terms (\ref{eqn:def of harmonic
potential in rep in StabNLS}) are added.


\subsection{Repulsive case}
\label{sec:Repulsive Case in StabNLS}

We deal with the repulsive case, namely $g = +1$, 
and show that the wavefunction $\phi (x,t)$ 
in Eq.~(\ref{eqn:NLS under external potential in StabNLS}) 
is absolutely stable for $D \geq 1$, 
when the wavefunction is confined in the harmonic
potential~(\ref{eqn:def of harmonic potential in rep in StabNLS}).

   Equation (\ref{eqn:NLS under external potential in StabNLS}) 
has two constants of motion,
\begin{eqnarray}
N & = & \int | \phi |^{2} {\rm d} x ,
\label{eqn:number of particles in StabNLS}
\\
H 
& = & 
\int
\left( |\nabla \phi  |^{2} 
+ V(x) |\phi |^{2} + \frac{g}{2}  |\phi |^{4} \right) {\rm d} x ,
\label{eqn:energy of field in StabNLS}
\end{eqnarray}
corresponding to the conservations of particle number $N$ and energy $H$, 
respectively. 
Here and hereafter, ${\rm d}x = {\rm d}x_{1}\cdot \cdot \cdot {\rm d}x_{D}$.
We define a norm, $\| \cdot \|_{p}$, as 
\begin{equation}
\| F \|_{p} 
\equiv
\left(
\int_{{\bf R}^{D}} 
|F(x)|^{p} {\rm d} x 
\right)^{1/p}. 
\label{eqn:def of norm in StabNLS}
\end{equation}
For $g = +1$,
(\ref{eqn:energy of field in StabNLS}) with
(\ref{eqn:def of harmonic potential in rep in StabNLS}) reads
\begin{eqnarray}
H 
& = & 
\int
\left(
\left| \nabla \phi  \right|^{2}
+ 
\sum_{n = 1}^{D}
\nu_{n}^{2} x_{n}^{2} | \phi |^{2}
+ 
\frac{1}{2}  |\phi |^{4}  
\right)
{\rm d} x 
\nonumber \\
& = &
\| \nabla \phi \|_{2}^{2} 
+ \sum_{n = 1}^{D} \nu_{n}^{2} \| x_{n} \phi \|_{2}^{2} 
+ \frac{1}{2} \| \phi \|_{4}^{4}.
\label{eqn:energy of field in harmonic potential case in StabNLS}
\end{eqnarray}
Because the third term in the second line of 
Eq.~(\ref{eqn:energy of field in harmonic potential case in StabNLS}) 
is positive, we have inequalities,  
\begin{equation}
H 
 >   
\| \nabla \phi  \|_{2}^{2}
+ 
\sum_{n = 1}^{D}
\nu_{n}^{2} \| x_{n} \phi \|_{2}^{2}
 \geq 
\| \nabla \phi  \|_{2}^{2}
+
\Lambda^{2} 
\| x \phi \|_{2}^{2},
\label{eqn:inequality from energy of field in harmonic potential case
in StabNLS}
\end{equation}
where the constant $\Lambda$ is the smallest among $\{\nu_{n}\}$,
\begin{equation}
\Lambda \equiv \mbox{min}(\{\nu_{n}\}).
\label{eqn:def of Lambda in StabNLS}
\end{equation}
Further, by applying the Cauchy-Schwarz inequality
(the uncertainty relation in physics), 
\begin{equation}
\| \nabla \phi  \|_{2}^{2}
\,
\| x \phi \|_{2}^{2}
\geq
\left(
D \| \phi \|_{2}^{2} /2
\right)^{2}
= 
(D N/2)^{2} ,
\label{eqn:uncertainty relation in StabNLS}
\end{equation}
to (\ref{eqn:inequality from energy of field in harmonic potential
case in StabNLS}), we obtain
\begin{equation}
H
> 
\| \nabla \phi  \|_{2}^{2} 
+ (\Lambda D N / 2)^{2} \| \nabla \phi  \|_{2}^{-2}
\geq 
\Lambda D N .
\label{eqn:inequality from energy of field in harmonic potential case
No2 in StabNLS}
\end{equation}
From (\ref{eqn:inequality from energy of field in harmonic potential case
No2 in StabNLS}), we see that 
$\| \nabla \phi  \|_{2}$ is bounded, 
and thus the wavefunction $\phi (x,t)$ is absolutely stable
for the repulsive case.


\subsection{Attractive case}
\label{sec:Attractive Case in StabNLS}

Next, we consider the attractive case, $g = -1$.
To investigate the time-evolution of the wavefunction $\phi (x,t)$,
we introduce the expectation value of the square of the radius,
$|x|^{2} = x_{1}^{2} + x_{2}^{2}+ \cdot \cdot \cdot + x_{D}^{2}$,
with respect to $\phi $,
\begin{equation}
\langle |x|^{2} (t) \rangle
\equiv
\| x \phi (t) \|_{2}^{2}
=
\int
|x|^{2} | \phi (x,t) |^{2}
{\rm d} x .
\label{eqn:expectation value of the square of the radius in StabNLS}
\end{equation}
The role of this quantity, which should be positive by definition, 
is fundamental in the Zakharov's theory of the stability~\cite{Z,ZS}.
From (\ref{eqn:NLS under external potential in StabNLS})
and (\ref{eqn:expectation value of the square of the radius in StabNLS}), 
we get
\begin{eqnarray}
\frac{{\rm d}}{{\rm d} t } \langle |x|^{2} \rangle 
& = &
\frac{1}{{\rm i}} 
\int 
\left( 
\phi^{*} x \cdot \nabla \phi - \phi  \, x \cdot \nabla \phi^{*}
\right) 
{\rm d} x 
\nonumber \\
& = &
2 \,\, \mbox{Im}
\int 
\phi^{*} x \cdot \nabla \phi 
\, {\rm d} x ,
\label{eqn:first derivative of variance in StabNLS}
\end{eqnarray} 
where the superscript $*$ means the complex conjugate.
By differentiating (\ref{eqn:first derivative of variance in StabNLS})
with respect to  $t$ 
and using Eq.~(\ref{eqn:NLS under external potential in StabNLS}) again, 
we obtain~\cite{Pita}
\begin{equation}
\frac{{\rm d}^{2}}{{\rm d}t^{2}} \langle |x|^{2} \rangle 
 =  
2 H - \int 
(2 V(x) + x \cdot \nabla V(x)) |\phi |^{2} 
{\rm d} x 
- \frac{1}{2} (D - 2) \, \| \phi \|_{4}^{4}.
\label{eqn:eq of motion for variance general version in StabNLS}
\end{equation}
Substitution of (\ref{eqn:def of harmonic potential in rep in StabNLS}) 
into (\ref{eqn:eq of motion for variance general version in StabNLS}) yields
the equation of motion for $\langle |x|^{2} \rangle$,
\begin{equation}
\frac{{\rm d}^{2}}{{\rm d}t^{2}} \langle |x|^{2} \rangle 
= 2 H - 4 \Lambda^{2} \langle |x|^{2} \rangle - f(t),
\label{eqn:eq of motion for variance in harmonic potential in StabNLS}
\end{equation}
where
\begin{equation}
f(t) 
\equiv 
4 \sum_{n = 1}^{D} 
(\nu_{n}^{2} - \Lambda^{2}) \| x_{n} \phi (t) \|_{2}^{2} 
+ \frac{1}{2} (D - 2) \, \| \phi (t) \|_{4}^{4},
\label{eqn:def of f(t) in StabNLS}
\end{equation}
with $\Lambda$ being defined by (\ref{eqn:def of Lambda in StabNLS}).
Note that $f(t) \geq 0$, if $D \geq 2$. 
A general solution of the differential equation
(\ref{eqn:eq of motion for variance in harmonic potential in StabNLS}) is
\begin{equation}
\langle |x|^{2} \rangle 
 =   
\frac{H}{2 \Lambda^{2}} 
+ A \sin (2 \Lambda t + \theta_{0})  
-\frac{1}{2 \Lambda} \int_{0}^{t} f(u) \sin [2 \Lambda (t - u)]
{\rm d}u ,
\label{eqn:sol of variance in StabNLS}
\end{equation}
where $A$ and $\theta_{0}$ are constants.
Without loss of generality, we assume that 
\begin{equation}
A > 0, \,\,\,\,\,\, 0 \leq \theta_{0} < 2 \pi .
\label{eqn:Assumption for A and theta0 in StabNLS}
\end{equation}
For the later convenience, we define a function, $l(\theta)$,  
\begin{equation}
l(\theta ) \equiv \frac{H}{2 \Lambda^{2}} + A \sin \theta ,
\label{eqn:def of g(theta) in StabNLS}
\end{equation}
with
\begin{equation}
\theta \equiv 2 \Lambda t + \theta_{0}.
\label{eqn:def of theta in StabNLS}
\end{equation}
Then, (\ref{eqn:sol of variance in StabNLS}) becomes
\begin{equation}
\langle |x|^{2} \rangle 
=  
l( \theta (t) )
-\frac{1}{2 \Lambda} \int_{0}^{t} f(u) \sin [2 \Lambda (t - u)]
{\rm d}u .
\label{eqn:sol of variance No. 2 in StabNLS}
\end{equation}

     The constants $A$ and $\theta_{0}$ are related to
the initial conditions of the wavefunction.
Using  (\ref{eqn:expectation value of the square of the
radius in StabNLS}) and (\ref{eqn:sol of variance in StabNLS})
in (\ref{eqn:sol of variance No. 2 in StabNLS}), we get
\begin{eqnarray}
\langle |x|^{2} \rangle |_{t=0} 
& = &   
\int |x|^{2} | \phi_{0} (x)|^{2} {\rm d} x 
\nonumber \\
& = &
l(\theta_{0})
= 
\frac{H}{2 \Lambda^{2}} + A \sin \theta_{0},  
\label{eqn:value of variance at t = 0 in StabNLS}
\end{eqnarray}
where
\begin{equation}
\phi_{0}(x) \equiv \phi (x,t = 0).
\label{eqn:def of phi0 in StabNLS}
\end{equation}
Differentiation of (\ref{eqn:sol of variance in StabNLS}) 
with respect to $t$ gives
\begin{equation}
\frac{{\rm d}}{{\rm d}t} \langle |x|^{2} \rangle 
=  
2 \Lambda A \cos (2 \Lambda t + \theta_{0})  
-\int_{0}^{t} f(u) \cos [2 \Lambda (t - u)] {\rm d}u .
\label{eqn:1st derivative of sol of variance in StabNLS}
\end{equation}
Then, from (\ref{eqn:first derivative of variance in StabNLS}) 
and (\ref{eqn:1st derivative of sol of variance in StabNLS}), we get 
\begin{eqnarray}
\left. \frac{{\rm d}}{{\rm d}t} \langle |x|^{2} \rangle \right|_{t = 0}
& = &
\frac{1}{{\rm i}} 
\int 
\left( 
\phi_{0}^{*} x \cdot \nabla \phi_{0} - \phi_{0}  \, x
\cdot \nabla \phi_{0}^{*} 
\right) 
{\rm d} x 
\nonumber
\\
& = &  
2 \Lambda A \cos \theta_{0}.  
\label{eqn:value of 1st derivative of sol of variance at t = 0 in StabNLS}
\end{eqnarray}
Solving (\ref{eqn:value of variance at t = 0 in StabNLS}) and 
(\ref{eqn:value of 1st derivative of sol of variance at t = 0 in StabNLS}),
we obtain
\begin{equation}
A 
= 
\left[
\left( \langle |x|^{2} \rangle|_{t = 0} - \frac{H}{2 \Lambda^{2}} \right)^{2}
+ \left( \frac{1}{2 \Lambda} \left. \frac{{\rm d}}{{\rm d}t} \langle
|x|^{2} \rangle \right|_{t = 0} \right)^{2}
\right]^{1/2},
\label{eqn:value of A in StabNLS}
\end{equation}
\begin{eqnarray}
\sin \theta_{0}
& = &
\left(
\langle |x|^{2} \rangle|_{t = 0} - \frac{H}{2 \Lambda^{2}}
\right)/A,
\label{eqn:value of sin theta0 in StabNLS}
\\
\cos \theta_{0}
& = &
\left(
\frac{1}{2 \Lambda} \left. \frac{{\rm d}}{{\rm d}t} \langle
|x|^{2} \rangle \right|_{t = 0}
\right)/A.
\label{eqn:value of cos theta0 in StabNLS}
\end{eqnarray}

      In the following, we analyze the solution (\ref{eqn:sol of
variance in StabNLS}) or equivalently
(\ref{eqn:sol of variance No. 2 in StabNLS})
in the case that the spatial dimension is larger than or equal to two, namely 
$D \geq 2$.


\noindent{\it 1) $H \leq 0$ case}

     We consider the case that the total energy of the system $H$ 
is less than or equal to zero. 
In this case, we show that the wavefunction surely
collapses in a finite time.

    From (\ref{eqn:value of variance at t = 0 in StabNLS}), 
$l (\theta_{0})$ should be positive
because of the positivity of $\langle |x|^{2} \rangle$ defined by  
(\ref{eqn:expectation value of the square of the radius in StabNLS}), 
and therefore the constant $A$ should satisfy 
\begin{equation}
A > \left| \frac{H}{2 \Lambda^{2}} \right| = - \frac{H}{2 \Lambda^{2}}.
\label{eqn:condition for A in H < 0 Case in StabNLS}
\end{equation}
When (\ref{eqn:condition for A in H < 0 Case in StabNLS}) is satisfied,
there exist two zero points of $l(\theta)$ which lie in $0 \leq \theta
\leq \pi$. We write them as $\theta_{+}$ and $\theta_{-}$ 
($0 \leq \theta_{-} < \pi /2 < \theta_{+} \leq \pi$). 
Explicitly, $\theta_{\pm}$ are given by
\begin{eqnarray}
\theta_{+} & = & \pi - \mbox{Arcsin} \left( \frac{-H}{2\Lambda^{2}A}
\right), 
\label{eqn:def of theta+ in StabNLS}
\\
\theta_{-} & = & \mbox{Arcsin} \left( \frac{-H}{2\Lambda^{2}A}
\right).
\label{eqn:def of theta- in StabNLS}
\end{eqnarray}
Here and hereafter, the region (branch) 
of $\mbox{Arcsin} \, x$ is defined to be 
$[-\pi /2,+\pi /2]$. Then, the constant $\theta_{0}$ should be
\begin{equation}
\theta_{-} < \theta_{0} < \theta_{+}.
\label{eqn:theta- < theta0 < theta+ in StabNLS}
\end{equation}

     Since the function $f (t)$ defined by (\ref{eqn:def of f(t) in StabNLS}) 
is larger than or equal to zero when $D \geq 2$,
and $\sin [2 \Lambda (t - u)]$ 
appearing in (\ref{eqn:sol of variance in StabNLS}) 
is also larger than or equal to zero for $0 \leq t \leq \frac{\pi}{2\Lambda}$, 
we obtain an inequality,
\begin{equation} 
\langle |x|^{2} \rangle \leq l(\theta (t)),  
\label{eqn:inequality for variance and g(theta) in StabNLS}
\end{equation}
for $0 \leq t \leq \frac{\pi}{2\Lambda}$
or equivalently $\theta_{0} \leq \theta \leq \theta_{0} + \pi$. 
The function $l(\theta (t))$ becomes surely 
negative for some value of
$t$ which lies in $0 \leq t \leq \frac{\pi}{2\Lambda}$, 
when (\ref{eqn:condition for A in H < 0 Case in StabNLS}) and 
(\ref{eqn:theta- < theta0 < theta+ in StabNLS}) are satisfied. 
Then, from the inequality (\ref{eqn:inequality
for variance and g(theta) in StabNLS}), $\langle |x|^{2} \rangle$
should also be negative for some value of $t$, $0 \leq t \leq
\frac{\pi}{2\Lambda}$. 
This contradicts to the definition of $\langle |x|^{2} \rangle$
(\ref{eqn:expectation value of the square of the radius in StabNLS}) and
implies the development of the singularity, characterized by the divergence
of $\| \nabla \phi \|_{2}$, in a finite time defined as $t_{0}$. 
In fact, from the Cauchy-Schwarz inequality (\ref{eqn:uncertainty
relation in StabNLS}), we see that $\| \nabla \phi \|_{2}$ tends to infinity 
when $\langle |x|^{2} \rangle$ goes to zero.
It is known that the instant $t_{0}$ may differ from the moment 
at which $\langle |x|^{2} \rangle$ vanishes, defined as $t_{{\rm c}}$, 
and that $t_{0}$ is less than or equal to $t_{{\rm c}}$, 
in general~\cite{Berge}. 
We call  $t_{{\rm c}}$ the collapse-time in this paper.

   We can show that $\sup_{x \in {\bf R}^{D}} |\phi (x,t)|$ tends to
infinity when $\| \nabla \phi \|_{2}$ diverges.
This problem has been discussed also in Refs.~\cite{Glassey,NTT}.  
For $g = -1$, the energy (\ref{eqn:energy of field in StabNLS}) with
the harmonic potential (\ref{eqn:def of harmonic potential in rep in
StabNLS}) is written as  
\begin{eqnarray}
H 
& = & 
\int
\left(
\left| \nabla \phi  \right|^{2}
+ 
\sum_{n = 1}^{D}
\nu_{n}^{2} x_{n}^{2} | \phi |^{2}
- 
\frac{1}{2}  |\phi |^{4}  
\right)
{\rm d} x 
\nonumber \\
& = &
\| \nabla \phi \|_{2}^{2} 
+ \sum_{n = 1}^{D} \nu_{n}^{2} \| x_{n} \phi \|_{2}^{2}
- \frac{1}{2}\| \phi \|_{4}^{4} 
.
\label{eqn:energy of field in g = -1 and harmonic potential case in StabNLS}
\end{eqnarray}
Then, the following inequalities hold,
\begin{eqnarray}
\hspace{-0.5cm}
\| \phi \|_{4}^{4} 
& = &
-2 H   
+ 2 \| \nabla \phi \|_{2}^{2}
+ 2 \sum_{n = 1}^{D} \nu_{n}^{2} \| x_{n} \phi \|_{2}^{2}
\nonumber 
\\
& \geq &
- 2 H   
+ 2 \| \nabla \phi \|_{2}^{2}
+ 2 \Lambda^{2} \langle |x|^{2} \rangle 
\nonumber 
\\
& \geq &
- 2 H   
+ 2  \| \nabla \phi \|_{2}^{2} 
+ 2 (\Lambda DN/2)^{2} \| \nabla \phi \|_{2}^{-2}.
\label{eqn:inequality for interacting energy in StabNLS}
\end{eqnarray}
In deriving the third line of (\ref{eqn:inequality for
interacting energy in StabNLS}), we have used (\ref{eqn:uncertainty
relation in StabNLS}).  
Further, combining an inequality 
\begin{equation}
\| \phi \|_{4}^{4} 
\leq 
\left( \sup_{x \in {\bf R}^{D}} |\phi (x,t)| 
\right)^{2} \| \phi \|_{2}^{2} 
= 
\left( \sup_{x \in {\bf R}^{D}} |\phi (x,t)| \right)^{2} N ,
\label{eqn:inequality for sup phi in StabNLS}
\end{equation}
with (\ref{eqn:inequality for interacting energy in StabNLS}), we find that
$\sup_{x \in {\bf R}^{D}} |\phi (x,t)|$ goes to infinity when 
$\| \nabla \phi \|_{2}$ diverges.

      From the inequality (\ref{eqn:inequality for variance and g(theta) in
StabNLS}), we see that $\langle |x|^{2} \rangle$ vanishes
 before $l(\theta(t))$ reaches zero at $\theta = \theta_{+}$.
Thus, by using (\ref{eqn:def of theta in StabNLS}), 
we can obtain an upper bound of the collapse-time, $t_{{\rm c},M}$, 
\begin{equation}
t_{{\rm c},M} = \frac{1}{2\Lambda}(\theta_{+} - \theta_{0}).
\label{eqn:formula for t0M in StabNLS}
\end{equation}
When $\theta_{0} \leq \pi /2$, namely 
$\left. \frac{{\rm d}}{{\rm d}t} \langle |x|^{2} \rangle \right|_{t =
0} \geq 0$, from (\ref{eqn:value of sin theta0 in StabNLS}) we get
\begin{equation}
\theta_{0} = \mbox{Arcsin} \left( 
\frac{\langle |x|^{2} \rangle|_{t = 0}}{A} - \frac{H}{2 \Lambda^{2} A} 
\right).
\label{eqn:theta0 when H < 0 and theta- < theta0 =< pi /2 in StabNLS}
\end{equation}
Substituting (\ref{eqn:def of theta+ in StabNLS}) and
(\ref{eqn:theta0 when H < 0 and theta- < theta0 =< pi /2 in StabNLS})
into (\ref{eqn:formula for t0M in StabNLS}), we have
\begin{eqnarray}
t_{{\rm c},M} 
& = &
\frac{1}{2\Lambda}
\left[
\pi 
- \mbox{Arcsin} 
\left(\frac{-H}{2\Lambda^{2} A }\right)
-\mbox{Arcsin} \left( 
\frac{\langle |x|^{2} \rangle|_{t = 0}}{A} - \frac{H}{2 \Lambda^{2} A} \right)
\right]
\nonumber \\
& = &
\frac{1}{2\Lambda}
(
\pi /2 
+ \mbox{Arcsin} \, \xi
),
\label{eqn:t0M when H < 0 and theta- < theta0 =< pi /2 in StabNLS}
\end{eqnarray}
where $\xi$ is defined as
\begin{eqnarray}
\xi 
& \equiv &
\left[
\left( \langle |x|^{2} \rangle|_{t = 0} - \frac{H}{2 \Lambda^{2}} \right)^{2}
+ 
\left( \frac{1}{2 \Lambda} \left. \frac{{\rm d}}{{\rm d}t} \langle
|x|^{2} \rangle \right|_{t = 0} \right)^{2}
\right]^{-1}
\nonumber \\
& & \times 
\left[
\frac{1}{2 \Lambda} \left. \frac{{\rm d}}{{\rm d}t} \langle
|x|^{2} \rangle \right|_{t = 0}
\sqrt{
\langle |x|^{2} \rangle|_{t = 0}
\left( \langle |x|^{2} \rangle|_{t = 0} - \frac{H}{\Lambda^{2}} \right)
+ \left( \frac{1}{2 \Lambda} \left. \frac{{\rm d}}{{\rm d}t} \langle
|x|^{2} \rangle \right|_{t = 0} \right)^{2}
}
\right.
\nonumber \\
& & 
\,\,\,\,\,\,\,\, 
+ 
\left.
\frac{H}{2 \Lambda^{2}}
\left( \langle |x|^{2} \rangle|_{t = 0} - \frac{H}{2 \Lambda^{2}} \right)
\right]
.
\label{eqn:def of z1 in StabNLS}
\end{eqnarray}
In deriving (\ref{eqn:t0M when H < 0 and theta- < theta0 =< pi /2 in StabNLS}),
we have used a formula,
\begin{eqnarray}
\lefteqn{
\mbox{Arcsin} \, \eta - \mbox{Arcsin} \, \zeta 
}
\nonumber \\
& = & 
\mbox{Arcsin} \left(
\eta \sqrt{1 - \zeta^{2}}  
- \zeta \sqrt{1 - \eta^{2}}
\right),
\label{eqn:formula for Arcsin in StabNLS}
\end{eqnarray}
for $0 \leq \eta \leq 1$ and  $0 \leq \zeta \leq 1$.

     When $\theta_{0} \geq \pi /2$ namely 
$\left. \frac{{\rm d}}{{\rm d}t} \langle
|x|^{2} \rangle \right|_{t = 0} \leq 0$, $\theta_{0}$ becomes
\begin{equation}
\theta_{0} 
= 
\pi - 
\mbox{Arcsin} \left( 
\frac{\langle |x|^{2} \rangle|_{t = 0}}{A} - \frac{H}{2 \Lambda^{2} A}
\right),
\label{eqn:theta0 when H < 0 and pi /2 =< theta0 < theta+ in StabNLS}
\end{equation}
instead of (\ref{eqn:theta0 when H < 0 and theta- < theta0 =< pi
/2 in StabNLS}). In spite of this change, 
we obtain $t_{{\rm c},M}$ just the same as (\ref{eqn:t0M when H < 0 and
theta- < theta0 =< pi /2 in StabNLS}).


\noindent {\it 2) $H > 0$ case}

For the $H \leq 0$ case,
we have shown that the wavefunction surely collapses in a finite time, 
regardless of the initial condition on the wavefunction $\phi_{0} (x)$.
Here we prove that even when the total energy $H$
is positive, the collapse of wavefunction occurs in a finite time for
a certain class of the initial conditions.
We investigate such conditions according to the
value of $\theta_{0}$, as follows.   

     We first consider the case that $0 \leq \theta_{0}
\leq \pi /2$, which is equivalent to 
\begin{equation}
\langle |x|^{2} \rangle|_{t = 0} 
\geq 
\frac{H}{2 \Lambda^{2}},
\,\,\,\,\,\,\,\,\,\,\,\,\,\,
\left. \frac{{\rm d}}{{\rm d}t} \langle
|x|^{2} \rangle \right|_{t = 0} 
\geq 
0.
\label{eqn:cond for variance and deri in 0 =< theta0 =< pi /2 in StabNLS} 
\end{equation} 
As mentioned in the $H \leq 0$ case,  
$f (t)$ defined by (\ref{eqn:def of f(t) in StabNLS}) 
is larger than or equal to zero when $D \geq 2$, and $\sin [2 \Lambda
(t - u)]$  appearing in (\ref{eqn:sol of variance in StabNLS}) 
is also larger than or equal to zero for $0 \leq t \leq
\frac{\pi}{2\Lambda}$ or equivalently $\theta_{0} \leq \theta \leq
\theta_{0} + \pi$. 
Then, if 
\begin{equation}
l(\theta_{0}+\pi) \leq 0 ,
\label{eqn:sufficient condition for collape in H > 0 and 0 =< theta0
=< pi /2 in StabNLS}
\end{equation}
$l(\theta (t))$ becomes negative for some value of $t$
in an interval $0 \leq t \leq \frac{\pi}{2\Lambda}$,
and thus due to (\ref{eqn:inequality for variance and g(theta) in StabNLS})
the collapse of the wavefunction happens.
By use of (\ref{eqn:value of variance at t = 0 in StabNLS}) and 
\begin{equation}
l(\theta_{0}+\pi) = H/\Lambda^{2} - l(\theta_{0}), 
\label{eqn:relation between g(theta0+pi) and g(theta0) in StabNLS} 
\end{equation}
the condition (\ref{eqn:sufficient condition for collape in H > 0 and 0
=< theta0 =< pi /2 in StabNLS}) is rewritten as
\begin{equation}
\langle |x|^{2} \rangle|_{t = 0} 
\geq 
H/\Lambda^{2}.
\label{eqn:sufficient condition for collape in H > 0 and 0 =< theta0 
=< pi /2 No. 2 in StabNLS}
\end{equation}
When the condition (\ref{eqn:sufficient condition for collape in H > 0
and 0 =< theta0 =< pi /2 No. 2 in StabNLS}) is satisfied, 
there exist zero points of $l(\theta)$,
of which we denote the positive and minimum one by $\theta_{*}$.
It is easily confirmed that 
$\theta_{*}$ lies in $\pi \leq \theta_{*} \leq 3 \pi /2$. 
Considering $0 \leq \theta_{0} \leq \pi /2$
and using (\ref{eqn:value of sin theta0 in StabNLS}), we get
\begin{equation}
\theta_{0} 
= 
\mbox{Arcsin} 
\left( 
\frac{\langle |x|^{2} \rangle|_{t = 0}}{A} - \frac{H}{2 \Lambda^{2} A} 
\right).
\label{eqn:theta0 when H > 0 and 0 =< theta0 =< pi /2 in StabNLS}
\end{equation}
Similarly, 
since $l(\theta_{*}) = 0$ and $\pi \leq \theta_{*} \leq 3 \pi /2$,
we have
\begin{equation}
\theta_{*} = 
\pi + \mbox{Arcsin} \left( 
\frac{H}{2 \Lambda^{2} A} 
\right).
\label{eqn:theta* when H > 0 and 0 =< theta0 =< pi /2 in StabNLS}
\end{equation}
Then, from (\ref{eqn:theta0 when H > 0 and 0 =< theta0 =< pi /2 in
StabNLS})  
and (\ref{eqn:theta* when H > 0 and 0 =< theta0 =< pi /2 in StabNLS}), 
we obtain the upper bound of the collapse-time $t_{{\rm c},M}$, 
\begin{eqnarray}
t_{{\rm c},M} 
& = & 
\frac{1}{2 \Lambda}(\theta_{*} - \theta_{0})
\nonumber \\
& = &
\frac{1}{2\Lambda}
(
\pi /2 
+ \mbox{Arcsin} \, \xi
),
\label{eqn:t0M when H > 0 and 0 =< theta0 =< pi /2 in StabNLS}
\end{eqnarray}
with $\xi$ being defined by (\ref{eqn:def of z1 in StabNLS}).

      Next, we consider the case that $\pi /2 < \theta_ {0} < 3\pi /2$, namely
\begin{equation}
\left. \frac{{\rm d}}{{\rm d}t} \langle
|x|^{2} \rangle \right|_{t = 0} < 0.
\label{eqn:cond for variance and derivative in pi/2 < theta0 =< pi in StabNLS} 
\end{equation} 
In this case, if
\begin{equation}
A \geq \frac{H}{2 \Lambda^{2}},
\label{eqn:sufficient condition for collape H > 0 and pi/2 < theta0 =<
pi in StabNLS}
\end{equation}
$l(\theta (t))$ becomes negative for some value of $t$
which lies in $0 \leq t \leq \frac{\pi}{2\Lambda}$.
Then, if (\ref{eqn:sufficient condition for collape H > 0 and pi/2 < theta0 =<
pi in StabNLS}) is satisfied, 
we see from (\ref{eqn:inequality for variance and g(theta) in StabNLS})
that $\phi$ collapses in a finite time, and 
the upper bound of the collapse time $t_{{\rm c},M}$ is calculated 
in the same way as
(\ref{eqn:t0M when H > 0 and 0 =< theta0 =< pi /2 in StabNLS}).

    We further investigate the collapse-condition 
(\ref{eqn:sufficient condition for collape H > 0 and pi/2 < theta0 =<
pi in StabNLS}).    
Substituting (\ref{eqn:value of A in StabNLS}) 
into (\ref{eqn:sufficient condition for collape H
> 0 and pi/2 < theta0 =< pi in StabNLS}), we have
\begin{equation}
\left( \frac{1}{2 \Lambda} \left. \frac{{\rm d}}{{\rm d}t} \langle
|x|^{2} \rangle \right|_{t = 0} \right)^{2}
\geq
\langle |x|^{2} \rangle|_{t = 0}
\left( \frac{H}{\Lambda^{2}} - \langle |x|^{2} \rangle|_{t = 0}  \right).
\label{eqn:sufficient condition for collape H > 0 and pi/2 < theta0 =<
pi No. 2 in StabNLS}
\end{equation}
When $H/\Lambda^{2} \leq \langle |x|^{2} \rangle|_{t = 0}$
which is equivalent to the collapse-condition for the $0 \leq \theta_{0}
\leq \pi /2$ case (\ref{eqn:sufficient condition for collape in H > 0
and 0 =< theta0 =< pi /2 No. 2 in StabNLS}), 
the right hand side of (\ref{eqn:sufficient condition for collape H >
0 and pi/2 < theta0 =< pi No. 2 in StabNLS}) is negative. 
Then, regardless of the value of 
$\frac{1}{2 \Lambda} \left. \frac{{\rm d}}{{\rm d}t} \langle
|x|^{2} \rangle \right|_{t = 0}$, (\ref{eqn:sufficient condition for
collape H > 0 and pi/2 < theta0 =< pi No. 2 in StabNLS}) holds.
When $H/\Lambda^{2} \geq \langle |x|^{2} \rangle|_{t = 0}$,
the right hand side of (\ref{eqn:sufficient condition for collape H >
0 and pi/2 < theta0 =< pi No. 2 in StabNLS}) is positive.
By taking the square roots of both sides of 
(\ref{eqn:sufficient condition for collape H > 0 and pi/2 < theta0 =<
pi No. 2 in StabNLS}) and noting that 
$\left. \frac{{\rm d}}{{\rm d}t} \langle |x|^{2} \rangle \right|_{t = 0}$ 
is negative, we get 
\begin{equation}
\left. \frac{{\rm d}}{{\rm d}t} \langle
|x|^{2} \rangle \right|_{t = 0}
\leq 
- 2 
\sqrt{
\langle |x|^{2} \rangle|_{t = 0}
\left( 
H 
- \Lambda^{2} \langle |x|^{2} \rangle|_{t = 0}
\right)
}
.
\label{eqn:sufficient condition for collape H > 0 and pi/2 < theta0 =<
pi No. 3 in StabNLS}
\end{equation}
The condition (\ref{eqn:sufficient condition for collape H > 0 and
pi/2 < theta0 =< pi No. 3 in StabNLS})
can be regarded as an extention of   
the one obtained by Weinstein in the case (iii) of {\bf Theorem 4.2} 
in Ref.~\cite{Weinstein}, 
where the ``free'' or the  conventional (i.e. without external potential) 
nonlinear Schr\"odinger field is considered.
In fact, if $\nu_{n} = 0$ for all $n$, (\ref{eqn:sufficient condition for
collape H > 0 and pi/2 < theta0 =< pi No. 3 in StabNLS}) becomes the same
(apart from the discord of the factor, 2) as the one for the free 
nonlinear Schr\"odinger (NLS) field~\cite{Weinstein}. We note that, if 
\begin{equation}
\nu_{n} = \nu \,\,\,\,\, \mbox{(for all $n$),} 
\label{eqn:lambdan = lambda in StabNLS}
\end{equation}
we have
\begin{equation}
H - \Lambda^{2} \langle |x|^{2} \rangle|_{t = 0}
=
\|\nabla \phi_{0}\|_{2}^{2} - \| \phi_{0} \|_{4}^{4}/2,
\label{eqn:H - Lambda 2 variance at t = 0 for the same lambda
in StabNLS}
\end{equation} 
and thus
the collapse-condition (\ref{eqn:sufficient condition for collape H >
0 and pi/2 < theta0 =< pi No. 3 in StabNLS}) becomes formally the same 
as in the free NLS field case.

     Finally, we consider the case that $3 \pi /2 \leq \theta_ {0} < 2 \pi$,
 namely 
\begin{equation}
\langle |x|^{2} \rangle|_{t = 0} 
<
\frac{H}{2 \Lambda^{2}},
\,\,\,\,\,\,\,\,\,\,\,\,\,\,
\left. \frac{{\rm d}}{{\rm d}t} \langle
|x|^{2} \rangle \right|_{t = 0} 
\geq 
0.
\label{eqn:cond for variance and derivative in 3 pi/2 < theta0 =< 2 pi
in StabNLS}  
\end{equation} 
From (\ref{eqn:value of variance at t = 0 in StabNLS}),
$l(\theta_{0})$ should be positive, yielding that
the function $l(\theta )$ in (\ref{eqn:def of g(theta) in StabNLS})
is always positive for $\theta_{0} \leq \theta \leq
\theta_{0} + \pi$ or equivalently $0 \leq t \leq \frac{\pi}{2 \Lambda}$.
Then, we cannot see from (\ref{eqn:inequality for variance and
g(theta) in StabNLS}) whether $\langle |x|^{2} \rangle$ becomes
negative for $0 \leq t \leq \frac{\pi}{2 \Lambda}$. Further, for $t \geq  
\frac{\pi}{2 \Lambda}$, the inequality (\ref{eqn:inequality for variance and
g(theta) in StabNLS}) does not hold any more, because the integral
\begin{equation}
\int_{0}^{t} f(u) \sin [2 \Lambda (t - u)]
{\rm d}u ,
\label{eqn:integral of f(u) and sin [2 Lambda (t - u)]}
\end{equation}
appearing in the right hand side of (\ref{eqn:sol of variance in StabNLS})
is not necessarily positive for $t \geq \frac{\pi}{2 \Lambda}$.
Accordingly, we cannot see whether $\langle |x|^{2} \rangle$ becomes
negative for $t \geq \frac{\pi}{2 \Lambda}$, either.
Therefore, in the case that $3 \pi /2 \leq \theta_ {0} < 2 \pi$,
we cannot conclude that the wavefunction $\phi$ collapses in a finite time.

     It is interesting to consider a special case that 
the space is two-dimensional ($D = 2$), and the constants $\{ \nu_{n} \}$ 
satisfy (\ref{eqn:lambdan = lambda in StabNLS}).
In this case, $f(t)$ defined by (\ref{eqn:def of f(t) in StabNLS}) 
is identically zero, and (\ref{eqn:sol of variance in StabNLS})
reduces to
\begin{equation}
\langle |x|^{2} \rangle 
=  
l(\theta (t))
=
\frac{H}{2 \nu^{2}} 
+ A \sin (2 \nu t + \theta_{0}).
\label{eqn:sol of variance in StabNLS for D = 2 and lambda n = lambda}
\end{equation}
Then, we can trace the exact time-evolution of $\langle |x|^{2} \rangle$. 
If the collapse of $\phi$ happens which can be proved in the
same manner as in the previous cases, the collapse-time $t_{{\rm c}}$ is 
equal to its upper bound $t_{{\rm c},M}$. 

\bigskip


   To conclude this sub-section, some comments are in order.
First, what we have presented in this section is a sufficient
condition for the collapse. There remains a possibility that,
even when the collapse-condition in this section is not satisfied,
the expectation value $\langle |x|^{2} \rangle$ may go to zero
after several damped oscillations,
which is due to the effect of the negative non-homogeneous term
$-f(t)$ ($D \geq 2$) in the ordinary differential equation
(\ref{eqn:eq of motion for variance in harmonic potential in StabNLS}).
In this case, the upper bound of the collapse-time $t_{{\rm c},M}$ 
(\ref{eqn:t0M when H > 0 and 0 =< theta0 =< pi /2 in StabNLS})
will be replaced by larger ones.
A quantitative analysis of $f(t)$ is required to make clear this 
damped-oscillating phenomenon, which will be a future problem.

   Second, different from the discussion
in the repulsive case ($g = +1$),
we have not considered the case $D = 1$ here.
If $D = 1$, the positivity of $f(t)$ 
defined by (\ref{eqn:def of f(t) in StabNLS}) 
is not guaranteed, and thus we cannot examine the stability
of the wavefunction by using the above mentioned method.
The one-dimensional and self-focusing NLS
equation with no external potential is integrable and
has stable solition solutions~\cite{ZakharovShabat}. 
The external harmonic potentials destroy the integrability. 
However, since the equation has no singularity at the origin, 
we conjecture that the effect 
of the potentials is only a deformation of solitary waves.
  
\section{Collapse of the condensate}
\label{sec:BecPL}
\setcounter{equation}{0}
\setcounter{figure}{0}

We predict the collapse of the Bose-Einstein condensate in a magnetic trap
where the atoms have a negative $s$-wave scattering 
length~\cite{TW1}--\cite{WT}.
We prove that the singularity of wavefunction emerges in a
finite time even when the total energy of the system is positive. 
In addition, we present a refined formula for a critical number of
atoms above which the collapse of the condensate occurs. 
This number for $^{7}{\rm Li}$ atoms can be the same as the one in the
recent experiment.


\subsection{Time-evolution of the condensate}
\label{sec:Time-Evolution of the Condensate}
        
  We consider the Bose-Einstein condensate confined by magnetic
fields and investigate the time evolution of wavefunction when the
interactions between atoms are effectively attractive. 
Since the magnetic trap has an axial symmetry and the
system is near the ground state, 
we assume that the wavefunction is axially symmetric.  
The axis of the symmetry is chosen to be the $z$-axis.
Let $t$ denote time and $r_{\bot}$ the radius of the projection 
of the position vector $\bm{r}$ on the $xy$-plane.
Time-evolution of macroscopic wavefunction $\Psi (r_{\bot}, z, t)$ of
the condensate at very low temperature is described by the 
Gross-Pitaevskii equation for the system~\cite{P}--\cite{Gross2},
\begin{equation}
{\rm i} \hbar \frac{\partial \Psi }{\partial t }
= 
- \frac{\hbar^{2}}{2m}\frac{1}{r_{\bot}}\frac{\partial }{\partial r_{\bot}}
\left( r_{\bot} \frac{\partial \Psi }{\partial r_{\bot}}\right)
-\frac{\hbar^{2}}{2m} \frac{\partial^{2} \Psi }{\partial z^{2}} 
+ \frac{1}{2} m \omega_{\bot}^{2}r_{\bot}^{2}\Psi
+ \frac{1}{2} m \omega_{z}^{2}z^{2}\Psi 
+ \frac{4\pi \hbar^{2}a}{m}  |\Psi |^{2} \Psi .
\label{eqn:GP eq in 3D case}
\end{equation}
Here $m$ is the atomic mass, $\omega_{z}$ and $\omega_{\bot}$ 
are the trap (angular) frequencies along the $z$-axis and in the $xy$-plane,
respectively,
and $a$ is the $s$-wave scattering length. 
We take $a$ to be  negative.
By introducing a characteristic length $r_{0}$, 
\begin{equation}
r_{0} \equiv \left( \frac{\hbar }{2 m \omega_{\bot}}\right)^{1/2},
\label{eqn:def of r0}
\end{equation}
we prepare dimensionless variables,
\begin{equation}
\rho = r_{\bot}/r_{0}, \,\,\,\,\,\,\, \xi = z/r_{0}, 
\,\,\,\,\,\,\, \tau = \omega_{\bot} t, \,\,\,\,\,\,\, \psi = r_{0}^{3/2} \Psi .
\label{eqn:scaling in 3Dcase}
\end{equation}
In terms of these variables, Eq.~(\ref{eqn:GP eq in 3D case}) is
written in the dimensionless form 
\begin{equation}
{\rm i} \frac{\partial \psi }{\partial \tau } 
= -\frac{1}{\rho}
\frac{\partial }{\partial \rho}\left( \rho \frac{\partial \psi
}{\partial \rho }\right) 
- \frac{\partial^{2} \psi}{\partial \xi^{2}}
 + \frac{1}{4} (\rho{^2} + \delta^{2}\xi^{2})\psi +  c | \psi |^{2} \psi ,  
\label{eqn:GP eq after scaling in 3D case}
\end{equation}
where 
\begin{equation}
\delta \equiv \omega_{z}/\omega_{\bot}, \,\,\,\,\,\, c \equiv 8 \pi a/r_{0}.
\label{eqn:def of c in 3D case}
\end{equation}
Note that the constant $c$ is dimensionless and negative. 
In the terminology of the soliton theory,
we refer to Eq.~(\ref{eqn:GP eq after scaling in 3D case}) as axially symmetric
attractive nonlinear Schr\"odinger (NLS) equation with harmonic potentials. 
The NLS equation (\ref{eqn:GP eq after scaling
in 3D case}) has two integrals of motion, 
\begin{equation}
I_{1} = 2 \pi \int_{-\infty}^{\infty}{\rm d}\xi
\int_{0}^{\infty} \rho \, {\rm d}\rho | \psi |^{2},
\label{eqn:number of particles in 3D case}
\end{equation}
\begin{equation}
I_{2} = 2 \pi \int_{-\infty}^{\infty}{\rm d}\xi \int_{0}^{\infty} \rho
\, {\rm d}\rho \left [ |\partial_{\rho} \psi  |^{2} 
+ |\partial_{\xi} \psi  |^{2} 
+ \frac{1}{4} (\rho^{2} + \delta^{2}\xi^{2}) |\psi |^{2} 
+ \frac{1}{2} c |\psi |^{4} \right ] . 
\label{eqn:energy of field in 3D case}
\end{equation}
The total number of atoms
$N$ and the total energy $E$ are related to the constants $I_{1}$ and
$I_{2}$ as 
\begin{equation}
N = I_{1}, \,\,\,\,\,\, E = \hbar \omega_{\bot} I_{2}. 
\label{eqn:relation between I1 and N in 3D case}
\end{equation} 
By definition, $I_{1}$ is positive. We define
\begin{equation}
\langle \eta^{2}\rangle = 
2 \pi \int_{-\infty}^{\infty}{\rm d}\xi \int_{0}^{\infty}
\rho \, {\rm d}\rho (\rho^{2} + \xi^{2}) |\psi |^{2},
\label{eqn:average of rho in 3D case}
\end{equation}
which measures the geometrical extent of the wavefunction $\psi$ 
and plays a key role in the Zakharov's theory~\cite{Z,ZS}.

  Differentiating (\ref{eqn:average of rho in 3D case}) 
by $\tau$ twice and using (\ref{eqn:GP eq after scaling in 3D case}), we have
\begin{equation}
\frac{{\rm d}^{2}}{{\rm d}\tau^{2}} \langle \eta^{2} \rangle 
= 8 I_{2} - 4 \Omega^{2} \langle \eta^{2} \rangle - f(\tau ),
\label{eqn:eq of motion for ave of rho in 3D case}
\end{equation}
where $\Omega$ and $f(\tau )$ stand for 
\begin{equation}
\Omega \equiv {\rm min}(1,\delta ),
\label{eqn:def of Omega}
\end{equation}
\begin{equation}
f(\tau ) 
\equiv  
4 \pi \int_{-\infty}^{\infty}{\rm d}\xi \int_{0}^{\infty}
\rho \, {\rm d}\rho [ 2(\delta^{2} - 1)\xi^{2} |\psi |^{2}
+ |c| |\psi |^{4} ] \,\,\,\,\,\, \mbox{(for $\delta > 1$)},
\label{eqn:def of f for lambda > 1}
\end{equation}
\begin{equation}
f(\tau ) 
\equiv 
4 \pi \int_{-\infty}^{\infty}{\rm d}\xi \int_{0}^{\infty}
\rho \, {\rm d}\rho [ 2(1 - \delta^{2})\rho^{2} |\psi |^{2}
+ |c| |\psi |^{4} ] \,\,\,\,\,\, \mbox{(for $0 < \delta < 1$)}.
\label{eqn:def of f for 0 < lambda < 1}
\end{equation}
A differential equation (\ref{eqn:eq of motion for ave of rho in 3D case})
for $\langle \eta^{2} \rangle$ is readily solved to give 
\begin{equation}
\langle \eta^{2} \rangle 
=  
A \sin (2 \Omega \tau + \theta_{0}) + \frac{2}{\Omega^{2}} I_{2} 
-\frac{1}{2\Omega} \int_{0}^{\tau} f(u) \sin [2 \Omega (\tau - u)]
{\rm d}u.
\label{eqn:sol of ave of rho in 3D case}
\end{equation}
Constants $A$ and $\theta_{0}$ are to be determined 
by the initial conditions on $\psi$. 
Without loss of generality, we assume $A$ to be positive.

   We first consider the case that $I_{2}$ and therefore the total
energy of the condensate $E$ are negative. Because $\langle \eta^{2} \rangle$
is a positive quantity, the right hand side of (\ref{eqn:sol of ave of rho
in 3D case}) should also be positive for $\tau = 0$. Accordingly,
$\theta_{0}$ in (\ref{eqn:sol of ave of rho in 3D case}) must be 
$0 < \theta_{0} < \pi$. 
Let us define a function, $G(\tau )$, as 
\begin{equation}
G(\tau ) \equiv A \sin (2 \Omega \tau + \theta_{0}) +
\frac{2}{\Omega^{2}} I_{2}.  
\label{eqn:def of G+ in 3D case} 
\end{equation}
Since the function $f(\tau )$ is always positive and 
$\sin [2 \Omega (\tau - u)]$ in (\ref{eqn:sol of ave of rho in 3D case})
is positive for $0 \leq \tau \leq \pi /2 \Omega$, the last term in
(\ref{eqn:sol of ave of rho in 3D case}) is negative for $0 \leq \tau
\leq \pi /2 \Omega$. We thus obtain an inequality,
\begin{equation}
\langle \eta^{2} \rangle \,\, \leq \,\, G(\tau ),  
\label{eqn:ave of rho is smaller than G+}
\end{equation}
for $0 \leq \tau \leq \pi /2 \Omega$.
From the definition (\ref{eqn:def of G+ in 3D case}), 
we see that  $G(\tau )$
becomes negative for some value of $\tau$ which lies in $0 \leq \tau
\leq \pi /2\Omega$, if $0 < \theta_{0} < \pi$. 
 Then, from the inequality 
(\ref{eqn:ave of rho is smaller than G+}), $\langle \eta^{2} \rangle $
should become negative for some value of $\tau$, 
$0 \leq \tau \leq \pi /2\Omega$. This contradicts to the
definition of $\langle \eta^{2} \rangle $
(\ref{eqn:average of rho in 3D case}), which implies the development
of the singularity.
Thus, when the total energy of the condensate $E$ is
negative, the wavefunction $\psi$ 
surely collapses in a finite time, regardless of the initial condition
on $\psi$. 

  Next, we  prove that, 
even when the total energy $E$ is positive, 
the collapse of wavefunction occurs in a finite time for a certain class 
of the initial conditions on $\psi$.
Actually, in the experiments of the Bose-Einstein condensation under a magnetic
trap, the total energy $E$ and therefore $I_{2}$ are positive. 
We assume the initial shape of $\psi$ to be gaussian, 
\begin{equation}
\psi (\rho ,\xi , \tau = 0)
= \frac{(q^{2}k\delta^{1/2}I_{1})^{1/2}}{(2 \pi )^{3/4}}
\exp [-(q^{2}\rho^{2} + \delta k^{2} \xi^{2})/4] ,
\label{eqn:initial shape of psi in 3D case}
\end{equation} 
where $q$ and $k$ are positive parameters.
From Eqs.~(\ref{eqn:GP eq after
scaling in 3D case}), (\ref{eqn:average of rho in 3D case})  
and (\ref{eqn:initial shape of psi in 3D case}), we can show that 
\begin{equation}
\frac{{\rm d}}{{\rm d}\tau }\langle \eta^{2} \rangle  = 0 
\,\,\,\,\,\,  \mbox{(for $\tau = 0$)},  
\label{eqn:initial cond of first derivative of ave of rho in 3D case}
\end{equation}
which yields $\theta_{0} = \pi /2$ or $\theta_{0} = 3\pi /2$
in (\ref{eqn:sol of ave of rho in 3D case}).
Correspondingly, the solution (\ref{eqn:sol of ave of rho in 3D
case}) now yields
\begin{equation}
\langle \eta^{2} \rangle  = \langle \eta^{2} \rangle_{+} 
\equiv 
A \cos (2 \Omega \tau ) +
\frac{2}{\Omega^{2}} I_{2}  
-\frac{1}{2\Omega} \int_{0}^{\tau} f(u) \sin [2\Omega (\tau - u)]
{\rm d}u,
\label{eqn:sol of ave of rho in 3D case for theta0 = pi/2}
\end{equation}
when $\theta_{0} = \pi /2$, and
\begin{equation}
\langle \eta^{2} \rangle = \langle \eta^{2} \rangle_{-}
\equiv  - A \cos (2\Omega \tau ) + 
\frac{2}{\Omega^{2}} I_{2} 
-\frac{1}{2\Omega} \int_{0}^{\tau} f(u) \sin [2\Omega (\tau - u)] \,
{\rm d}u,
\label{eqn:sol of ave of rho in 3D case for theta0 = -pi/2}
\end{equation}
when $\theta_{0} = 3\pi /2$.  
Substituting the initial shape (\ref{eqn:initial shape of psi in 3D
case}) into  $\langle \eta^{2} \rangle |_{\tau = 0}$ and $I_{2}$, we get 
\begin{equation}
\langle \eta^{2} \rangle |_{\tau = 0} = (2 q^{-2} + \delta^{-1} k^{-2}) I_{1},
\label{eqn:ave of rho for tau = 0 from initial condition in 3D case}
\end{equation}
\begin{equation}
I_{2} =  \left[ \frac{1}{2}(q^{2} + q^{-2}) 
+ \frac{\delta}{4}(k^{2} + k^{-2}) \right] I_{1}
+ \frac{c q^{2} k \delta^{1/2}}{16 \pi^{3/2}} I_{1}^{2}.
\label{eqn:I2 from initial condition in 3D case}
\end{equation}
From Eq.~(\ref{eqn:I2 from initial condition in 3D case}), we see that 
$I_{2}$ is positive when
\begin{equation}
I_{1} < I_{1,a}(q,k).
\label{eqn: I1  < I1n for Gaussian in 3D case}
\end{equation}
Here
\begin{equation}
I_{1,a}(q,k) \equiv  \frac{4 \pi^{3/2}}{|c|q^{2} k \delta^{1/2}}
[ 2 (q^{2} + q^{-2}) 
+ \delta (k^{2} + k^{-2}) ] .
\label{eqn:def of I10 in 3Dcase}
\end{equation}
On the other hand, Eqs.~(\ref{eqn:ave of rho for tau = 0 from initial
condition in 3D case})  and (\ref{eqn:I2 from initial condition in 3D
case}) give
\begin{equation}
\langle \eta^{2} \rangle |_{\tau = 0} - \frac{2}{\Omega^{2}} I_{2}
= \frac{|c| q^{2}k\delta^{1/2}}{8 \pi^{3/2}\Omega^{2}} I_{1}
(I_{1} - I_{1,b}(q,k)), 
\label{eqn:ave of rho for tau = 0 - 2 I2 No.1}
\end{equation}
where $I_{1,b}(q,k)$ is defined as
\begin{equation}
I_{1,b}(q,k) \equiv \frac{4 \pi^{3/2}}{|c|q^{2}k\delta^{1/2}} g(q,k),
\label{eqn:def of I1+- in 3Dcase}
\end{equation}
with
\begin{equation}
g(q,k) \equiv  2 [ q^{2} + (1 - 2\Omega^{2}) q^{-2} ]
+ \delta [ k^{2} + (1 - 2\Omega^{2}/\delta^{2}) k^{-2} ] .
\label{eqn:def of h+-}
\end{equation}
Depending on whether $I_{1}$ is larger or smaller than $I_{1,b}(q,k)$,
we can determine which of $\langle \eta^{2} \rangle_{\pm}$ should 
be chosen as the solution of Eq.~(\ref{eqn:eq of motion for ave of rho
in 3D case}). When $I_{1} > I_{1,b}(q,k)$ ($I_{1} < I_{1,b}(q,k)$), 
$\langle \eta^{2} \rangle_{+}$ 
($\langle \eta^{2} \rangle_{-}$) is the solution of  
Eq.~(\ref{eqn:eq of motion for ave of rho in 3D case}). Remark that
the function $g(q,k)$ may take both positive and negative values, 
because $1-2\Omega^{2} = -1 < 0$ for
$\delta > 1$ and $1 - 2(\Omega /\delta )^{2} = -1 < 0$ for $0 < \delta < 1$. 

     In the case that $g(q,k) > 0$ and therefore $I_{1,b}(q,k) > 0$, 
if $I_{1} < I_{1,b}(q,k)$, $\langle \eta^{2} \rangle_{-}$
is an appropriate solution of Eq.~(\ref{eqn:eq of motion for ave of rho
in 3D case}) and satisfies
\begin{equation}
\langle \eta^{2} \rangle_{-} \,\, \leq \,\, -A \cos (2\Omega \tau ) +
\frac{2}{\Omega^{2}}I_{2},
\label{eqn:ave of rho is smaller than -cos}
\end{equation}
for $0 \leq \tau \leq \pi /2\Omega$.
The right hand side of the inequality (\ref{eqn:ave of rho is smaller than
-cos}) is always positive because $A = \frac{2}{\Omega^{2}} I_{2} 
- \langle \eta^{2} \rangle |_{\tau = 0} < 2 I_{2}/\Omega^{2}$.
Then, we cannot see from (\ref{eqn:ave of rho is
smaller than -cos}) whether $\langle \eta^{2} \rangle _{-}$ surely
becomes negative for $0 \leq \tau \leq \pi /2\Omega$.
Further, for $\tau \geq \pi /2\Omega$, 
the inequality (\ref{eqn:ave of rho is smaller than -cos})
does not hold any more, because the integral,
\begin{equation}
\int_{0}^{\tau} f(u) \sin [2 \Omega (\tau - u)]{\rm d}u,
\label{eqn:integral of f and sin}
\end{equation}
is not necessarily positive for $\tau \geq \pi /2\Omega$.
Accordingly, we cannot see whether $\langle \eta^{2} \rangle_{-}$
becomes negative for $\tau \geq \pi /2\Omega$, either.
Therefore, when $I_{1} < I_{1,b}(q,k)$, we do not conclude that 
the wavefunction $\psi$ collapses in a finite time.
This conclusion is consistent with that 
in the case $3 \pi /2 \leq \theta_ {0} < 2 \pi$ 
in the sub-section \ref{sec:Attractive Case in StabNLS}.
On the other hand, if $I_{1} > I_{1,b}(q,k)$,  
the behavior of $\langle \eta^{2} \rangle$ may change drastically.
As the solution of (\ref{eqn:eq of motion
for ave of rho in 3D case}), $\langle \eta^{2} \rangle_{+}$ 
defined by (\ref{eqn:sol of ave of rho in 3D case for theta0 = pi/2}) is
appropriate and satisfies
\begin{equation}
\langle \eta^{2} \rangle_{+} \,\, \leq \,\, A \cos (2\Omega\tau) 
+ \frac{2}{\Omega^{2}} I_{2},
\label{eqn:ave of rho is smaller than G+ for theta0 = pi/2}
\end{equation}
for $0 \leq \tau \leq \pi /2\Omega$.
Using (\ref{eqn:sol of ave of rho in 3D case for theta0 = pi/2}) 
and (\ref{eqn:ave of rho for tau = 0 - 2 I2 No.1}), we have 
\begin{equation}
A = 
\langle \eta^{2} \rangle |_{\tau = 0} - \frac{2}{\Omega^{2}} I_{2}  
= 
-\frac{1}{2\Omega^{2}} g(q,k) I_{1} 
+ \frac{|c| q^{2}k\delta^{1/2}}{8 \pi^{3/2}\Omega^{2}} I_{1}^{2}.
\label{eqn:A for Gaussian in 3D case}
\end{equation}
Suppose that  
\begin{equation}
A \geq 2I_{2}/\Omega^{2},
\label{eqn:A < 2 I2 in 3D case} 
\end{equation} 
which is possible as seen from (\ref{eqn:A for Gaussian in 3D case}). 
When the inequality (\ref{eqn:A < 2 I2 in 3D case}) holds, 
the right hand side of (\ref{eqn:ave of rho is smaller than G+ for
theta0 = pi/2}) becomes negative for some value of $\tau$ which lies $0
\leq \tau \leq \pi /2\Omega$. Then, from (\ref{eqn:ave of rho is smaller
than G+ for theta0 = pi/2}), $\langle \eta^{2} \rangle_{+}$ also 
becomes negative for some value of $\tau$, $0 \leq \tau \leq \pi /2\Omega$. 
This is contradictory to the positivity of $\langle \eta^{2} \rangle_{+}$, 
implying that the singularity of the wavefunction emerges in a finite time. 
In other words, the wavefunction $\psi$ collapses in a finite time.
Let us rewrite the condition (\ref{eqn:A < 2 I2 in 3D case}) so as to
make its physical significance clear. 
Substituting (\ref{eqn:I2 from initial
condition in 3D case})  and (\ref{eqn:A for Gaussian in 3D case}) 
into (\ref{eqn:A < 2 I2 in 3D case}), we get
\begin{equation}
I_{1} \geq I_{1,c}(q,k),
\label{eqn:A < 2 I2 for Gaussian in 3D case}
\end{equation}
where
\begin{equation}
I_{1,c}(q,k) \equiv \frac{4 \pi^{3/2}}{|c|q^{2}k\delta^{1/2}} h(q,k),
\label{eqn:def of I1c in 3Dcase}
\end{equation}
with
\begin{equation}
h(q,k) \equiv  2 [ q^{2} + (1 - \Omega^{2}) q^{-2} ]
+ \delta [ k^{2} + (1 - \Omega^{2}/\delta^{2}) k^{-2} ] .
\label{eqn:def of hc}
\end{equation}
Contrary to $g(q,k)$, the function $h(q,k)$ is always positive. 
From (\ref{eqn:def of I10 in 3Dcase}), (\ref{eqn:def of I1+- in 3Dcase}) 
and (\ref{eqn:def of I1c in 3Dcase}), we find  that
\begin{equation}
I_{1,b} < I_{1,c} < I_{1,a} \,\,\,\,\,\,\,\, \mbox{(for all $q$ and $k$)}. 
\label{eqn:I1+- < I1c < I1n}
\end{equation} 
Therefore, if the initial shape of $\psi$ is gaussian
(\ref{eqn:initial shape of psi in 3D case}),
the collapse occurs when $I_{1} > I_{1,c}(q,k)$. The inequality
(\ref{eqn:I1+- < I1c < I1n}) assures the consistency of the analysis.

    In the case that $g(q,k) < 0$, and therefore $I_{1,b}(q,k) < 0$, 
$\langle \eta^{2} \rangle_{+}$ 
(\ref{eqn:sol of ave of rho in 3D case for theta0 = pi/2})
is the only solution of (\ref{eqn:eq of motion for ave of rho in 3D case})
because $I_{1}$ should be positive. 
Except for this, the discussion for the collapse goes in the
same manner for the $I_{1,b}(q,k) > 0$ case. Thus, we conclude that
the solution of the NLS equation (\ref{eqn:GP eq after scaling in 3D
case}) with the initial condition (\ref{eqn:initial shape of psi in 3D
case}) surely collapses in a finite time for $I_{1} \geq I_{1,c}(q,k)$ even
when the total energy is positive.  

    The above result predicts an interesting phenomenon, the collapse of 
the Bose-Einstein condensate, for the assembly of bosonic atoms with a 
negative $s$-wave scattering length. Using (\ref{eqn:def of c in 3D
case}), (\ref{eqn:relation between I1 and N in 3D case}) and
(\ref{eqn:def of I1c in 3Dcase}) in (\ref{eqn:A < 2 I2 for Gaussian in
3D case}), we obtain a critical number of atoms $N_{{\rm c}}$,  
\begin{equation}
N_{{\rm c}}(q,k) = N_{0}\frac{h(q,k)}{q^{2} k},
\label{eqn:Nc in 3D case}
\end{equation}
with
\begin{equation}
N_{0} \equiv (\pi /\delta )^{1/2} r_{0}/(2 |a|) .
\label{eqn:def of N0}
\end{equation}
The collapse of the wavefunction occurs when the number of the trapped 
atoms, $N$, exceeds $N_{{\rm c}}(q,k)$.
The formula (\ref{eqn:Nc in 3D case}) gives the critical number of atoms
$N_{{\rm c}}$ as a function of two parameters $q$ and $k$ 
in (\ref{eqn:initial shape of psi in 3D case}). 
The extents of the initial wavefunction are proportional to $q^{-1}$ for 
the $\rho$-direction and $k^{-1}$ for the $\xi$-direction. 
When $q$ and $k$ equal 1, the initial wavefunction (\ref{eqn:initial shape
of psi in 3D case}) represents 
the ground state for the Schr\"odinger equation under harmonic potentials,
\begin{equation}
{\rm i} \frac{\partial \psi }{\partial \tau } 
= -\frac{1}{\rho}
\frac{\partial }{\partial \rho}\left( \rho \frac{\partial \psi
}{\partial \rho }\right) 
- \frac{\partial^{2} \psi}{\partial \xi^{2}}
 + \frac{1}{4} (\rho{^2} + \delta^{2}\xi^{2})\psi .  
\label{eqn:Schrodinger equation under harmonic potentials}
\end{equation}
It is interesting that
the formula includes another anisotropy effect of the magnetic
trap through $\Omega = {\rm min}(1,\delta \equiv \omega_{z}/\omega_{\bot})$.


\subsection{Application to $^{7}{\rm Li}$ system}
\label{sec:AppToLi}

     We estimate the critical number for the assembly of 
$^{7}{\rm Li}$ atoms which have a negative $s$-wave scattering length.
In order to apply the above discussion to this system, 
we first improve quantitatively the evaluation 
for $\langle \eta^{2} \rangle_{+}$
(\ref{eqn:ave of rho is smaller than G+ for theta0 = pi/2}) as follows. 
If there exists a constant, $\bar{f}$, such that $0 < \bar{f} \leq f(\tau )$,
the inequality (\ref{eqn:ave of rho is smaller
than G+ for theta0 = pi/2}) can be replaced by  
\begin{equation}
\langle \eta^{2} \rangle_{+} \,\, \leq \,\, \left( A + 
\frac{\bar{f}}{4 \Omega^{2}} \right) \cos (2\Omega\tau) 
+ \left( \frac{2}{\Omega^{2}} I_{2} - \frac{\bar{f}}{4 \Omega^{2}}
\right) ,
\label{eqn:new estimate for eta2+}
\end{equation}
for $0 \leq \tau \leq \pi /2\Omega$.
Then, the condition for the collapse of $\psi$ is modified as
\begin{equation}
A \geq  \frac{2}{\Omega^{2}} I_{2} - \frac{\bar{f}}{2 \Omega^{2}},
\label{eqn:new condition for the collapse of psi} 
\end{equation} 
instead of (\ref{eqn:A < 2 I2 in 3D case}).
Here, we assume that the collapse of the wavefunction occurs
without oscillations, which means that $\langle \eta^{2} \rangle$ goes to zero
monotonously and the integral,
\begin{equation}
2 \pi \int_{-\infty}^{\infty}{\rm d}\xi \int_{0}^{\infty}
\rho \, {\rm d}\rho |\psi |^{4},
\label{eqn:integral of psi4}
\end{equation}
is a monotone increasing function of a (scaled) time $\tau$. 
Under this assumption, from (\ref{eqn:def of f for lambda > 1})
or (\ref{eqn:def of f for 0 < lambda < 1}), we can estimate $\bar{f}$ as
\begin{equation}
\bar{f} = 4 \pi |c| \int_{-\infty}^{\infty}{\rm d}\xi \int_{0}^{\infty}
\rho \, {\rm d}\rho |\psi (\rho, \xi, \tau = 0)|^{4}.
\label{eqn:estimate of delta}
\end{equation}
Using the initial condition on $\psi$ (\ref{eqn:initial shape
of psi in 3D case}) in (\ref{eqn:estimate of delta}), we get
\begin{equation}
\bar{f} = I_{1}^{2}k q^{2}/N_{0}.
\label{eqn:estimate of delta after substituting gaussian}
\end{equation}
Thus, substituting (\ref{eqn:I2 from initial condition in 3D case}),
(\ref{eqn:A for Gaussian in 3D case}) and (\ref{eqn:estimate of delta 
after substituting gaussian}) 
into (\ref{eqn:new condition for the collapse of psi}), 
we arrive at a renewed critical number of atoms, $N_{{\rm c, new}}$, 
\begin{equation}
N_{{\rm c, new}} = N_{0}\frac{h_{{\rm new}}(q,k)}{q^{2} k},
\label{eqn:new Nc}
\end{equation}
where
\begin{equation}
h_{{\rm new}}(q,k) \equiv \frac{2}{3}h(q,k).
\label{eqn:def of hnew}
\end{equation}
Apparently, $N_{{\rm c, new}}$ is smaller than $N_{{\rm c}}$.

   In the experiment for $^{7}{\rm Li}$~\cite{BSTH}, 
the values of frequencies $\omega_{z}$ and $\omega_{\bot}$ were
$\omega_{z} /2 \pi = 117 \, {\rm Hz}$ and $\omega_{\bot} /2 \pi = 163
\, {\rm Hz}$, which gives $\delta = 0.718$.
The $s$-wave scattering length of $^{7}{\rm Li}$ was 
observed to be $a = -27.3 a_{0}$ ($a_{0}$ : Bohr radius)~\cite{AMSH}. 
Substituting these experimental values into (\ref{eqn:def of N0}) and 
(\ref{eqn:def of hnew}), we get
\begin{equation}
N_{{\rm c, new}}(q,k)/N_{0} = 4 k^{-1}(1 + 0.485 q^{-4})/3
+ 0.479 q^{-2} k,
\label{eqn:new Nc for lambda = 0.71779}
\end{equation}
with $N_{0} = 1520$. 

\begin{figure}[htb]
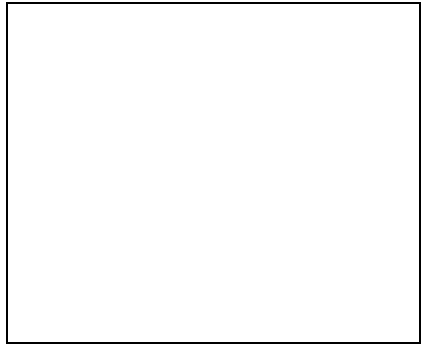

\begin{center}
\framebox[55mm]{\rule[-21mm]{0mm}{43mm}}
\end{center}
\caption{
$N_{{\rm c, new}}/N_{0}$ as a function of
$k$ for the case that $\delta = 0.718$~\protect\cite{BSTH}.
}
\label{fig:Li}
\end{figure}

    To fix a relation between $q$ and $k$, we use the minimal condition 
for $I_{2}$ (\ref{eqn:I2 from initial condition in 3D case}).
From $\partial_{q} I_{2}= 0$ and $\partial_{k} I_{2}= 0$, we have
\begin{equation}
q^{2} - q^{-2} = \delta (k^{2} - k^{-2}),
\label{eqn:minimal condition on I2}
\end{equation}
which yields $q$ as a function of $k$,
\begin{equation}
q = Q(k) \equiv \left[ \delta (k^{2} - k^{-2})/2 
+ \sqrt{ 1 + \delta^{2} (k^{2} - k^{-2})^{2}/4 } \right]^{1/2}.
\label{eqn:def of Q}
\end{equation}
Substitution of $\delta = 0.718$ into (\ref{eqn:def of Q}) gives
\begin{equation}
q = Q(k) \equiv \left[ 0.359 (k^{2} - k^{-2}) 
+ \sqrt{ 1 + 0.129 (k^{2} - k^{-2})^{2}} \right]^{1/2}.
\label{eqn:def of Q, lambda = 0.72}
\end{equation}
We plot $N_{{\rm c, new}}(Q(k),k)/N_{0}$ 
as a function of $k$ for $\delta = 0.718$ (Fig.~\ref{fig:Li}). 
The refined critical number $N_{{\rm c, new}}(Q(k),k)$ is a monotone
decreasing function of $k$. 
This behavior is reasonable: For smaller (larger) $k$, the wavefunction 
is more spread (confined), and such configuration imposes more (less)
severe restriction on the collapse.
In what follows, we consider the case $k > 1$ since the gaussian
wavefunction (\ref{eqn:initial shape of psi in 3D case}) 
with $q = k = 1$
does not include the effect of attractive
inter-atomic interaction which makes the exact ground state more confined.
This has been considered in the sub-section~\ref{sec:Attractive Case in AsymBec}.
For instance, in the case of almost spherical trap, the result 
(\ref{eqn:sol of LamdaBot in spherical approx in AsymBec}) gives $k=1.59$ and $q=1.45$.

  If we set $k = 1.20$, $Q(k)$ and $N_{{\rm c, new}}(Q(k),k)$ are 
calculated as $Q(1.20) = 1.14$ and $N_{{\rm c, new}}(Q(1.20),1.20) = 2840$. 
In the case that $k = 2.20$, $Q(k)$ and $N_{{\rm c,
new}}(Q(k),k)$ are estimated as $Q = 1.90$ and $N_{{\rm c, new}} = 1400$.
The latter value of $N_{{\rm c, new}}$ agrees well with the one in the
experiment, which is about $10^{3}$~\cite{BSH2}.
It is also consistent with the theoretical ones obtained by different
approaches, which are 
$(1 \sim 3) \times 10^{3}$~\cite{RHBE}--\cite{Stoof}, 
\cite{NTT}--\cite{KiZu}.

\section{Summary}
\label{sec:Dynamical Concl}
\setcounter{equation}{0}

 We have studied the dynamical properties of 
Bose-Einstein condensates.
In Sec.~\ref{sec:StabNLS}, 
as a preliminary to the analysis of the collapse of the condensate, 
we have considered the stability of the wavefunction of the
$D$-dimensional nonlinear Schr\"odinger equation (\ref{eqn:NLS under
external potential in StabNLS}) with harmonic potential
terms (\ref{eqn:def of harmonic potential in rep in StabNLS}), for
both repulsive ($g = +1$) and attractive ($g = -1$) cases. 
This is an extention of the Zakharov's theory~\cite{Z,ZS}.
In the repulsive case, 
we have shown that
the solution of Eq.~(\ref{eqn:NLS under external potential in StabNLS})
is absolutely stable for any spatial dimension
when confined in the harmonic potential.
In the attractive case for $D \geq 2$,
by solving the equation of
motion for the expectation value of the square of the radius 
(\ref{eqn:eq of motion for variance in harmonic potential in StabNLS}), 
we have investigated the time-evolution of the wavefunction $\phi$.
When the total energy of the system $H$ is equal to or less than zero,
the wavefunction $\phi$ surely collapses in a finite time,
regardless of the initial condition.
On the other hand, when $H > 0$, we have obtained conditions 
that the collapse of the wavefunction occurs, according to the value
of the constant $\theta_{0}$ appearing in (\ref{eqn:sol of variance in
StabNLS}). When $0 \leq \theta_{0} \leq \pi /2$ which is equivalent to 
(\ref{eqn:cond for variance and deri in 0 =< theta0 =< pi /2 in StabNLS}),
the collapse-condition is obtained as (\ref{eqn:sufficient condition
for collape in H > 0 and 0 =< theta0 =< pi /2 No. 2 in StabNLS}).  
When $\pi /2 < \theta_ {0} < 3\pi /2$, equivalent to (\ref{eqn:cond for
variance and derivative in pi/2 < theta0 =< pi in StabNLS}),
the collapse-condition is (\ref{eqn:sufficient condition for
collape H > 0 and pi/2 < theta0 =< pi No. 3 in StabNLS}).
This result is an extention of Weinstein's 
where the nonlinear Schr\"odinger (NLS) field without external potential,
is considered~\cite{Weinstein}. 
We have also obtained 
an upper bound of the collapse-time of the wavefunction
$t_{{\rm c},M}$, represented by the formula 
(\ref{eqn:t0M when H < 0 and theta- < theta0 =< pi /2 in StabNLS})
with (\ref{eqn:def of z1 in StabNLS}).
In a special case that $D = 2$ and $\nu_{n} = \nu$, 
since the function $f(t)$ defined by (\ref{eqn:def of f(t) in StabNLS}) 
is identically zero,
the time-evolution of $\langle |x|^{2} \rangle$ can be traced
exactly by (\ref{eqn:sol of variance in StabNLS for D = 2 and lambda n
= lambda}), and therefore 
the collapse-time $t_{{\rm c}}$ is just equal to its upper bound
$t_{{\rm c},M}$.   
Improvements of the collapse-conditions and the upper bound
of the collapse-time require a more detailed analysis of $f(t)$.
We, however, emphasize that the collapse time is the order of $1/\nu$.


     In Sec.~\ref{sec:BecPL}, 
we have considered the instability of the
Bose-Einstein condensate with effectively attractive inter-atomic interactions
under a magnetic trap. The Gross-Pitaevskii equation 
(\ref{eqn:GP eq in 3D case}) for the system is essentially the same as
the attractive nonlinear Schr\"odinger equation
with harmonic potential terms (\ref{eqn:GP eq after scaling in 3D case}). 
First, we have shown that the collapse of the wavefunction surely happens
in a finite time when the total energy of the condensate $E$ is negative.
Second, by assuming the initial condition on the wavefunction to
be gaussian (\ref{eqn:initial shape of psi in 3D case}) 
with two  parameters $q$ and $k$, we have proved that the collapse can
happen even when $E$ is positive.
This situation corresponds to the experiments on the Bose-Einstein condensate.
We have also proved that the condensate collapses when the number
of atoms $N$ exceeds $N_{{\rm c}}$ (\ref{eqn:Nc in 3D case}). 
   Further, we have
presented a refined
formula for the critical number $N_{{\rm c, new}}$ (\ref{eqn:new Nc}), 
which is smaller than $N_{{\rm c}}$.
Using a relation between $q$ and $k$ which is obtained from
the minimal condition for the energy (\ref{eqn:def of Q}), 
we have plotted  $N_{{\rm c, new}}/N_{0}$ 
as a function of $k$ for $\delta = 0.718$ (Fig.~\ref{fig:Li}).  
Our estimated number for the assembly of $^{7}{\rm Li}$, which is
about 1400 for $q = 1.90$ and  $k = 2.20$, agrees well with the one in the
experiment~\cite{BSH2}. 
We have simplified the analysis by assuming 
(i)~axial symmetry of the system, (ii)~zero temperature 
and (iii)~$s$-wave scattering. 
All the assumptions are more or less consistent in
the sense that we only consider the properties of the dilute condensate at
the very low energy region. 
In addition, we have not taken account (iv)~three-body collisions and 
(v)~the short-range repulsive part of the inter-atomic potential,
which may play important roles at the final state of the collapse. 
Detail examinations of those neglected effects are left for future studies.
On the other hand, it is extremely interesting to observe the
time-development of the wavefunction near the collapse.
The asymptotic analysis has been given in \cite{TW4}.

\clearpage

\chapter{Summary and Concluding Remarks}
\label{chap:concl}
\setcounter{equation}{0}

We have reviewed recent theoretical research 
on the properties of Bose-Einstein condensation, in particular, 
static and dynamical stabilities 
of Bose-Einstein condensates confined in magnetic trap.
It has been shown that the variational approach 
and the analysis of the time-dependent nonlinear Schr\"odinger equation 
give significant information on the stability of the condensates.

    Bose-Einstein condensation was discovered in 
liquid helium about 70 years ago, 
and has been ``rediscovered'' in alkali atomic gases confined in trap,
due to the developments of technology in atomic physics.
There are many advantages in experiments using atomic vapors 
under magnetic trap, which are summarized as follows.

\noindent 1)~The atom-atom interactions are weak 
(the $s$-wave scattering length is about $10^{-8} {\rm m}$, whereas at 
the required densities the inter-atomic spacing is about $10^{-6} {\rm
m}$, for the $^{87}{\rm  Rb}$ case~\cite{AEMWC}).
Therefore,  the purely quantum-statistical nature 
of the BEC transition can be captured.

\noindent 2)~We can choose
various species of bosonic (or fermionic) atoms and prepare even their 
mixtures with different number of each atoms. 

\noindent 3)~We have controllable parameters 
such as density, temperature and the frequencies of the magnetic trap.
In particular, the number of atoms and the temperature are now independent.
This offers an idealistic situation to study the quantum statistical mechanics.

\noindent 4)~The interaction strength can be varied by use of the
Feshbach resonance~\cite{InAnStMiStKe,StInAnMiStKe}. The $s$-wave
scattering length depends on the applied magnetic field and even the
sign of it changes.

\noindent These new aspects give rich dynamical behaviors of the condensate 
at microscopic, mesoscopic and macroscopic levels.
Understanding of those systems is fundamental 
in quantum statistical mechanics, condensed matter physics
and atom optics.
The trapped Bose-Einstein condensates may serve as a source of the
coherent matter waves.
In atom optics, we expect many applications such as atom
holography~\cite{Shimizu,ZoGoMe},  
atom interferometer~\cite{LeHaSmChRuPr} 
and atom laser~\cite{MeAnKuDuToKe}--\cite{HaDeKoWeHeRoPh}. 
In principle, Bose-Einstein condensation is expected to occur for 
all bosonic atoms and molecules with and without charges, and thus
we believe that those studies will continue to be interesting 
in coming years.

\clearpage

\chapter*{Acknowledgements}
\label{sec:Acknow in AsymBec}
\setcounter{equation}{0}
One of the authors (MW) likes to thank V. E. Zakharov,
L. P. Pitaevskii and Y. Kagan for useful comments.
This work was partially supported by Grant-in-Aid for Scientific
Research from the Ministry of Education, Science and Culture, Japan.
One of the authors (TT) acknowledges JSPS Research Fellowships
for Young Scientists.

\clearpage

\appendix

\renewcommand{\theequation}
{\Alph{chapter}.\arabic{equation}} 

\chapter{Bose-Einstein Condensation of a Free Boson Gas Confined in
Harmonic Potentials}
\label{chap:App3}
\setcounter{equation}{0}

  We discuss the Bose-Einstein condensation of a free gas under
harmonic potentials.
The $N$-body Hamiltonian (first quantization) is 
\begin{eqnarray}
H & = & \sum_{j=1}^{N} H_{j}
\nonumber \\
& = & \sum_{j=1}^{N} 
\left[
\frac{1}{2m}\left(p_{jx}^{2} + p_{jy}^{2} + p_{jz}^{2}\right)
+ \frac{m}{2}\left(\omega_{x}^{2}x_{j}^{2}+\omega_{y}^{2}y_{j}^{2}+\omega_{z}^{2}z_{j}^{2}
\right)
\right],
\label{eqn:Hamiltonian of N-HO in App}
\end{eqnarray}
where $\bm{p}_{j} \equiv \left(p_{jx},p_{jy},p_{jz}\right)$ is the
momentum operator of particle-$j$, $\bm{p}_{j} 
\equiv -{\rm i}\hbar \partial /\partial\bm{r}_{j}$. 
Since there is no interaction between particles, the eigenstates are 
expressed as
\begin{equation}
\Psi (\bm{r}_{1},\bm{r}_{2},\cdot \cdot \cdot ,\bm{r}_{N})
=
\prod_{j=1}^{N}\phi (\bm{r}_{j}),
\label{eqn:wavefunc. of HO system in App}
\end{equation}
and a single-particle state $\phi (\bm{r})$ satisfies
\begin{equation}
\left[
-\frac{\hbar^{2}}{2m}\left(
 \frac{\partial^{2}}{\partial x^{2}}
+\frac{\partial^{2}}{\partial y^{2}}
+\frac{\partial^{2}}{\partial z^{2}}
\right)
+ \frac{m}{2}\left(
\omega_{x}^{2}x^{2}+\omega_{y}^{2}y^{2}+\omega_{z}^{2}z^{2}
\right)
\right]\phi (\bm{r})
= 
\epsilon \phi (\bm{r}).
\label{eqn:Schroedinger eq. for HO system in App}
\end{equation}
The eigenvalues of (\ref{eqn:Schroedinger eq. for HO system in App}) 
is well-known; 
\begin{equation}
\epsilon_{n_{x}n_{y}n_{z}} 
= \left(n_{x} + \frac{1}{2}\right) \hbar \omega_{x}
+ \left(n_{y} + \frac{1}{2}\right) \hbar \omega_{y}
+ \left(n_{z} + \frac{1}{2}\right) \hbar \omega_{z},
\label{eqn:eigenvalues of HO system in App}
\end{equation}
where $n_{x}, n_{y}, n_{z} = 0, 1, 2, \cdot \cdot \cdot$.

  We adopt the grand canonical ensemble. The total 
particle number $N$  and the total energy $E$ are
given by
\begin{eqnarray}
N & = & \sum_{n_{x}, n_{y}, n_{z}} \left[
\exp \beta \left(\epsilon_{n_{x}n_{y}n_{z}} - \mu \right)
- 1
\right]^{-1},
\label{eqn:N of HO in App}
\\
E & = & \sum_{n_{x}, n_{y}, n_{z}} 
\epsilon_{n_{x}n_{y}n_{z}}\left[
\exp \beta \left(\epsilon_{n_{x}n_{y}n_{z}} - \mu \right)
- 1
\right]^{-1},
\label{eqn:E of HO in App}
\end{eqnarray}
where $\beta = \left(k_{{\rm B}} T\right)^{-1}$ and $\mu$
is the chemical potential. For later discussions,
we shift the chemical potential as 
$\mu - \hbar \left( \omega_{x}+\omega_{y}+\omega_{z}\right)/2
\rightarrow \mu$. 
At very low temperature, (\ref{eqn:N of HO in App}) 
is written in the following form, 
\begin{equation}
N = N_{0} + N_{1},
\label{eqn:N=N0+N1 in App}
\end{equation}
where
\begin{eqnarray}
N_{0} & = & 1/\left(e^{-\beta \mu} - 1\right),
\label{eqn:N0 of HO in App}
\\
N_{1} & = & \sum_{n_{x}, n_{y}, n_{z} \neq 0} \left[
\exp \beta \left( 
n_{x}\hbar \omega_{x} + n_{y} \hbar \omega_{y} +
n_{z} \hbar \omega_{z}
\right)
- 1
\right]^{-1}.
\label{eqn:N1 of HO in App}
\end{eqnarray}
Discussions, here and in what follows, are essentially the same 
as those for a free boson gas in a box. 
The Bose-Einstein condensation is the situation
where $N_{0}$ becomes macroscopic, that is, $N_{0} \sim O (N)$. 
Equation (\ref{eqn:N0 of HO in App}) implies $\mu
\leq 0$ and $-\beta \mu \sim O (1/N)$.
We have set $\mu =0$ in (\ref{eqn:N1 of HO in App}). 
Therefore, $N_{1}$ in (\ref{eqn:N1 of HO in App})
gives the maximum of the contribution from the
cxcited states at low temperature.
 
   We replace the summations in (\ref{eqn:N1 of HO in App})
by an integral over the single-particle energy
$\epsilon$ with the density of states $D(\epsilon )$,
\begin{equation}
D(\epsilon ) {\rm d}\epsilon 
= \frac{\epsilon^{2}}{2(\hbar \bar{\omega})^{3}}{\rm d}\epsilon .
\label{eqn:DOS in App}
\end{equation}
The formula (\ref{eqn:DOS in App}) can be derived
as follows. We estimate the number of the states,
$N(\epsilon)$, whose energies are smaller than $\epsilon$,
which is equivalent to the number of positive
integer sets $\{n_{x}, n_{y}, n_{z}\}$ satisfying 
$0 < n_{x}\hbar \omega_{x} + n_{y} \hbar \omega_{y} +
n_{z} \hbar \omega_{z} \leq \epsilon$.
Geometrically, this corresponds to a volume, 
$(\epsilon /\hbar \omega_{x}) (\epsilon /\hbar \omega_{y}) 
(\epsilon /\hbar \omega_{z})/6 $.
Therefore, we obtain $N(\epsilon) = \frac{1}{6}
\epsilon^{3}/(\hbar \bar{\omega})^{3}$, where 
$\bar{\omega}^{3} \equiv \omega_{x}\omega_{y}\omega_{z}$.
A relation $D(\epsilon) = {\rm d}N(\epsilon)/{\rm d}\epsilon$
gives (\ref{eqn:DOS in App}).

   Using (\ref{eqn:DOS in App}) to rewrite the
summations in (\ref{eqn:N1 of HO in App}), we obtain
\begin{eqnarray}
N_{1} & = & \int_{0}^{\infty}\frac{D(\epsilon){\rm
    d}\epsilon}{\exp (\beta \epsilon) - 1}
\nonumber \\
& = & \left(\frac{k_{{\rm B}}T}{\hbar
    \bar{\omega}}\right)^{3} \zeta (3),
\,\,\,\,\,\, T \leq T_{{\rm c}},
\label{eqn:N1 of HO No 2 in App}
\end{eqnarray}
where
\begin{equation}
\zeta (3)  =  \int_{0}^{\infty}\frac{x^{2}{\rm
    d}x}{e^{x} - 1}
= 1.202\cdot \cdot \cdot . 
\label{eqn:zeta(3) in App}
\end{equation}
The transition temperature is defined by 
\begin{equation}
N = 
 \zeta (3)
\left(\frac{k_{{\rm B}} T_{{\rm c}}}{\hbar
    \bar{\omega}}\right)^{3}.
\label{eqn:Tc in App}
\end{equation}
For $T \leq T_{{\rm c}}$, we have
\begin{equation}
N = N_{0} + \zeta (3) \left(\frac{k_{{\rm B}} T}{\hbar
    \bar{\omega}}\right)^{3}, 
\label{eqn:N=N0+N1 No2 in App}
\end{equation}
and therefore
\begin{equation}
\frac{N_{0}}{N} 
= 1 - \left(\frac{T}{T_{{\rm c}}}\right)^{3},
\,\,\,\,\,\, T \leq T_{{\rm c}}.
\label{eqn:N0 vs Tc in App}
\end{equation}
The exponent is changed from 3/2 (confined in a box)
to 3 (confined in harmonic potentials).

  The total energy $E$ in (\ref{eqn:E of HO in App})
is calculated in the thermodynamic limit~\cite{GrHoSe};
$N \rightarrow \infty$, $\bar{\omega} \rightarrow 0$,
$N \bar{\omega}^{3} = {\rm fixed}$. The result is 
\begin{eqnarray}
E & = & 3 N k_{{\rm B}} T\frac{g_{4}(z)}{g_{3}(z)},
\,\,\,\,\,\, T > T_{{\rm c}},
\nonumber \\
& = & 3 \frac{(k_{{\rm B}}T)^{4}}{(\hbar \bar{\omega})^{3}} \zeta (4),
\,\,\,\,\,\, T \leq T_{{\rm c}},
\label{eqn:E of HO No 2 in App}
\end{eqnarray}
where with $z = \exp (\beta \mu )$
\begin{equation}
g_{n} (z) \equiv \sum_{l = 1}^{\infty}\frac{z^{l}}{l^{n}},
\label{eqn:def of gn in App}
\end{equation}
and $\zeta (4) = \pi^{4}/90 = 1.082 \cdot \cdot \cdot$.
The specific heat $C= {\rm d}E/{\rm d}T$ has a
discontinuity, $\Delta C/N k_{{\rm B}} 
= 9\zeta (3)/\zeta (2) = 6.6$. The transition is of 
second order. It is well-known that for a free
boson in a box the derivative of the specific heat
is discontinuous and the transition is of 3rd order
(see for instance, Ref.~\cite{Huang}).

\clearpage

\chapter{Pseudopotential}
\label{chap:App1}
\setcounter{equation}{0}
   In this appendix, we give some details of deriving 
the pseudopotential. The argument  follows Refs.~\cite{Huang,HY}.

    We consider the two-body problem. Each particle has the 
mass, $m$, and an inter-particle potential, $v(\bm{r})$, is 
the ``hard-sphere'' one, 
\begin{equation}
v(\bm{r}) = \left\{
\begin{array}{ll}
0 & (r > a) \\
\infty & (r \leq a),
\end{array}
\right.
\label{eqn:hard sphere inter-particle potential in BecStabRev}
\end{equation}
where $a$ is the hard-sphere diameter with $\bm{r}$ the relative
position vector between two particles and $r=|\bm{r}|$.
The Schr\"odinger equation in the center-of-mass system is
\begin{equation}
\frac{\hbar^{2}}{2\mu}\left(\Delta + k^{2}\right) \psi (\bm{r}) 
=
v(\bm{r}) \, \psi (\bm{r}) 
,
\label{eqn:Schroedinger eq for 2-body system in App}
\end{equation}
where  $\mu$ means the reduced mass, 
\begin{equation}
\mu = m/2.
\label{eqn:reduced mass in App}
\end{equation}
Obviously, $\psi (\bm{r})$ is the wavefunction 
in the center-of-mass coordinate system,
and $(\hbar k)^{2}/(2\mu)$ is the energy of the relative motion. 
Substituting 
(\ref{eqn:hard sphere inter-particle potential in BecStabRev}) 
into (\ref{eqn:Schroedinger eq for 2-body system in App}), we have
\begin{eqnarray}
\left(\Delta + k^{2}\right) \psi (\bm{r}) & = & 0 \,\,\,\,\, (r > a),
\nonumber \\
\psi (\bm{r}) & = & 0 \,\,\,\,\, (r \leq a).
\label{eqn:Schroedinger eq for 2-body system with hard sphere in App}
\end{eqnarray}
In terms of  the spherical coordinate,
\begin{equation}
\bm{r} 
= 
\left(
r \sin \theta \cos \phi ,
r \sin \theta \sin \phi ,
r \cos \theta
\right),
\label{eqn:def of spherical cooridinate in App}
\end{equation}
the solution of Eq.~(\ref{eqn:Schroedinger eq for 2-body system with
  hard sphere in App}) for $r > a$ can be written as 
\begin{equation}
\psi (\bm{r}) =
\sum_{l = 0}^{\infty} \sum_{m = -l}^{+l} Y_{lm}(\theta,\phi)
A_{lm}\left(j_{l}(kr) - \tan \eta_{l} \, n_{l}(kr)\right),
\label{eqn:sol of Schroedinger eq for 2-body system with hard sphere in App}
\end{equation}
with the boundary condition,
\begin{equation}
\psi (\bm{r})|_{r = a}  =  0.
\label{eqn:bound. cond. for psi in App}
\end{equation}
Here $Y_{lm}(\theta,\phi)$ is a normalized spherical harmonic function, 
$j_{l}(x)$ and $n_{l}(x)$ the spherical Bessel and Neumann
functions respectively, and $A_{lm}$ and $\eta_{l}$ constants.
We note that the constant $\eta_{l}$ is determined by the condition 
(\ref{eqn:bound. cond. for psi in App}) as 
\begin{equation}
\tan \eta_{l} = j_{l}(ka)/n_{l}(ka).
\label{eqn:determination of etal in App}
\end{equation}
The scattering length $a_{l}$ for the partial $l$-wave is defined by 
\begin{equation}
a_{l} \equiv - \lim_{k \rightarrow 0} \tan \eta_{l}(k)/k.
\label{eqn:al in App}
\end{equation}
In what follows,
we assume that the energy of the relative motion 
$(\hbar k)^{2}/(2\mu)$ is sufficiently small, 
and thus we consider a spherically symmetric ($s$-wave) solution, 
\begin{equation}
\psi (\bm{r}) 
=
A\left(j_{0}(kr) - \tan \eta_{0} \, n_{0}(kr)\right),
\label{eqn:s-wave sol for 2-body system with hard sphere in App}
\end{equation}
where 
\begin{eqnarray}
j_{0}(x) &=& \sin x/x,
\label{eqn:j0 in App}
\\
n_{0}(x) &=& -\cos x/x,
\label{eqn:n0 in App}
\\
A &\equiv& A_{00}/\sqrt{4 \pi}.
\label{eqn:rel A and A00 in App}
\end{eqnarray}
From (\ref{eqn:determination of etal in App}),
(\ref{eqn:j0 in App}) and (\ref{eqn:n0 in App}), we have
\begin{equation}
\tan \eta_{0} = -\tan (ka),
\label{eqn:rel between eta0 and etaka in App}
\end{equation}
leading to 
\begin{equation}
\eta_{0} = -ka.
\label{eqn:rel between eta0 and etaka No. 2 in App}
\end{equation}
Thus, $a$ is identified with the $s$-wave scattering length.

    An idea of the pseudopotential is as follows: we find an equation with 
some ``potential'' such that (\ref{eqn:s-wave sol for 2-body system
  with hard sphere in App}) is the solution everywhere.
For sufficiently small $x$, $j_{0}(x)$ and $n_{0}(x)$ behave like
\begin{equation}
j_{0}(x) \approx 1, \,\,\,\,\, 
n_{0}(x) \approx -1/x \,\,\,\,\, (x \ll 1).
\label{eqn:n0 for small x in App}
\end{equation}
Thus, from (\ref{eqn:s-wave sol for 2-body system with hard sphere in App}),
for sufficiently small $kr$, we get
\begin{equation}
r \psi (\bm{r}) 
=
A\left(r + \frac{\tan \eta_{0}}{k}\right),
\label{eqn:s-wave sol for small kr  with hard sphere in App}
\end{equation}
which gives
\begin{equation}
A
=
\frac{\partial}{\partial r}\left(r \psi (\bm{r})\right).
\label{eqn:Value of A in App}
\end{equation}
Remark that the relation (\ref{eqn:Value of A in App}) is used only 
at $\bm{r} = 0$.
Because  $j_{0}(x)$ is regular at $x=0$, $j_{0}(kr)$ satisfies
\begin{equation}
(\Delta + k^{2}) j_{0}(kr) = 0,
\label{eqn:eq. for j0 in App}
\end{equation}
for all $r$. 
On the other hand, $n_{0}(x)$ is singular at $x=0$.
Then, we calculate 
\begin{equation} 
F_{0}(r)
\equiv
(\Delta + k^{2}) n_{0}(kr),
\label{eqn:def of F0 in App}
\end{equation}
with care.
We integrate $F_{0}(r)$ over a sphere $V$ of radius
$\epsilon$ about the origin.
From (\ref{eqn:def of F0 in App}), we have
\begin{equation}
\int_{V} {\rm d}^{3} \bm{r} \, F_{0}(r)
= 
\int_{V} {\rm d}^{3}\bm{r} \, \Delta n_{0}(kr)
+ k^{2}\int_{V} {\rm d}^{3}\bm{r} \, n_{0}(kr).
\label{eqn:integration of F0 in App}
\end{equation}
By applying the divergence theorem to the first term in the right hand side 
of Eq.~(\ref{eqn:integration of F0 in App}), we get
\begin{eqnarray}
\int_{V} {\rm d}^{3}\bm{r} \, \Delta n_{0}(kr)
& = &
\int_{\partial V} {\rm d}\bm{S} \cdot \nabla n_{0}(kr)
\nonumber \\
& = &
4 \pi \epsilon^{2}\left.
\frac{\partial}{\partial r} n_{0}(kr)
\right|_{r=\epsilon} 
\nonumber \\
& = &
4 \pi \epsilon \sin (k\epsilon) + \frac{4 \pi}{k} \cos (k\epsilon).
\label{eqn:integration of 1st term of F0 in App}
\end{eqnarray}
The second term in Eq.~(\ref{eqn:integration of F0 in App}) gives
\begin{eqnarray}
k^{2}\int_{V} {\rm d}^{3}\bm{r} \, n_{0}(kr)
& = & 4 \pi \int_{0}^{\epsilon} r^{2} {\rm d}r \, 
\left(\frac{-\cos (kr)}{kr}\right)
\nonumber \\
& = & -4\pi \epsilon \sin (k\epsilon) 
- \frac{4\pi}{k}\cos (k\epsilon) 
+ \frac{4\pi}{k}.
\label{eqn:integration of 2nd term of F0 in App}
\end{eqnarray}
Substituting (\ref{eqn:integration of 1st term of F0 in App})
and (\ref{eqn:integration of 2nd term of F0 in App}) into 
(\ref{eqn:integration of F0 in App}), we obtain
\begin{equation}
\int_{V} {\rm d}^{3} \bm{r} \, F_{0}(r)
= \frac{4\pi}{k}.
\label{eqn:integration of F0 No. 2 in App}
\end{equation}
Noting that $F_{0}(r)$ is identically equal to zero for $r \neq 0$,
we conclude from (\ref{eqn:integration of F0 No. 2 in App}) that
\begin{equation}
F_{0}(r)
=
(\Delta + k^{2}) n_{0}(kr)
=\frac{4\pi}{k}\delta (\bm{r}).
\label{eqn:rel between F0 and delta-func. in App}
\end{equation}
Using (\ref{eqn:rel between eta0 and etaka in App}),
(\ref{eqn:Value of A in App}),
(\ref{eqn:eq. for j0 in App})
and (\ref{eqn:rel between F0 and delta-func. in App})
in Eq.~(\ref{eqn:s-wave sol for 2-body system with hard sphere in App}),
we have an equation that the solution 
(\ref{eqn:s-wave sol for 2-body system with hard sphere in App})
satisfies everywhere,
\begin{equation}
(\Delta + k^{2}) \psi (\bm{r})
=\frac{4\pi}{k}  \tan (ka) \, \delta (\bm{r}) \, 
\frac{\partial}{\partial r}\left(r \psi (\bm{r})\right). 
\label{eqn:Eq for psi for all r in App}
\end{equation}
For sufficiently small $ka$, we can replace $\tan (ka)$
by $ka$. Then, by dividing the both sides 
of Eq.~(\ref{eqn:Eq for psi for all r in App}) by $\hbar^{2}/(2 \mu)$,
we finally arrive at 
\begin{equation}
-\frac{\hbar^{2}}{2\mu} \Delta \psi (\bm{r}) 
+ \tilde{v} (\bm{r}) \, \psi (\bm{r}) 
= 
\frac{\hbar^{2}}{2\mu} k^{2} \psi (\bm{r}) 
,
\label{eqn:Eq for psi for all r No. 2 in App}
\end{equation}
where 
\begin{equation}
\tilde{v} (\bm{r})
\equiv
\frac{4\pi\hbar^{2}a}{m} \delta (\bm{r}) \, 
\frac{\partial}{\partial r}\left(r \, \cdot \right). 
\label{eqn:def of pseudopotential in App}
\end{equation}
The operator $\tilde{v} (\bm{r})$
(\ref{eqn:def of pseudopotential in App}) is known
as the  pseudopotential~\cite{Huang,HY}.
We note that $\partial/\partial r\left(r \, \cdot \right)$
appearing in (\ref{eqn:def of pseudopotential in App})
is not a hermitian operator.
But, if $\psi (\bm{r})$ is well behaved, namely 
differentiable at the origin,
we can replace $\partial/\partial r\left(r \, \cdot \right)$ by unity. 
So far, we have considered $a$ to be positive.
In general, however, the ``diameter'' of the hard-sphere $a$ 
can  be extended to be negative. 
This occurs when we may replace the low energy scattering 
from an attractive inter-particle potential 
of finite range by that from a hard-sphere one,
known as the ``shape-independent approximation.''

\clearpage

\chapter{The Ground State Energy under the Thomas-Fermi Approximations}
\label{chap:App2}
\setcounter{equation}{0}

  We give some details of calculations of the ground state energy
of the condensate
under the one-, two- and three-dimensional Thomas-Fermi approximations.

  First, we consider the one-dimensional case. We start from 
the one-dimensional
Ginzburg-Pitaevskii-Gross equation~\cite{GinzPita}--\cite{Gross2}
with a harmonic potential term,
\begin{equation}
-\frac{\hbar^{2}}{2m}\frac{\partial^{2} \Psi}{\partial x^{2}}  
+ \frac{m}{2}\omega^{2}x^{2}\Psi 
+ g |\Psi |^{2} \Psi = \mu \Psi.
\label{eqn:1D GP eq in AsymBec}
\end{equation}
Here and hereafter, $g (>0)$ means the strength of the inter-atomic 
interaction, and $\mu$ is the chemical potential.
In the Thomas-Fermi approximation, the first term (the kinetic term)
in the left hand
side of Eq.~(\ref{eqn:1D GP eq in AsymBec}) is neglected, which gives
the number density of the condensate, $|\Psi|^{2} $, as
\begin{equation}
|\Psi|^{2} = \frac{1}{g}\left(\mu - \frac{1}{2}m\omega^{2}x^{2}\right). 
\label{eqn:1D density in AsymBec}
\end{equation}
By integrating the density (\ref{eqn:1D density in AsymBec})
in the interval $[-x_{0},x_{0}]$, where $x_{0} \equiv 
\sqrt{2\mu/(m \omega^{2})}$, we get the number of the particles $N$,
\begin{equation}
N = 
\int_{-x_{0}}^{x_{0}} {\rm d} x 
|\Psi|^{2}
=
\frac{4}{3g}\left(\frac{2}{m \omega^{2}}\right)^{1/2}
\mu^{3/2},
\label{eqn:N in 1D in AsymBec}
\end{equation}
leading to 
\begin{equation}
\mu = \left(\frac{3gN}{4}\right)^{2/3}
\left(\frac{m \omega^{2}}{2}\right)^{1/3}.
\label{eqn:mu in 1D in AsymBec}
\end{equation}
Substituting (\ref{eqn:mu in 1D in AsymBec}) into 
the thermodynamic identity,
\begin{equation}
\mu = \frac{\partial E}{\partial N},
\label{eqn:thermodynamic identity in AsymBec}
\end{equation}
we have the energy, $E_{\rm{1D}}$, given by
\begin{equation}
E_{\rm{1D}} = 
\frac{3}{5}
\left(\frac{3g}{4}\right)^{2/3}
\left(\frac{m \omega^{2}}{2}\right)^{1/3}
N^{5/3}.
\label{eqn:E in 1D in AsymBec}
\end{equation}


    Next, we consider the two-dimensional case.
The two-dimensional equation is 
\begin{equation}
-\frac{\hbar^{2}}{2m}
\left(
\frac{\partial^{2}}{\partial x^{2}}  
+ \frac{\partial^{2}}{\partial y^{2}}
\right)
\Psi
+ \frac{m}{2}
(\omega_{x}^{2}x^{2} + \omega_{y}^{2}y^{2})
\Psi 
+ g |\Psi |^{2} \Psi = \mu \Psi.
\label{eqn:2D GP eq in AsymBec}
\end{equation}
According to the Thomas-Fermi approximation,
we ignore the kinetic energy terms in the left hand side of 
(\ref{eqn:2D GP eq in AsymBec}) and get
\begin{equation}
|\Psi|^{2} = \frac{1}{g}\left( \mu - \frac{1}{2}m\omega_{x}^{2}x^{2}
- \frac{1}{2}m\omega_{y}^{2}y^{2} \right). 
\label{eqn:2D density in AsymBec}
\end{equation}
By integrating (\ref{eqn:2D density in AsymBec}) in the region 
$\mu - \frac{1}{2}m\omega_{x}^{2}x^{2} - \frac{1}{2}m\omega_{y}^{2}y^{2} > 0$, 
we have the number of particles $N$,
\begin{equation}
N =  
\int_{-x_{0}}^{x_{0}} {\rm d} x 
\int_{-y_{0}}^{y_{0}} {\rm d} y 
|\Psi|^{2},
\label{eqn:N in 2D No1 in AsymBec}
\end{equation}
where $x_{0}$ and $y_{0}$ are defined as
\begin{equation}
x_{0}  \equiv  \left(\frac{2\mu}{m \omega_{x}^{2}}\right)^{1/2},
\,\,\,\,\,\,\,
y_{0}  \equiv  \left(\frac{2}{m \omega_{y}^{2}}\right)^{1/2}
\left(\mu - \frac{1}{2}m\omega_{x}^{2}x^{2}\right)^{1/2}.
\label{eqn:def of x0 and y0 in 2D in AsymBec}
\end{equation}
Substituting (\ref{eqn:2D density in AsymBec}) 
and (\ref{eqn:def of x0 and y0 in 2D in AsymBec})
into (\ref{eqn:N in 2D No1 in AsymBec}), we have
\begin{equation}
N
=
\frac{4}{3g}\left(\frac{2}{m \omega_{y}^{2}}\right)^{1/2}
\left(\frac{2}{m \omega_{x}^{2}}\right)^{1/2}
\frac{3 \pi}{8} \mu^{2},
\label{eqn:N in 2D in AsymBec}
\end{equation}
from which we get
\begin{equation}
\mu = \left(\frac{3gN}{4}\right)^{1/2}
\left(\frac{m \omega_{x}^{2}}{2}\right)^{1/4}
\left(\frac{m \omega_{y}^{2}}{2}\right)^{1/4}
\left(\frac{8}{3 \pi}\right)^{1/2}.
\label{eqn:mu in 2D in AsymBec}
\end{equation}
Using (\ref{eqn:mu in 2D in AsymBec}) 
in (\ref{eqn:thermodynamic identity in AsymBec}), we obtain 
the energy in the two-dimensional case, $E_{\rm{2D}}$,
\begin{equation}
E_{\rm{2D}} 
= 
\frac{2}{3}
\left(\frac{8}{3 \pi}\right)^{1/2}
\left(\frac{3g}{4}\right)^{1/2}
\left(\frac{m \omega_{x}^{2}}{2}\right)^{1/4}
\left(\frac{m \omega_{y}^{2}}{2}\right)^{1/4}
N^{3/2}.
\label{eqn:E in 2D in AsymBec}
\end{equation}


    Finally, we consider the three-dimensional case.
The three-dimensional equation is 
\begin{equation}
-\frac{\hbar^{2}}{2m}
\left(
\frac{\partial^{2}}{\partial x^{2}}  
+ \frac{\partial^{2}}{\partial y^{2}}
+ \frac{\partial^{2}}{\partial z^{2}}
\right)
\Psi
+ \frac{m}{2}
(\omega_{x}^{2}x^{2} + \omega_{y}^{2}y^{2} + \omega_{z}^{2}z^{2})
\Psi 
+ g |\Psi |^{2} \Psi = \mu \Psi.
\label{eqn:3D GP eq in AsymBec}
\end{equation}
As in the previous cases, we approximate the number density 
of the condensate as
\begin{equation}
|\Psi|^{2} = \frac{1}{g}\left( \mu - \frac{1}{2}m\omega_{x}^{2}x^{2}
- \frac{1}{2}m\omega_{y}^{2}y^{2}
- \frac{1}{2}m\omega_{z}^{2}z^{2} \right). 
\label{eqn:3D density in AsymBec}
\end{equation}
The number of particles $N$ is given by
\begin{equation}
N =  
\int_{-x_{0}}^{x_{0}} {\rm d} x 
\int_{-y_{0}}^{y_{0}} {\rm d} y 
\int_{-z_{0}}^{z_{0}} {\rm d} z 
|\Psi|^{2},
\label{eqn:N in 3D No1 in AsymBec}
\end{equation}
where $x_{0}$, $y_{0}$ and $z_{0}$ are defined as
\begin{eqnarray}
x_{0} & \equiv &
\left(\frac{2\mu}{m \omega_{x}^{2}}\right)^{1/2},
\,\,\,\,\,\,\,
y_{0}  \equiv  \left(\frac{2}{m \omega_{y}^{2}}\right)^{1/2}
\left(\mu - \frac{1}{2}m\omega_{x}^{2}x^{2}\right)^{1/2},
\nonumber
\\
z_{0} & \equiv & 
\left(\frac{2}{m \omega_{z}^{2}}\right)^{1/2}
\left(\mu - \frac{1}{2}m\omega_{x}^{2}x^{2}
- \frac{1}{2}m\omega_{y}^{2}y^{2}\right)^{1/2}.
\label{eqn:def of x0, y0 and z0 in 3D in AsymBec}
\end{eqnarray}
Substituting (\ref{eqn:3D density in AsymBec}) and 
(\ref{eqn:def of x0, y0 and z0 in 3D in AsymBec}) into
(\ref{eqn:N in 3D No1 in AsymBec}), we obtain
\begin{equation}
N
=
\frac{8\pi}{15g}
\left(\frac{2}{m \omega_{z}^{2}}\right)^{1/2}
\left(\frac{2}{m \omega_{y}^{2}}\right)^{1/2}
\left(\frac{2}{m \omega_{x}^{2}}\right)^{1/2}
\mu^{5/2}
,
\label{eqn:N in 3D in AsymBec}
\end{equation}
which gives
\begin{equation}
\mu =
\left(\frac{15g}{8\pi}\right)^{2/5}
\left(\frac{m \omega_{x}^{2}}{2}\right)^{1/5}
\left(\frac{m \omega_{y}^{2}}{2}\right)^{1/5}
\left(\frac{m \omega_{z}^{2}}{2}\right)^{1/5}
N^{2/5}.
\label{eqn:mu in 3D in AsymBec}
\end{equation}
From (\ref{eqn:thermodynamic identity in AsymBec})
and (\ref{eqn:mu in 3D in AsymBec}), we have 
the energy in the three-dimensional case, denoted by $E_{\rm{3D}}$,
\begin{equation}
E_{\rm{3D}} 
 = 
\frac{5}{7}
\left(\frac{15g}{8\pi}\right)^{2/5}
\left(\frac{m \omega_{x}^{2}}{2}\right)^{1/5}
\left(\frac{m \omega_{y}^{2}}{2}\right)^{1/5}
\left(\frac{m \omega_{z}^{2}}{2}\right)^{1/5}
N^{7/5}.
\label{eqn:E in 3D in AsymBec}
\end{equation}

   To summarize, the ground state energy in the Thomas-Fermi
approximation for the $d$-dimensional Ginzburg-Pitaevskii-Gross
equation with harmonic potential terms has the particle number dependence as 
\begin{equation}
E_{d{\rm D}} \sim N^{(d+4)/(d+2)}.
\label{eqn:EdD in AsymBec}
\end{equation}
This $N$-dependence can be used to identify the effective dimensionality of 
the condensate under the anisotropic magnetic traps in the strongly
repulsive case.

\clearpage

\chapter*{Table and Figure Captions}
\label{chap:FigCap}
\setcounter{equation}{0}
\setcounter{figure}{0}

\bigskip
\noindent 
{\bf Table~\ref{tab:BEC}} \,\,\,
Experimental values of parameters. $a$; $s$-wave scattering
length, $T_{{\rm c}}$; critical temperature, $N_{{\rm t}}$; 
the total number of 
atoms in trap, $\omega$; trap frequency. Note that the total number of the
atoms is the sum of normal and condensed ones.
The values of $T_{{\rm c}}$, $N_{{\rm t}}$ and $\omega$ are taken from 
the reference in the right most column.

\bigskip
\noindent 
{\bf Fig.~\ref{fig:PD}} \,\,\,
Approximations for the repulsive inter-atomic interaction:
I. weak interaction case (\ref{eqn:cond for both gaussian approx No 2
in Asym Bec}), II. strong interaction case
(\ref{eqn:cond for both TFA approx No 2 in Asym Bec}), 
III. intermediate case-1 
(\ref{eqn:cond for bot gaussian and z TFA approx No 2 in Asym Bec}), 
IV. intermediate case-2 
(\ref{eqn:cond for bot TFA and z gaussian approx No 2 in Asym Bec}).
The abscissa and ordinate represent the anisotropy of the trap, 
$\delta$, and the strength of the interaction, $G_{\bot}$, respectively.

\bigskip
\noindent 
{\bf Fig.~\ref{fig:31}} \,\,\,
Stability of the two-component condensate:
$N_1=N_2$ case for (a) $\alpha > 0$, (b) $\alpha < 0$.
The boundaries of stable and unstable regions (the solid lines) are
determined by Eqs.~(\protect\ref{eq:hessian}) and (\protect\ref{eq:31}).
The dashed lines are discussed in \protect\ref{sec:Phase Separation in 2com}.

\bigskip
\noindent 
{\bf Fig.~\ref{fig:41}} \,\,\,
  Phase diagram for $\alpha_{11}=\alpha_{22}\equiv\alpha>0$ 
and $\alpha_{12}<-\alpha$.

\bigskip
\noindent 
{\bf Fig.~\ref{fig:42}} \,\,\,
Phase diagram for $\alpha_{11}=\alpha_{22}\equiv\alpha<0$. 
(a) $\alpha_{12}>0$, (b) $\alpha<\alpha_{12}<0$, 
(c) $\alpha_{12}<\alpha<0$.

\bigskip
\noindent 
{\bf Fig.~\ref{fig:43}} \,\,\,
  Function $\gamma(\beta)$ defined in (\protect\ref{eq:4310}).
We set $\alpha_{11} : \alpha_{22} : \alpha_{12} = 1 : -4 :2^{-\frac{3}{2}}$.

\bigskip
\noindent 
{\bf Fig.~\ref{fig:44}} \,\,\,
$\lim_{N_1 \to \infty} f(N_1)/N_1$ as a function of 
$2^{\frac{5}{2}}\alpha_{12}/\alpha_{11}$.
We set $\alpha_{11} : \alpha_{22} = 1 : -4$.

\bigskip
\noindent 
{\bf Fig.~\ref{fig:45}} \,\,\,
Phase diagram for $\alpha_{11}>0$ and $\alpha_{22}<0$.
(a) $\alpha_{12}>2^{-\frac{5}{2}}\alpha_{11}$, 
(b) $0<\alpha_{12}<2^{-\frac{5}{2}}\alpha_{11}$,
(c) $\alpha_{12}<0$.

\bigskip
\noindent 
{\bf Fig.~\ref{fig:Li}} \,\,\,
$N_{{\rm c, new}}/N_{0}$ as a function of
$k$ for the case that $\delta = 0.718$~\protect\cite{BSTH}.

\clearpage

\pagestyle{empty}

\vspace{5mm}

\begin{center}

\begin{tabular}{|c|c|c|c|c|c|} \hline

atom & $a$ [m] & $T_{\rm c}$ [K] & $N_{\rm t}$ & $\omega$ [Hz] & Reference \\ \hline

$^{1}{\rm H}$  
&   $7.2 \times 10^{-11}$   
& $5 \times 10^{-5}$ 
& $2 \times 10^{10}$          
& 
\begin{tabular}{c}
$\omega_{x}=2 \pi \times 3.90 \times 10^{3}$, \\
$\omega_{y}=2 \pi \times 3.90 \times 10^{3}$, \\
$\omega_{z}=2 \pi \times 10.2$
\end{tabular}
&\protect\cite{Hydrogen}
\\ \hline

$^{7}{\rm Li}$ 
& $-1.44 \times 10^{-9}$    
& $3 \times 10^{-7}$       
& $1 \times 10^{5}$          
& 
\begin{tabular}{c}
$\omega_{x}=2 \pi\times 150.6$, \\
$\omega_{y}=2 \pi \times 152.6$, \\
$\omega_{z}=2 \pi \times 131.5$ 
\end{tabular}
&\protect\cite{BSH2}
\\ \hline

$^{23}{\rm Na}$
&   $2.75 \times 10^{-9}$    
& $2 \times 10^{-6}$       
& $7 \times 10^{5}$          
& 
\begin{tabular}{c}
$\omega_{x}=2 \pi\times 745$, \\
$\omega_{y}=2 \pi \times 235$, \\
$\omega_{z}=2 \pi \times 410$ 
\end{tabular}
&\protect\cite{DMADDKK}
\\ \hline

$^{87}{\rm Rb}$ 
&
$5.77 \times 10^{-9}$
& $1.7 \times 10^{-7}$    
& $2 \times 10^{4}$          
& 
\begin{tabular}{c}
$\omega_{x}=2 \pi \times 120/\sqrt{8}$, \\
$\omega_{y}=2 \pi \times 120/\sqrt{8}$, \\
$\omega_{z}=2 \pi \times 120$ 
\end{tabular}
&\protect\cite{AEMWC}
\\  \hline

$^{87}{\rm Rb}$ 
&
$5.77 \times 10^{-9}$
& $4.3 \times 10^{-7}$    
& $1.5 \times 10^{6}$          
& 
\begin{tabular}{c}
$\omega_{x}=2 \pi \times 64$, \\
$\omega_{y}=2 \pi \times 64$, \\
$\omega_{z}=2 \pi \times 181$ 
\end{tabular}
&\protect\cite{HWCH}
\\  \hline

$^{87}{\rm Rb}$ 
&
$5.77 \times 10^{-9}$
&
$5.5 \times 10^{-7}$ 
& $1 \times 10^{6}$          
& 
\begin{tabular}{c}
$\omega_{x}=2 \pi \times 280$, \\
$\omega_{y}=2 \pi \times 280$, \\
$\omega_{z}=2 \pi \times 24$ 
\end{tabular}
&\protect\cite{ErMaScScRe}
\\  \hline

$^{87}{\rm Rb}$ 
&
$5.77 \times 10^{-9}$
& 
$5\times 10^{-7}$ 
& 
$5 \times 10^{5}$          
& 
\begin{tabular}{c}
$\omega_{x}=2 \pi \times 20$, \\
$\omega_{y}=2 \pi \times 200$, \\
$\omega_{z}=2 \pi \times 200$ 
\end{tabular}
&\protect\cite{EsBlHa}
\\  \hline

\end{tabular}

\end{center}


\end{document}